\documentclass[preprint2]{aastex}

\newcommand{\noprint}[1]{}

\newcommand{\degdot}{\ensuremath{^\circ\hspace{-.4em}{.}}\hspace{.08em}}
\newcommand{\arcseccomma}{\ensuremath{''\hspace{-.4em}{.}}\hspace{.11em}}

\newcommand{\subscript}[1]{\ensuremath{_{\mbox{\smaller[2]#1}}}}
\newcommand{\superscript}[1]{\ensuremath{^{\mbox{\smaller[2]#1}}}}

\usepackage{amsmath} 
\usepackage{relsize}
\usepackage{rotating}

\shorttitle{Improved variable star search}
\shortauthors{Fruth et al.}

\begin{document}

\title{IMPROVED VARIABLE STAR SEARCH \\IN LARGE PHOTOMETRIC DATA SETS -- \\
 NEW VARIABLES IN \textit{CoRoT} FIELD LRa02 \\DETECTED BY BEST II}

\author{\textsc{T.~Fruth\altaffilmark{1}, P.~Kabath\altaffilmark{2}, J.~Cabrera\altaffilmark{1}, R.~Chini\altaffilmark{3,4}, Sz.~Csizmadia\altaffilmark{1}, P.~Eigm\"uller\altaffilmark{1}, A.~Erikson\altaffilmark{1}, S.~Kirste\altaffilmark{1}, R.~Lemke\altaffilmark{3}, M.~Murphy\altaffilmark{5}, T.~Pasternacki\altaffilmark{1}, H.~Rauer\altaffilmark{1,6}, and R.~Titz-Weider\altaffilmark{1}}} 

\altaffiltext{1}{Institut f\"ur Planetenforschung, Deutsches Zentrum f\"ur Luft- und Raumfahrt, 
Rutherfordstr.~2, 12489~Berlin, Germany}
\altaffiltext{2}{European Southern Observatory, Alonso de C\'ordova~3107, Vitacura, Casilla~19001, Santiago~19, Chile}
\altaffiltext{3}{Astronomisches Institut, Ruhr-Universit\"at Bochum, 44780~Bochum, Germany}
\altaffiltext{4}{Instituto de Astronom\'{\i}a, Universidad Cat\'{o}lica del Norte, Antofagasta, Chile}
\altaffiltext{5}{Depto. F\'isica, Universidad Cat\'olica del Norte, PO~1280, Antofagasta, Chile}
\altaffiltext{6}{Zentrum f\"ur Astronomie und Astrophysik, Technische Universit\"at Berlin, 10623~Berlin, Germany}
\email{thomas.fruth@dlr.de}

\begin{abstract}
The CoRoT field LRa02 has been observed with the Berlin Exoplanet Search Telescope II (BEST II) during the southern summer 2007/2008. A first analysis of stellar variability led to the publication of 345 newly discovered variable stars. Now, a deeper analysis of this data set was used to optimize the variability search procedure. Several methods and parameters have been tested in order to improve the selection process compared to the widely used $J$~index for variability ranking. This paper describes an empirical approach to treat systematic trends in photometric data based upon the analysis of variance statistics that can significantly decrease the rate of false detections. 

Finally, the process of reanalysis and method improvement has virtually doubled the number of variable stars compared to the first analysis by Kabath et al. A supplementary catalog of 272 previously unknown periodic variables plus 52 stars with suspected variability is presented. Improved ephemerides are given for 19 known variables in the field. In addition, the BEST~II results are compared with CoRoT data and its automatic variability classification.
\end{abstract}


\section{INTRODUCTION}
During the last decade, ground- and space-based surveys have been very successful in detecting transiting exoplanets. In addition to their primary science goal, the large photometric data sets acquired by them allow studying millions of stars for variability. Numerous projects thus provide an exceedingly increasing number of detections that are collected by variable star catalogs such as the General Catalogue of Variable Stars \citep[GCVS;][]{GCVS} or the Variable Star Index\footnote{http://www.aavso.org/vsx/} (VSX). Such catalogs not only broaden the statistical sample of variable stars, but are also important to gain further knowledge about the different processes that cause stellar variability.

Several methods have been proposed to search for periodic signals in astronomical time series \citep[for a good overview, see, e.g.,][]{Schwarzenberg-Czerny1999}. One of the most widely applied algorithms is the analysis of variance (AoV) statistic \citep{czerny}, which provides an optimal period search in uneven sampled observations. It has been used very successfully by projects like HAT \citep{Bakos2004}, WASP \citep[e.g.,][]{Maciejewski2011}, or OGLE \citep[e.g.,][]{Soszynski2008}. In addition, the $J$~index \citep{stetson} is frequently used to quantify variability in general and/or for selecting candidate stars prior to a period search in order to minimize computation time \citep[e.g.,][]{zhang,Pepper2006,Pasternacki}.

However, both methods -- the AoV period search and Stetson's variability index -- are strongly affected by systematic trends present in ground-based data sets \citep[see, e.g.,][]{Pepper2006,LRc01,LRa02,Hartman2011}. Most dominant are diurnal systematics, introducing artificial variability with periods of one day or multiples thereof. Such trends generally yield a higher ranking of non-variable stars, thus increasing the false alarm rate. A common approach to account for candidates with systematic variability is to set limits, e.g., to exclude detections within certain period ranges. However, any such manual mechanism is usually not well applicable to other data sets or projects, and the number of missed detections (false negatives) is often unknown. Therefore, a more sophisticated treatment of systematic variability in combination with period search and ranking is needed.

The Berlin Exoplanet Search Telescope \citep[BEST; ][]{BEST} and BEST~II \citep{Corot1b2b} are used to perform ground-based support of the CoRoT space mission \citep{corot}. By obtaining high-precision and long-time series photometry of the CoRoT target fields prior to the satellite's observations, planetary candidates can quickly be checked in the BEST data archive \citep[e.g.,][]{deeg2009,Corot1b2b,Alonso2012}. In addition, the obtained light curves can be used to identify variable stars. The BEST project has already yielded the detection of several hundreds of new periodic variable stars \citep{LRc01,IR01,LRa01,LRc02,LRa02,Pasternacki}. 

A first characterization of periodic stellar variability in the CoRoT field LRa02 has been published by \citet{LRa02}. Following the detections from this first publication (denoted as Paper~I hereafter), it was possible to optimize the BEST~II reduction pipeline. The large, well-characterized data set LRa02 was used as a proxy to analyze and automatically include systematic biases in the variable star candidate selection and period search, yielding a significantly reduced false alarm rate.

The optimized procedure to rank periodic variable stars is described in this paper. In addition, we present a large catalog extension to the variable star classification in LRa02 that was obtained through the reanalysis and compare our results to publicly available CoRoT data.

\section{TELESCOPE AND OBSERVATIONS}
BEST II is located at the Observatorio Cerro Armazones, Chile. It is operated by the \textit{Institut f\"ur Planetenforschung} of the \textit{Deutsches Zentrum f\"ur Luft- und Raumfahrt} in robotic mode from Berlin since summer 2007. 

The system consists of a 25 cm Baker-Ritchey-Chr\'etien telescope with a focal ratio of $f/5.0$, yielding a wide field of view (FOV) of $1\degdot 7 \times 1\degdot 7$. It is equipped with a 4k $\times$ 4k, 16 bit Finger Lakes Imager CCD (KAF-16801E1) with a pixel size of 9~$\mu$m, and an angular resolution of $1\arcseccomma 5$ pixel$^{-1}$. BEST~II observes without any filter to maximize the photon yield -- the CCD sensitivity peaks at 650 nm and is roughly comparable to the Johnson $R$ band. 

The CoRoT long-run field LRa02 was observed by BEST II for 41 nights from 2007 November to 2008 February prior to the satellite observations. As the FOV of CoRoT is slightly larger than BEST II, we split the field into two subfields (called LRa02a and LRa02b, respectively) and pointed at them alternating. Paper~I indicates their corresponding center coordinates and shows the orientation with respect to the CoRoT FOV (Figure~1 in Paper~I).

\section{DATA SET AND VARIABILITY CRITERIA}
The acquired data set was calibrated and reduced using the BEST automated photometric data pipeline as outlined in \citet{Corot1b2b}. The actual reduction procedure for the field LRa02 has already been described in detail in Paper~I and below only those steps relevant to the new results are discussed. 

In Paper~I, Kabath et al.~reported the detection of 350 periodic variable stars (of which five were previously known). Their selection was based upon the variability index~$J$ \citep{stetson}, calculated for each star by
\begin{equation}\label{eq:jindex1}
J=\frac{ \sum_{k=1}^{n-1}{w_k\, \textnormal{sgn} (P_k) \sqrt{\left|P_k\right|} }}
  {\sum_{k=1}^{n-1}{w_k}},
\end{equation}
where $k$ is indexing individual data points. $P_k$ is calculated from each pair of subsequent magnitudes $m_k$ and $m_{k+1}$ using the corresponding normalized residuals $\delta_k$ and $\delta_{k+1}$:
\begin{equation}\label{eq:jindex2}
  P_k = \delta_k \delta_{k+1} \ \ \ \textnormal{with} \ \ \ 
  \delta_k= \sqrt{\frac{n}{n-1}} \left( \frac{m_k-\overline{m}}{\sigma_k} \right),
\end{equation}
where $\sigma_k$ denotes the uncertainty of measurement $k$, $\overline{m}$ the mean magnitude, and $n$ the number of measurements for the selected star. The weights $w_k$ in Equation~(\ref{eq:jindex1}) were calculated following the approach of \cite{zhang} as
\begin{equation}\label{eq:jindex3}
  w_k = \exp {\left( -\frac{t_{k+1}-t_k}{\Delta t} \right) },
\end{equation}
where $t_k$ is the time of observation $k$ and $\Delta t$ is the median of all pair time spans $(t_{k+1}-t_k)$.

In Paper~I, the limit of $J\geq 0.5$ was applied in order to distinguish variable from constant field stars. This preselection yielded 1{,}858 stars in LRa02a and 1{,}868 stars in LRa02b, respectively. For each star, the AoV statistic $\Theta$ \citep{czerny} was calculated for a period range of 0.1--35 days. The light curves were folded with the frequency $\omega\subscript{max}$ corresponding to the maximum AoV value,
\begin{equation}\label{eq:wmax}
  \Theta(\omega\subscript{max}) = \max \left( \Theta(\omega) \right),
\end{equation}
and then inspected visually. Most folded light curves showed no clear periodic variability or an artificial period of one day or multiples thereof, which can be caused by systematic effects due to the observational cycle. \citet{LRa02} finally identified 173 periodic variables in LRa02a and 177 in LRa02b.

\section{REANALYSIS}\label{sec:reanalysis}
\begin{figure*}[htpc]
  \includegraphics[width=\linewidth]{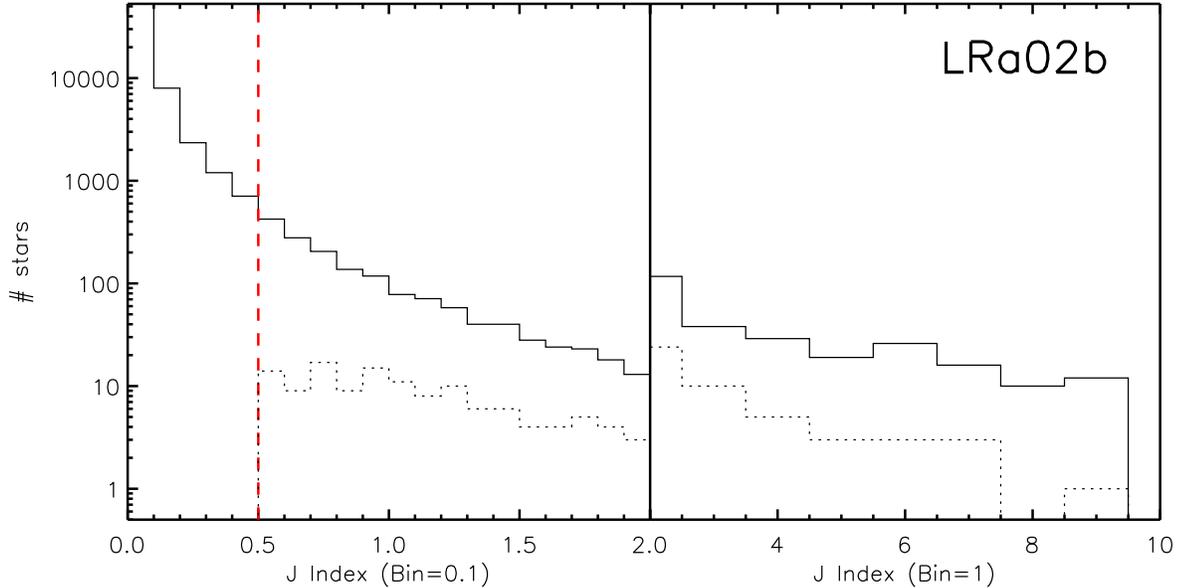}
  \figcaption{\footnotesize $J$~index histogram for subfield LRa02b (qualitatively equal to LRa02a). The solid line shows the total star count in bins of 0.1 ($J<2$) and 1 ($J\geq 2$), whereas the dotted line shows the number of variable star detections from Paper~I. The red dashed line denotes the cutoff limit of $J=0.5$ as applied in Paper~I. \label{fig:jhist}}
\end{figure*}

In the BEST archive, some stars are marked as clear variables with large $J$ indices (up to 10 and higher). Figure~\ref{fig:jhist} shows the count of all stars in subfield LRa02b and the corresponding number of variable star detections in Paper~I as a function of the $J$~index. Altogether, the large number of false positives shows that the $J$~index alone is not an effective criterion for selecting variable stars. Furthermore, a number of clear detections with low $J$ values indicates that several variables must have been missed in Paper~I due to the cutoff.

In order to improve the quality of our preselection process and to maximize the detection yield, we performed a deeper analysis of our data set LRa02. Field LRa02 was chosen for this purpose because it was observed with longest duration within the BEST project so far. 

The performed reanalysis consisted of three main steps. 

First, the best period was searched in all light curves from the initial data set without any preselection criterion, i.e., directly using the AoV multiharmonic algorithm for \textit{all} stars in both subsets. Each of the resulting 37{,}361 folded light curves in LRa02a and 66{,}974 in LRa02b was then examined visually for periodic stellar variability. In addition to the variable stars from Paper~I, this first step already revealed 189 \textit{additional} variable stars that were not detected in our first publication. During this step, we also discovered a bug in our implementation of the $J$~index that yielded systematically lower values especially for short periods, which is why many of these detections show periods of less than a day.

Second, the combined data set of variable stars from Paper~I and the additional manual detections were used to optimize the BEST~II selection process (see next Section~\ref{sec:limitimprove}). With a very good knowledge of this data set, it was possible to compare different selection methods and to adjust their corresponding parameters.  

Third, the new search algorithm was applied to the data set with optimized parameters. The improvements to the pipeline finally lead to an increase of detections by another 135 variable stars that were not found during all previous steps (see Section~\ref{sec:results:newmethod}).

\section{IMPROVEMENTS ON VARIABILITY SEARCH}\label{sec:limitimprove}
The deep analysis of the BEST II data set LRa02 gives us the opportunity to study the performance of variability search algorithms in detail. The aim is to recover all variable stars in the data set automatically and to minimize the number of false alarms (and manpower) at the same time.

In this section, we show the limitations of the $J$~index with regard to systematic trends (Section~\ref{sec:jlimit}), describe how to quantify the performance of a variability search (Section~\ref{sec:limitimprove1}), present the algorithms tested (Sections~\ref{sec:limitimprove2} and \ref{sec:limitimprove3}), and finally compare the performance of different approaches and parameters (Section~\ref{sec:limitimprove4}).

\subsection{Limitations of the Variability Index $J$}\label{sec:jlimit}
After the first step of the reanalysis, the visual inspection, the majority of new detections showed $J$~indices below the limit of 0.5 applied before, which is why they were not detected in Paper~I. 
\begin{figure}
  \includegraphics[width=\linewidth]{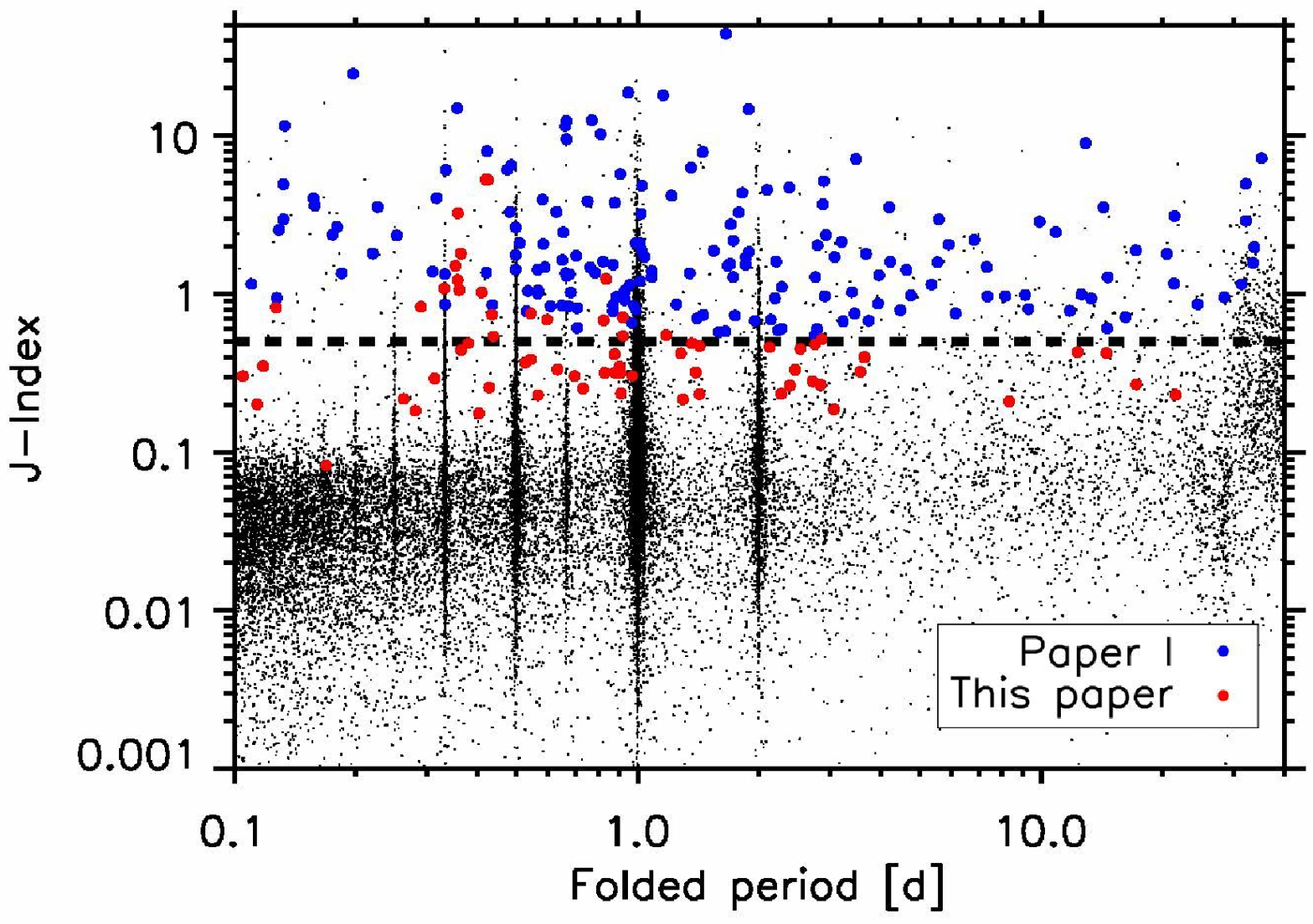}
  \includegraphics[width=\linewidth]{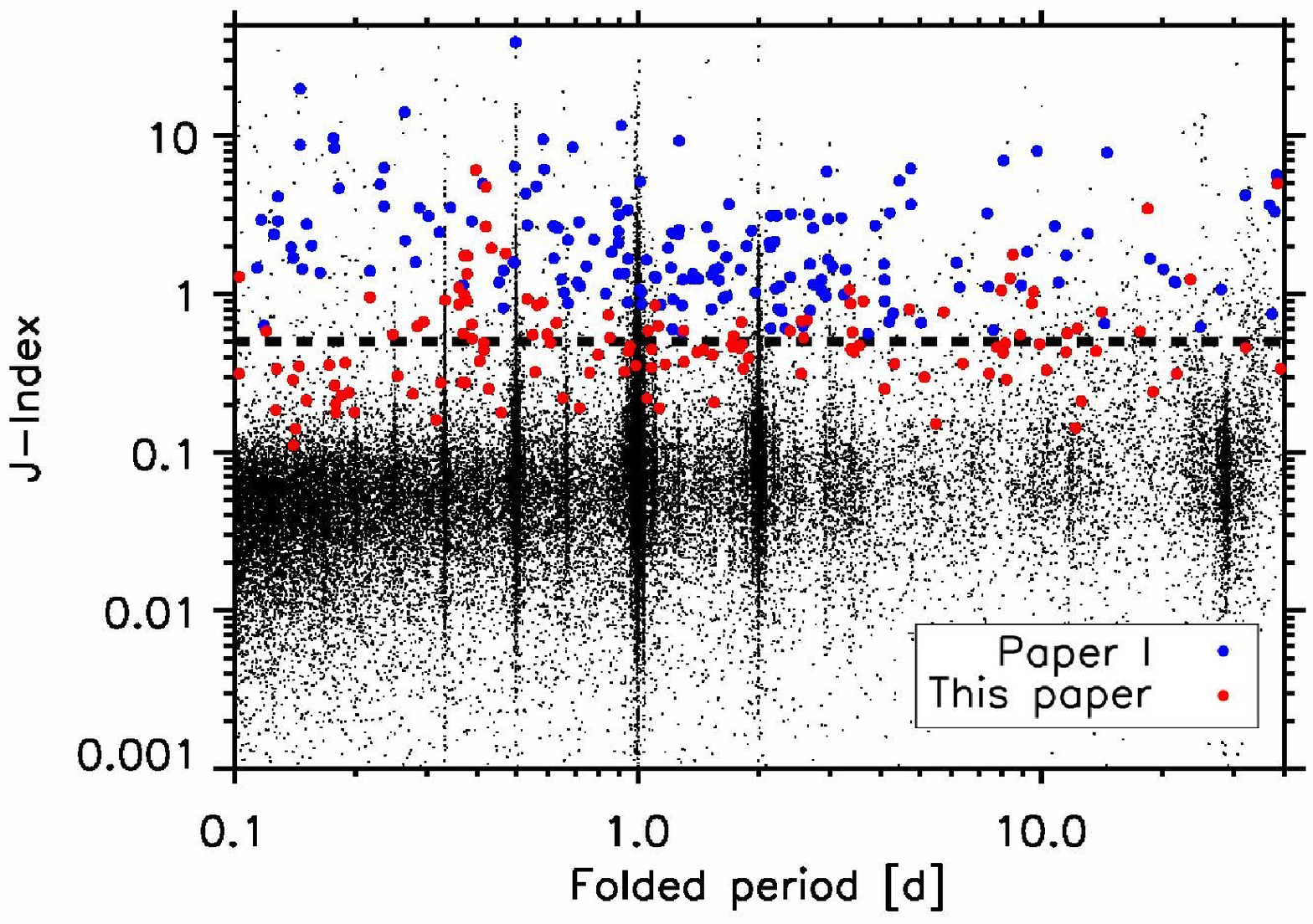}
\caption{\footnotesize Variability $J$~index plotted vs.~determined period without treatment of systematic effects for all stars in the BEST~II field LRa02a (top) and LRa02b (bottom). Variable stars identified in Paper~I are marked with blue dots, whereas variable stars from the manual reanalysis appear red. The dashed line shows the selection limit of $J=0.5$ as applied in Paper~I. \label{fig:jper}}
\end{figure}
The relation between periods and variability is shown in Figure~\ref{fig:jper}. A clear bulk of stars can be found at about $0.01\leq J \leq 0.1$, with the limits being widely period independent and populated by stars showing noise only. Most new variable star detections are found in the region between $J=0.1$ and the cutoff limit of $J=0.5$ from Paper~I. A small number of new variables with $J>0.5$ were not detected in Paper~I because their $J$ indices have been initially underestimated (see Section \ref{sec:reanalysis}). 

The dominant variation in many light curves is due to diurnal systematics, aliasing, or a combination of both. Figure~\ref{fig:jper} shows a large accumulation of stars having periods of one day or integral fractions/multiples thereof, often in combination with large $J$~indices. Consequently, this leads to a very high number of false alarms when using the Stetson index as the only criterion for variability selection. In the example of the data set LRa02, a cutoff limit of $J=0.1$ would be sufficient to include all variable stars in the selection sample, but only $74\%$ of all light curves would be sorted out. The remaining large sample of 31{,}000 stars is mainly affected by systematic effects and contains only 681 stars with real physical variability (see Section~\ref{sec:results}). The corresponding false alarm rate of about 98\% shows the need for an automated treatment of systematic variability, which is not part of the $J$~index.

\subsection{Quantitative Assessment of Period Search Algorithms}\label{sec:limitimprove1}
Two fundamental criteria are used to assess the quality of period search algorithms: the significance of the detection itself and the correct determination of the frequency of variability.

First, a quantity $\xi$ is introduced to evaluate the detection efficiency of any given search algorithm. Detection methods are usually based upon a single numerical value $q$ (e.g., the Stetson index, $q\equiv J$) that can be used to prioritize a candidate list. The success of ranking variable stars high in the list is measured with $\xi$ for each tested search algorithm. It ranges from 0 for the perfect algorithm (all previously identified variable stars listed first) to 1 (listed last). For details on the calculation of $\xi$ see the Appendix.

The second criterion is tested by comparing the frequency $\omega\subscript{correct}$ that was verified manually with the frequency of a tested algorithm. We consider a tolerance range of 2\% around $\omega\subscript{correct}$ for a correct determination. Also included are 2\% deviation around half or twice that value, because the distinction between these is often ambiguous from the light curve itself. The fraction $n_\omega$ of correctly identified frequencies can then be used for a quantitative comparison between tested algorithms.

\subsection{Frequency Determination and Exclusion of Systematics}\label{sec:limitimprove2}
\begin{figure}[t]\centering
  \includegraphics[width=\linewidth]{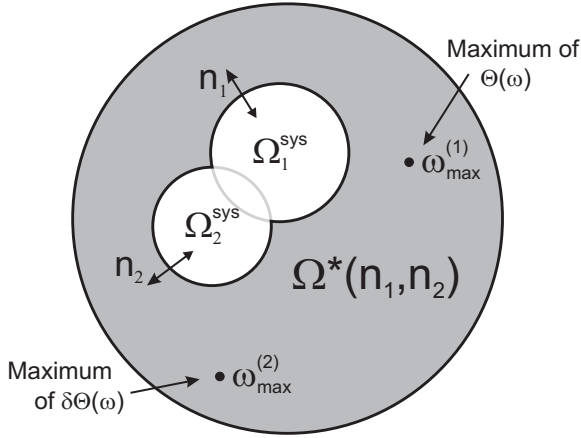}
\figcaption{\footnotesize Schematic view on the rejection of systematic frequencies. The set $\Omega^*(n_1,n_2)$ of non-systematic frequencies is obtained by excluding the subsets $\Omega\superscript{sys}_1$ (master power spectrum cut) and $\Omega\superscript{sys}_2$ (empty phases). The size of both can be adjusted with the parameters $n_1$ and $n_2$. Maxima of $\Theta(\omega)$ and $\delta\Theta(\omega)$ are searched within $\Omega^*(n_1,n_2)$ to obtain the frequencies $\omega\subscript{max}^{(1)}$ and $\omega\subscript{max}^{(2)}$, respectively. \label{fig:sysfreq-sets}}
\end{figure}

Because systematics and their aliases are usually limited to a set of few well-defined frequencies $\{\omega\subscript{sys}\}$, they can be excluded by searching the best frequency $\omega\subscript{max}$ only on a subset $\Omega^*=\{\omega^*\}=\{\omega\}\backslash \{\omega\subscript{sys}\}$ (Figure~\ref{fig:sysfreq-sets}). We tested three different methods to account for systematic frequencies, both independent of each other as well as in combination.

\begin{figure*}[htpc]\centering
  \includegraphics[width=.8\linewidth]{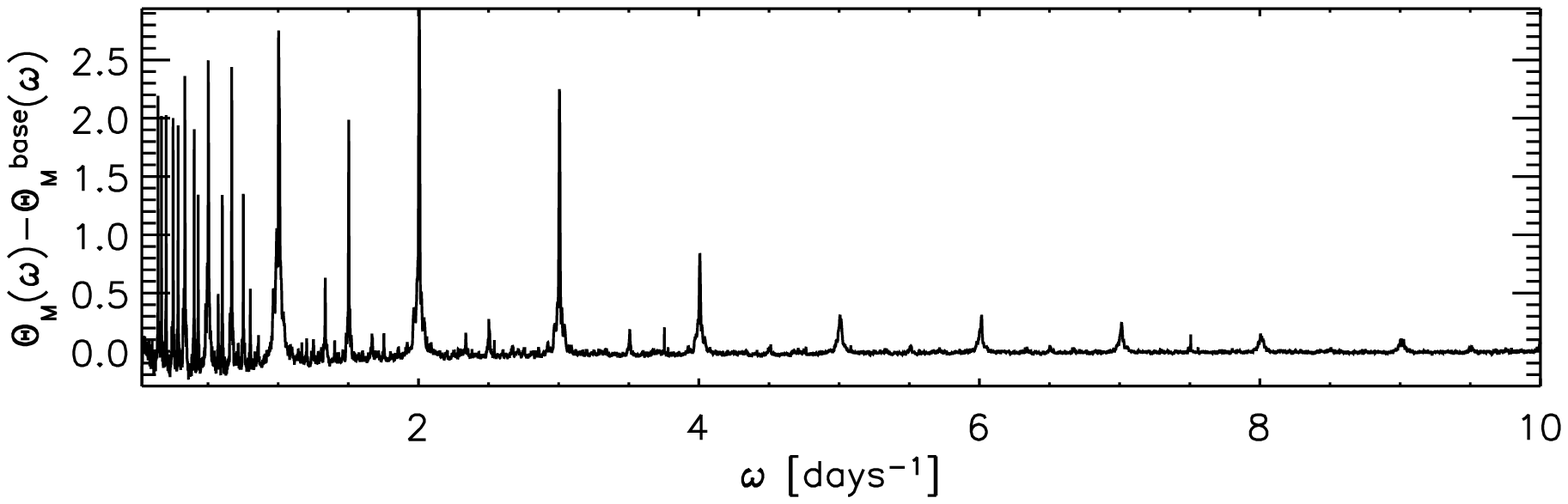}
  \includegraphics[width=.8\linewidth]{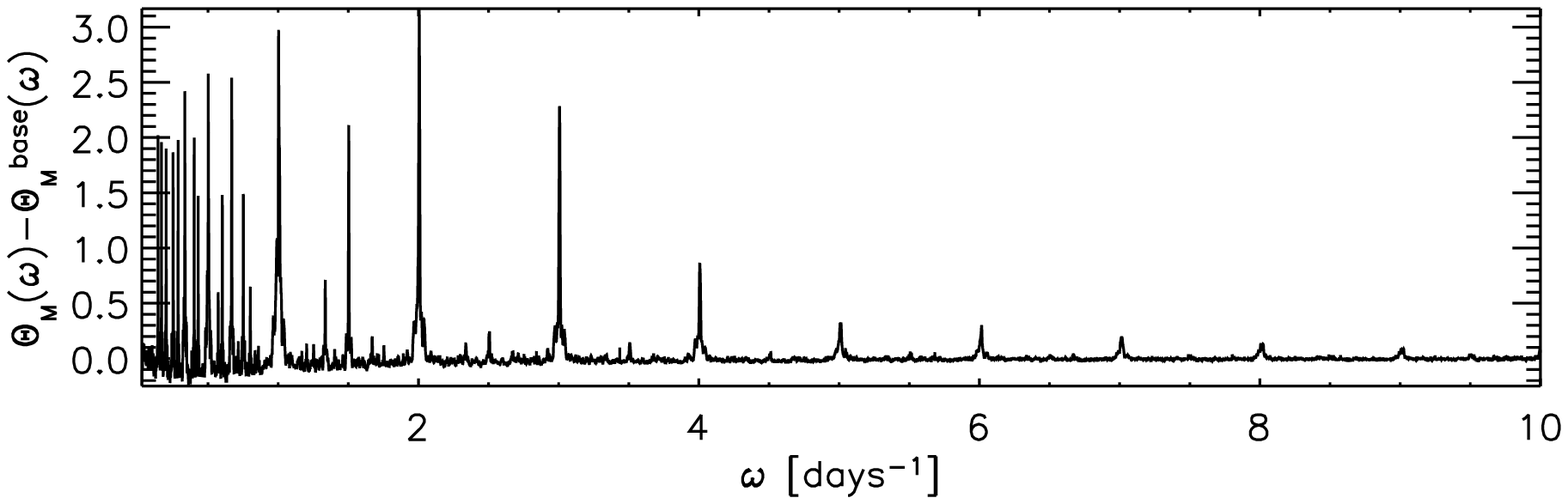}
\figcaption{\footnotesize Master power spectrum for data sets LRa02a (top) and LRa02b (bottom).\label{fig:masterspec}} 
\end{figure*}

1. \textit{Master power spectrum.}
Systematic periodic signals affect many light curves in a data set and can thus be distinguished from real stellar variability by analyzing many power spectra $\Theta_i(\omega)$ of individual stars $i$ statistically. We use the mean of all $N_*$ spectra to build a \textit{master} spectrum 
\begin{equation}
  \Theta_M(\omega) = \frac{1}{N_*}\sum_{i=1\ldots N_*}{\Theta_i(\omega)}.
\end{equation}
In order to identify significant peaks in the master spectrum, a baseline fit is determined. Among several investigated functions, a polynomial log-log fit, 
\begin{equation}
  \ln (\Theta_M\superscript{base}(\omega)) = \sum_{i=0}^{4}{c_i \cdot (\ln \omega)^i},
\end{equation}
with coefficients $c_i$ maps the baseline best and most reliable for various tested data sets. The baseline-subtracted spectrum can then be searched for systematics. Figure~\ref{fig:masterspec} shows how $\Theta_M-\Theta_M\superscript{base}$ peaks clearly at the diurnal frequencies that we aim to identify.
Finally, a simple cutoff using the standard deviation $\sigma_M\superscript{base}$ of the subtracted spectrum $\Theta_M-\Theta_M\superscript{base}$ is applied to filter systematics automatically. The corresponding set of frequencies having peaks at least $n_1\cdot \sigma_M\superscript{base}$ above the average power spectrum is defined by
\begin{equation}\label{eq:n1}
  \Omega\superscript{sys}_1(n_1) = \{ \omega \ |\ \Theta_M(\omega) > \Theta_M\superscript{base}(\omega) + n_1\cdot \sigma_M\superscript{base} \},
\end{equation}
where the parameter $n_1$ can be adjusted to quantify the degree of exclusion.

2. \textit{Frequencies with empty phases.}
Ground-based observations are strongly affected by periodic gaps in the data, most commonly due to the diurnal cycle. The incomplete phase coverage leads to aliasing and can often cause false positive detections. We use a simple model to exclude frequencies with poor phase coverage: first, the folded light curve is split into $N\subscript{boxes}=100$ intervals of the same length. The number of empty intervals $N\subscript{boxes}\superscript{empty}(\omega)$ is then counted for each sampled frequency. A frequency is considered systematic by this criterion if the fraction of empty intervals is larger than a relative threshold parameter $n_2$, i.e.,
\begin{equation}\label{eq:n2}
  \Omega\superscript{sys}_2(n_2) = \{ \omega \ |\ N\subscript{boxes}\superscript{empty}(\omega) > n_2 \cdot N\subscript{boxes} \}.
\end{equation}

Both criteria \ref{eq:n1} and \ref{eq:n2} are merged to exclude systematic frequencies determined in either way, i.e., the overall set of non-systematic frequencies (Figure~\ref{fig:sysfreq-sets}) is defined by
\begin{equation}\label{eq:n12}
  \Omega^*(n_1,n_2) = \{\omega\} \ \backslash\ \left( \Omega\superscript{sys}_1(n_1) \cap \Omega\superscript{sys}_2(n_2) \right)
\end{equation}
and the search for $\omega\subscript{max}$ is restricted from the whole frequency range (Equation~(\ref{eq:wmax})) to the subset $\Omega^*(n_1,n_2)$: 
\begin{equation}\label{eq:wmax1}
  \forall \omega \in \Omega^*(n_1,n_2): \Theta(\omega) \leq \Theta(\omega\subscript{max}^{(1)}).
\end{equation}

Note that the function $N\subscript{boxes}\superscript{empty}(\omega)$ is similar but not equivalent to the window function~$\gamma_N(\omega)$ \citep{deeming1975}. Limiting the latter was tested as an alternative criterion; it shows a very similar ranking performance, but slightly less correctly determined frequencies. Being more simple, the empty phase criterion was chosen for the final test setup.

3. \textit{Power spectrum scaling.}
In addition to the exclusion of systematic frequencies, we also investigated a method to include the information about systematics into the AoV results directly. Instead of searching for the maximum of $\Theta(\omega)$, an artificial spectrum 
\begin{equation}\label{eq:psscale}
  \delta\Theta(\omega)=\Theta(\omega)/\Theta_M(\omega)
\end{equation}
is created by dividing the AoV spectrum $\Theta$ of every star by the master power spectrum $\Theta_M$. Its maximum is found at the frequency $\omega\subscript{max}^{(2)}$ in analogy to Equation~(\ref{eq:wmax1}):
\begin{equation}\label{eq:wmax2}
  \forall \omega \in \Omega^*(n_1,n_2): \delta\Theta(\omega) \leq \delta\Theta(\omega\subscript{max}^{(2)})\ .
\end{equation}

\subsection{Variable Star Ranking}\label{sec:limitimprove3}
In addition to the $J$~index, two methods to prioritize variable star candidates were tested. 

The first method takes the AoV result directly, i.e.,
\begin{equation}\label{eq:q11}
  q_1^{(1)} = \Theta(\omega\subscript{max}^{(1)}).
\end{equation}
In its special case of no excluded systematic frequencies ($n_1 \rightarrow\infty$, $n_2=1$), this is a widespread method for prioritizing variable star candidates. Likewise, the maximum of the divided power spectrum $\delta\Theta$ could serve as a variability indicator:
\begin{equation}\label{eq:q12}
  q_1^{(2)} = \delta\Theta(\omega\subscript{max}^{(2)}).
\end{equation}

The AoV statistic $\Theta\equiv\frac{(n-n_\|)\|x_\|\|^2}{n_\|\|x-x_\|\|^2}$ compares the quadratic norm of a model~$x_\|$ (with $n_\|$ free parameters) with the residuals that remain after subtraction of the model from $n$ observations $x$ \citep{Schwarzenberg-Czerny1999}. Because it has optimum period detection properties \citep{czerny}, we expect $q_1$ to yield the best ranking. However, as an empirical alternative we also tested the light curve's standard deviation with and without the periodic signal obtained by the AoV multiharmonic fit:
\begin{equation}
  q_2^{(1)} = \frac{\sigma}{\sigma'(\omega\subscript{max}^{(1)})} \hspace{5mm} \mbox{and} \ \ \  q_2^{(2)} = \frac{\sigma}{\sigma'(\omega\subscript{max}^{(2)})}
\end{equation}
It is dependent upon the frequency $\omega\subscript{max}$ determined in the previous section, which is why different choices of $\omega\subscript{max}$ lead to different rankings. The quoted $\sigma'$ refers to the standard variation \textit{after} subtraction of the corresponding fit \citep[for calculation of coefficients, see][]{Schwarzenberg-Czerny1998}.

\begin{deluxetable}{ccp{5mm}ccp{5mm}ccp{5mm}c}
\tabletypesize{\footnotesize}
\tablewidth{0pt}
\tablecaption{Results of the tested variable star ranking methods and parameters.
\label{tab:alg-comparison}}
\tablehead{
  \multicolumn{3}{c}{} & \multicolumn{2}{c}{$\omega\subscript{max}^{(1)}$} && \multicolumn{2}{c}{$\omega\subscript{max}^{(2)}$} &&  $J$ \\\cline{4-5}\cline{7-8}\\[-0.9em]
  \multicolumn{3}{c}{} & $q_1^{(1)}$ &  $q_2^{(1)}$ &&  $q_1^{(2)}$ &  $q_2^{(2)}$ && Index
}
\startdata
$\stackrel{(n_1,n_2)}{\xi}$ 
      & LRa02a
      && $\stackrel{(0,\,0)}{0.0051}$ 
      & $\stackrel{(0,\,0)}{0.010}$ 
      && $\stackrel{(0.2,\,10\%)}{0.0035}$ 
      & $\stackrel{(0.2,\,\geq 70\%)}{0.010}$ 
      && $0.032$
\\    & LRa02b 
      && $\stackrel{(0,\,0)}{0.0035}$ 
      & $\stackrel{(0,\,0)}{0.0056}$ 
      && $\stackrel{(0,\,0)}{0.0027}$ 
      & $\stackrel{(0,\,0)}{0.0062}$ 
      && $0.034$
\\\tableline\\[-0.9em]
$\stackrel{(n_1,n_2)}{n_\omega}$ 
   & LRa02a
   && \multicolumn{2}{c}{$\stackrel{(5-10,\,10\%)}{65\%}$}  
   && \multicolumn{2}{c}{$\stackrel{(\geq 5,\,10\%)}{84\%}$} \\
   & LRa02b
   && \multicolumn{2}{c}{$\stackrel{(\geq 10,\,0)}{63\%}$}  
   && \multicolumn{2}{c}{$\stackrel{(\geq 5,\,10\%)}{81\%}$} \\
\enddata
\\[-2em]
\tablecomments{\footnotesize The quantities~$\xi$ and~$n_\omega$ are shown for each tested ranking parameter~$q$, both methods of determining the best frequency~$\omega\subscript{max}^{(k)}$ and both analyzed data sets. For clarity, only the best value achievable by variation of the parameters~$n_1$ (master power spectrum cut) and~$n_2$ (empty phases) is shown for each method. The corresponding values/ranges of~$n_1$ and~$n_2$ are shown in small brackets above each value.}
\end{deluxetable}

\subsection{Comparison of Variability Search Performance}\label{sec:limitimprove4}
The quantities $\xi$ and $n_\omega$ have been calculated for both frequencies $\omega\subscript{max}^{(1)}$ and $\omega\subscript{max}^{(2)}$ and both tested ranking methods $q_{1,2}^{(k)}$. For each run, the parameters $n_1$ and $n_2$ were varied independently on the following values:
\begin{eqnarray*}
  n_1 &=& \{0,0.01,0.05,0.1,0.2,0.5,1,2,5,10,1000\} \\
  n_2 &=& \{0\%,10\%,20\%,30\%,40\%,50\%, \\
       && \ 60\%,70\%,80\%,90\%,100\%\}
\end{eqnarray*}
The results of the comparison are summarized in Table~\ref{tab:alg-comparison}, which shows the best value for $\xi$ and $n_\omega$ achievable with each tested method. 

\subsubsection{Number of Harmonics}
Paper~I and the first step of the reanalysis (Section~\ref{sec:reanalysis}) determined the stellar variability using AoV periodograms with two harmonics, which was also used for this comparison. Furthermore, the number of harmonics was set to $N=7$ in a second test in order to increase the sensitivity on sharp signals that are, e.g., caused by eclipsing binaries \citep[for the sensitivity dependence on the number of harmonics, see, e.g.,][]{Schwarzenberg-Czerny1999}. Both results show very similar ranking performances (for the best ranking method $q_1$, we find $\xi_7\approx \xi_2\pm0.001$ in both data sets), but the $N=2$ test naturally yields a slightly better (12\%--15\%) frequency match with the initial run that was obtained using the same number of harmonics. However, the test with seven harmonics revealed a number of additional interesting eclipsing binaries that could not be detected using the smaller number of model parameters (see Sections~\ref{sec:results:newmethod} and~\ref{sec:results:cat}). Therefore, we prefer the latter for our improved variability search (Section~\ref{sec:newselmethod}) and focus in the following on the details of the search performance with $N=7$ harmonics.

\subsubsection{Ranking}
\begin{figure*}[htpc]\centering
  \includegraphics[width=.45\linewidth]{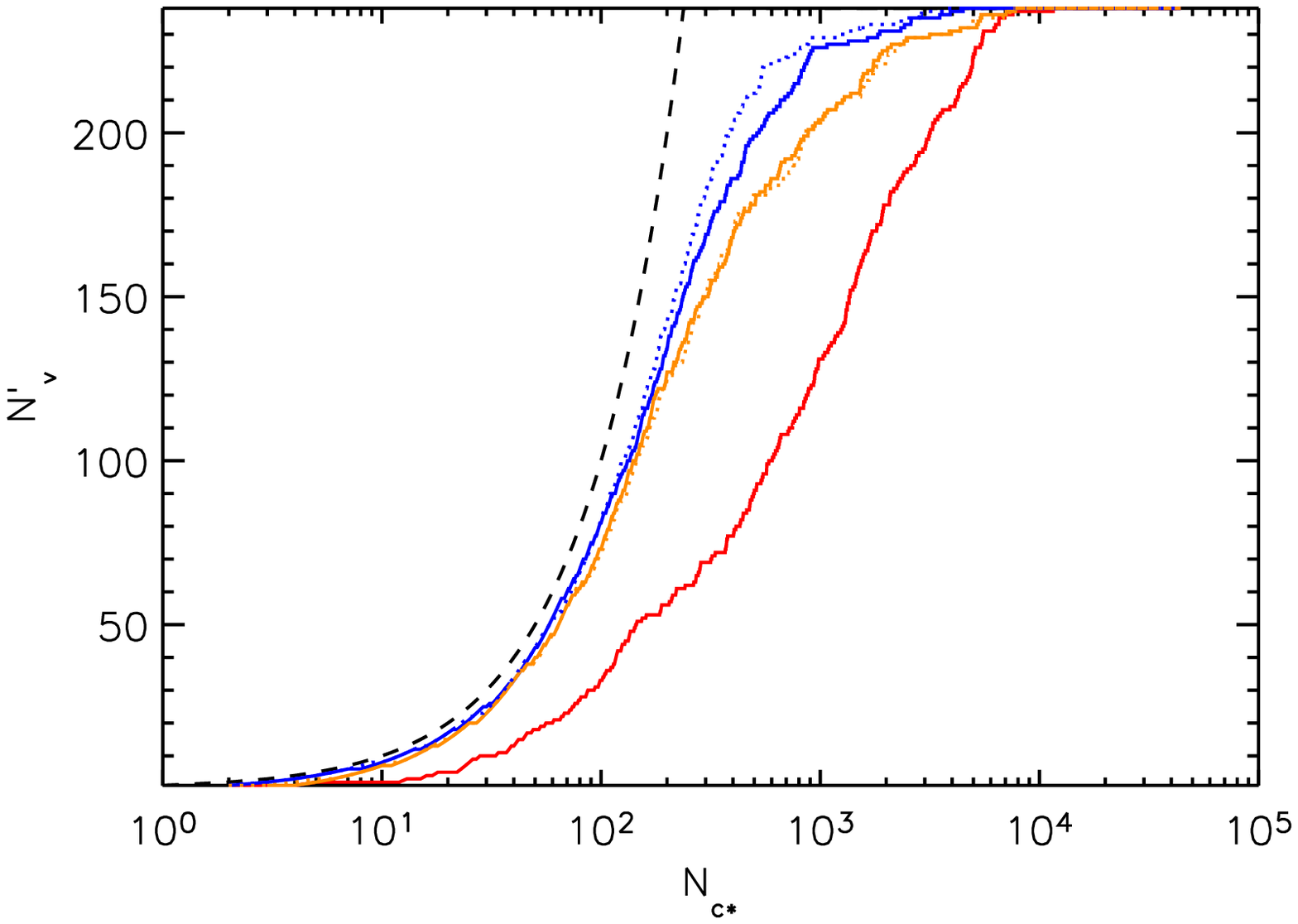}
  \includegraphics[width=.45\linewidth]{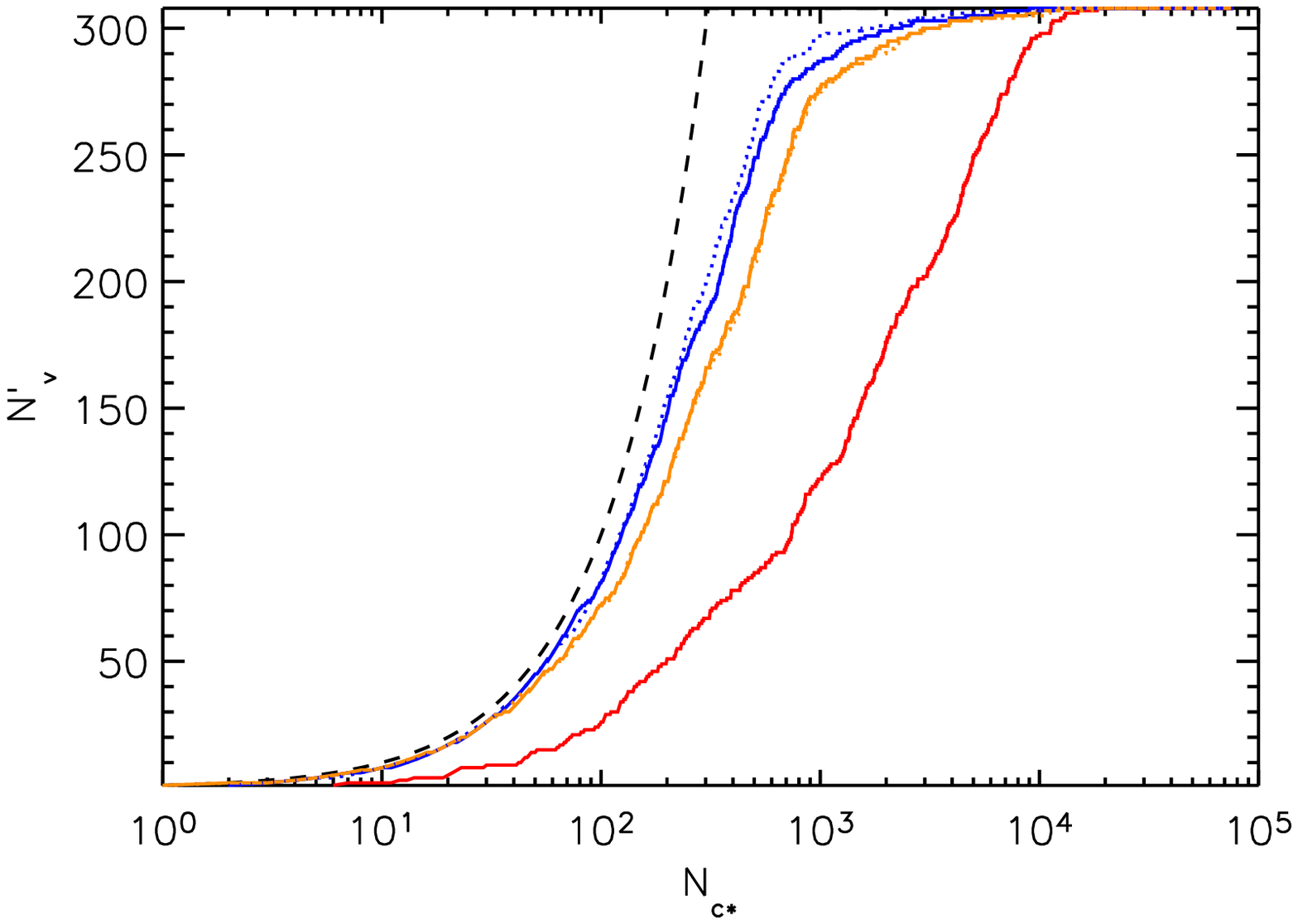}
\figcaption{\footnotesize Number of variable stars ${N}_v'$ as a function of the number of inspected stars $N_{c*}$ for LRa02a (left) and LRa02b (right). The different lines represent tested ranking methods: $q_1^{(k)}$, $q_2^{(k)}$, and the Stetson $J$~index (left to right). Solid lines represent the unweighted case ($k=1$), whereas dotted lines include the effect of master power spectrum division ($k=2$). Only the parameters $n_1$ and $n_2$ of the most successful sorting are used for each method (compare Table~\ref{tab:alg-comparison}). The black dashed line shows the optimal ranking for comparison. \label{fig:xicomparison}}
\end{figure*}

The performance of the ranking differs only slightly between the tested methods. Figure~\ref{fig:xicomparison} shows how the quantities $q_1$ (AoV) and $q_2$ ($\sigma$-ratio) both provide a sorting that lists stars with real variability first. The numerical quantity $\xi$ yields with $\approx 0.003$ for $q_1^{(2)}$ a minimum close to the optimal ranking ($\xi=0$). It is an order of magnitude lower than the previously used $J$~index ($\xi\approx 0.03$) and significantly lower than ranking the AoV power without exclusion of systematics ($\xi\approx 0.022$ for $q_1^{(1)}$ with $n_1=1000$ and $n_2=100\%$). In particular, this corresponds to a drastically decreased false alarm rate.

\begin{figure*}[htpc]\centering
  \includegraphics[width=.45\linewidth]{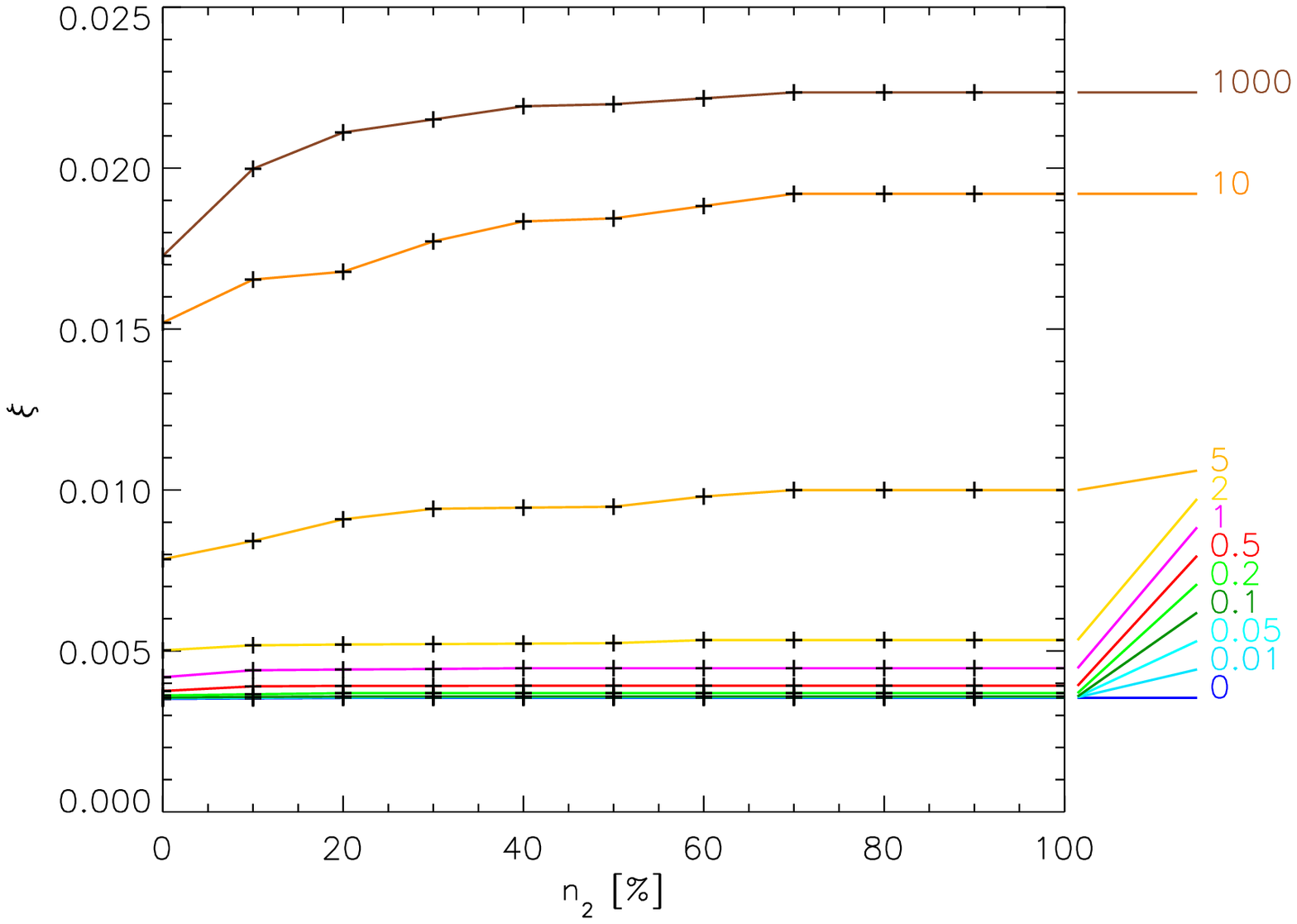}
  \includegraphics[width=.45\linewidth]{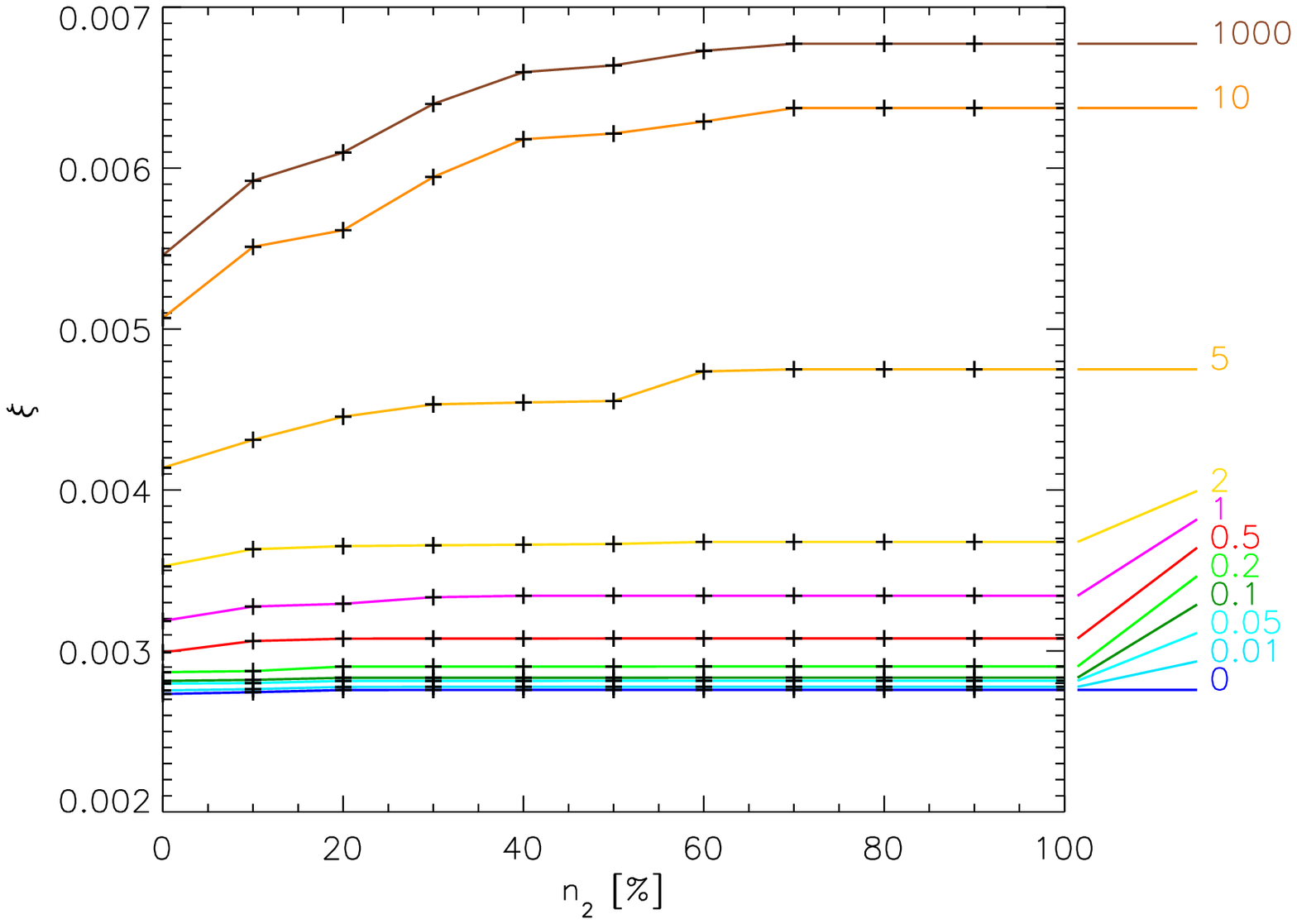}
\figcaption{\footnotesize Influence of systematic frequency exclusion method on variable star ranking efficiency for the data set LRa02b. The quantity $\xi$ is shown on the $y$-axis as a function of the model parameters $n_1$~(lines) and $n_2$~($x$-axis) for the best tested ranking method using $q_1$. The left plot shows the results with $\omega\subscript{max}^{(1)}$ as the maximum frequency, whereas the effect of division by the master power spectrum ($\omega\subscript{max}^{(2)}$) can be seen on the right. \label{fig:xi:n1n2}}
\end{figure*}

Figure~\ref{fig:xi:n1n2} shows the dependence of the ranking performance on the parameters $n_1$ and $n_2$ for the two best methods $q_1^{(1)}$ and $q_1^{(2)}$. In both cases, the most restrictive exclusion of systematic frequencies yields the best sorting. Thereby, the cut in the master power spectrum (Equation~(\ref{eq:n1})) has a slightly larger impact than the exclusion of empty phases (Equation~(\ref{eq:n2})). The minimum of $\xi$ is reached for $n_1=n_2=0$, but is almost independent of $n_2$, because the first criterion is more restrictive.

\subsubsection{Frequency Determination}

\begin{figure*}[htpc]\centering
  \includegraphics[width=.45\linewidth]{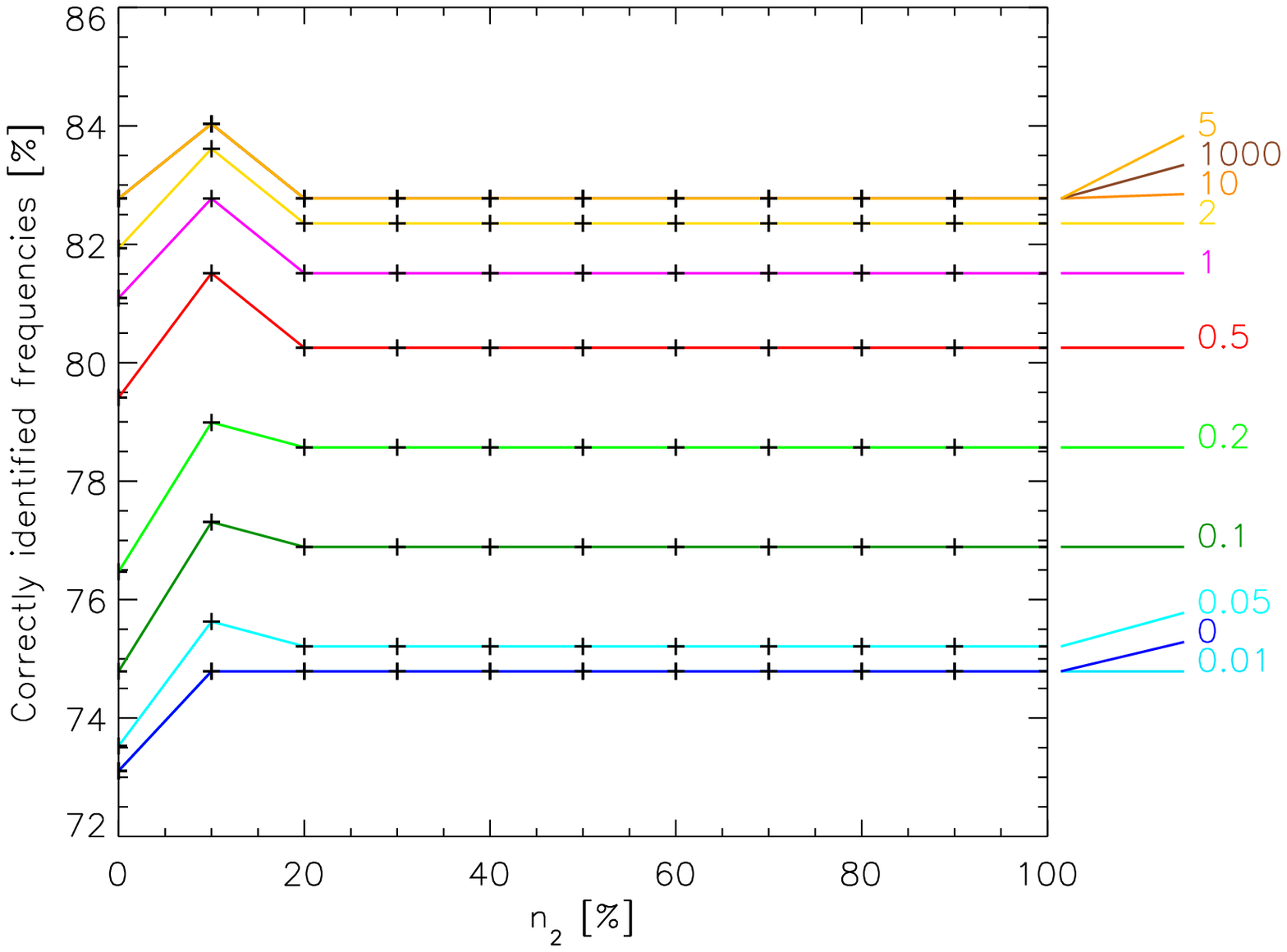}
  \includegraphics[width=.45\linewidth]{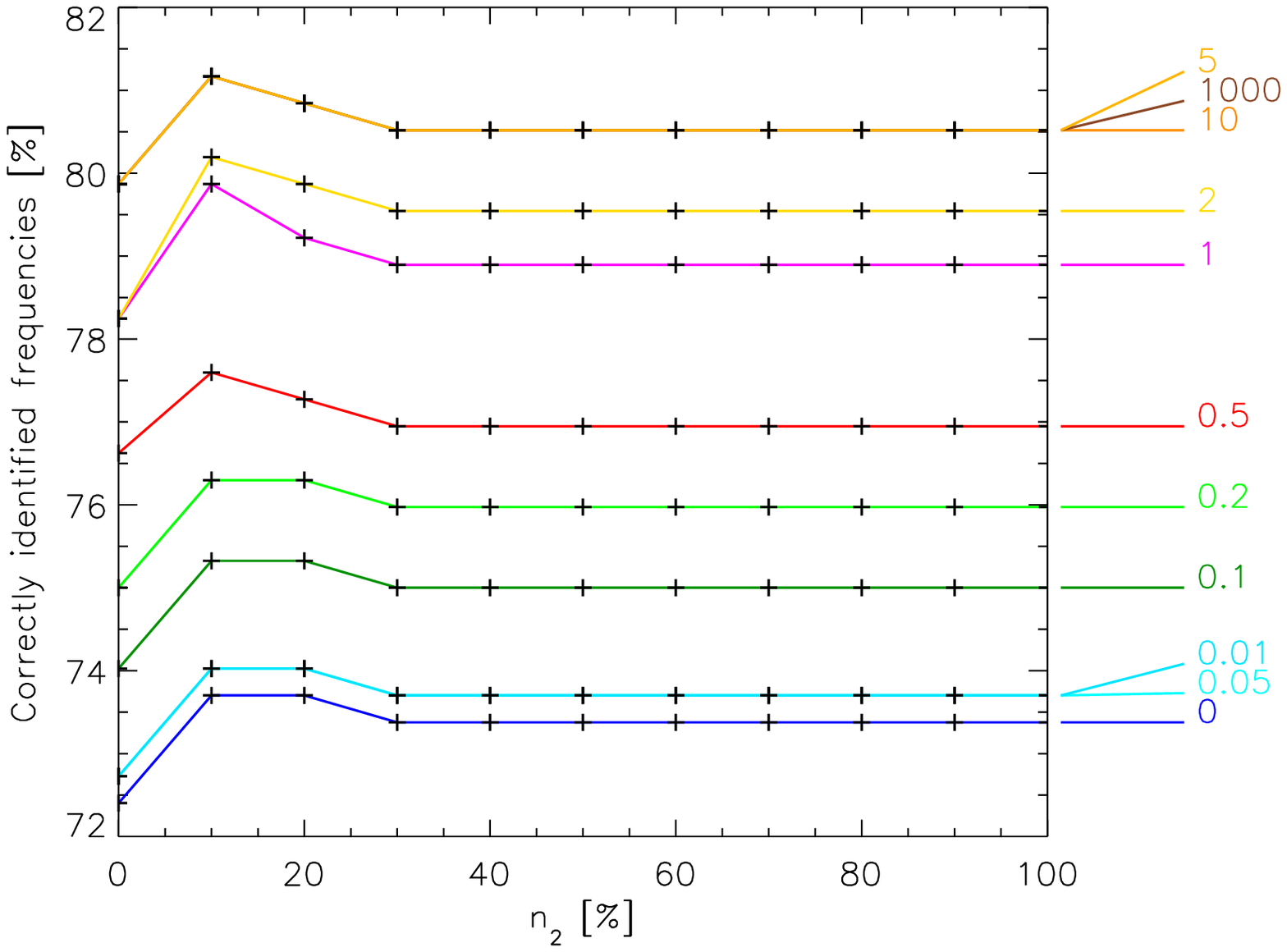}
\figcaption{\footnotesize Fraction $n_\omega$ of correctly identified frequencies $\omega\subscript{max}^{(2)}$ as a function of model parameters $n_1$~(lines) and $n_2$~($x$-axis) for variable stars in data set LRa02a (left) and LRa02b (right), respectively. \label{fig:foundfrac}}
\end{figure*}

The frequencies found in Paper~I and by manual reanalysis of the data set are in good agreement with the values of $\omega\subscript{max}^{(k)}$ (see Table~\ref{tab:alg-comparison} and Figure~\ref{fig:foundfrac}). Without master spectrum division ($k=1$), about two-thirds of the frequencies are recovered. However, the yield increases to about 80\% if the procedure is applied ($k=2$).

Interestingly, the frequency exclusion from the first criterion (Equation~(\ref{eq:n1})) now has the reverse effect -- the maximal agreement is reached if it is almost switched off by setting $n_1 \geq 5$. Smaller values of $n_1$ are too restrictive and can increase the number of wrong periods by up to about 10\%. On the other hand, the exclusion of empty phases has again a small influence, although a value of $n_2=10\%$ yields a slight improvement for the majority of tested scenarios (Figure~\ref{fig:foundfrac}).

The remaining small group of variable stars detected with a different period has been analyzed carefully. The majority of them shows multi-period variation and was identified with a rational multiple (e.g., 1/7, 2/5) of the original frequency. For some stars, the original period had to be revised during the reanalysis (see also Section~\ref{sec:results}). A small rest shows amplitudes close to the noise level, such that the period could not be determined unambiguously.

\subsection{New Selection Method}\label{sec:newselmethod}

\begin{figure*}[htpc]
\includegraphics[width=0.5\textwidth]{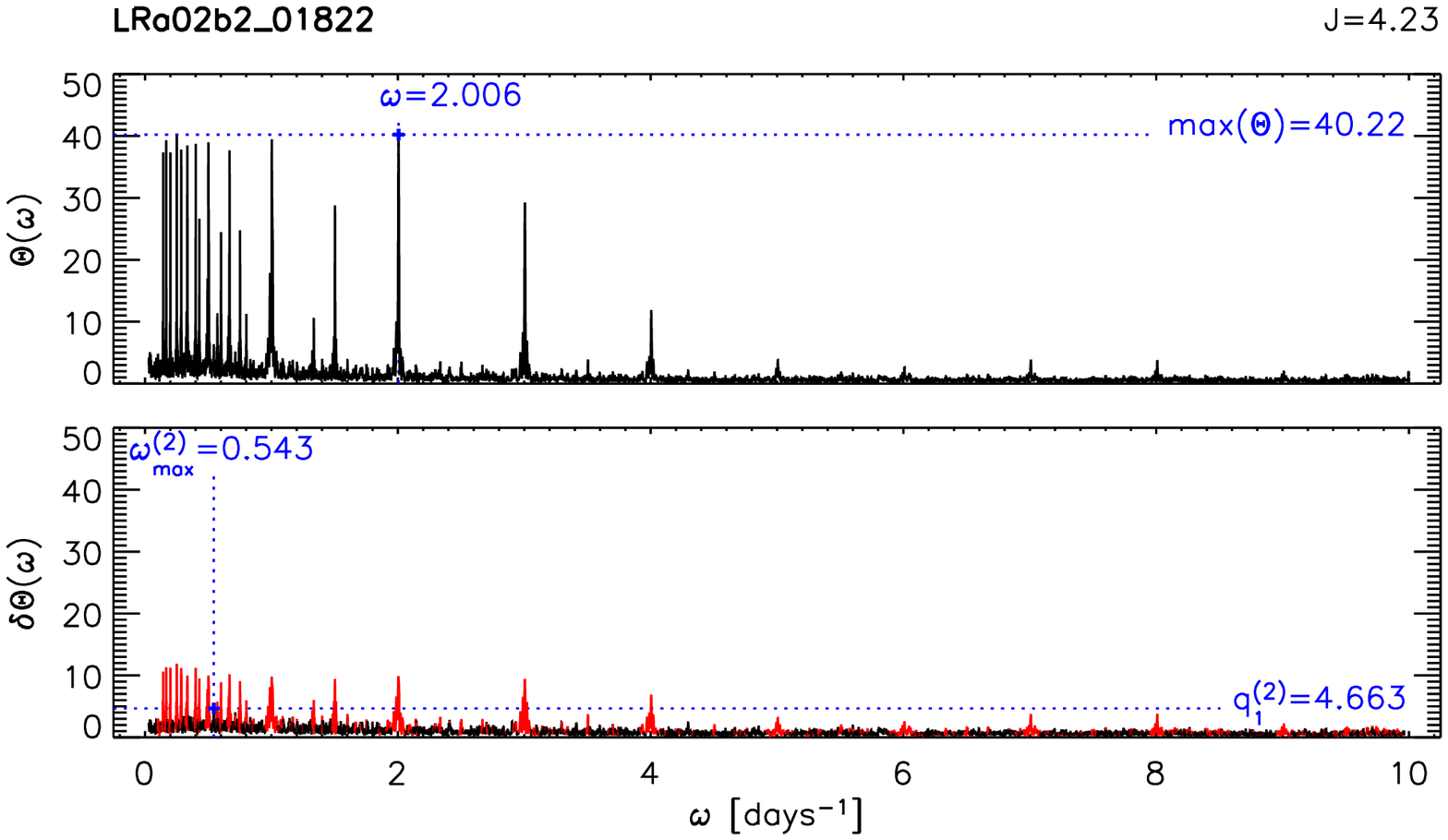}
\includegraphics[width=0.5\textwidth]{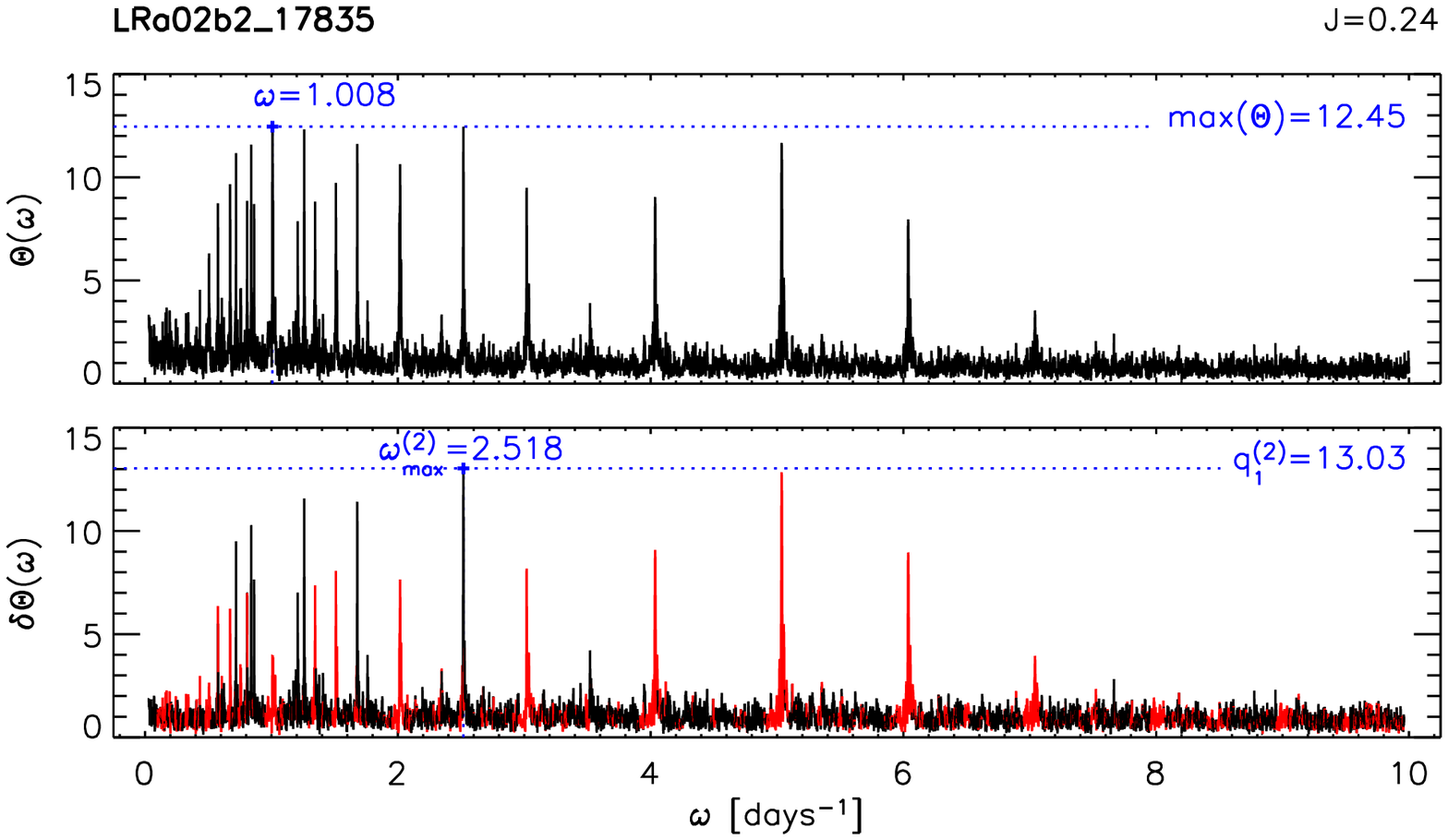}
\figcaption{\footnotesize Example for power spectra $\Theta(\omega)$ (upper plots) and $\delta\Theta(\omega)$ (lower plots, see Equation~(\ref{eq:psscale})) -- the star LRa02b2\_01822 (left plots) is strongly affected by systematic noise, while LRa02b2\_17835 (right plots) shows both physical and systematic variability. In each plain AoV spectrum $\Theta(\omega)$ (upper plots), the position of the overall maximum is marked -- it is found at systematic frequencies for both cases. Furthermore, the maximum of $\Theta(\omega)$ and the $J$~index are both much larger for the first star, leading to a false alarm when using these quantities to rank the variability. The functionality of the new variable star search algorithm is shown in the lower plots. In addition to the division by the master power spectrum, systematic frequencies are filtered out (Equations~(\ref{eq:n1})--(\ref{eq:n12}) with parameters $n_1=0$ and $n_2=10\%$, marked red in the spectrum), and the maximum $\omega\subscript{max}^{(2)}$ is determined on the non-systematic subset of frequencies (Equation~(\ref{eq:wmax2})). To use the corresponding maximum $q_1^{(2)}=\delta\Theta(\omega\subscript{max}^{(2)})$ for ranking is much more sensitive to real variability. The second star LRa02b2\_17835 is a new detection of this paper (Table~\ref{tab:vars1} and Figure~\ref{fig:lcs:new}).  \label{fig:specexamples}}
\end{figure*}

Based upon the results from the comparison, a new procedure was set up to search for variable stars within the BEST project.
\begin{enumerate}
	\item The $J$~index is used to exclude non-variable stars in order to save computation time. High values of $J$ can originate from either real variability or systematic trends, leading to a very high false alarm rate when being used as the only ranking criterion. However, low $J$~indices give a reliable criterion for non-variability, i.e., neither physical nor systematic variations. In the studied data set LRa02, \textit{no} star shows clear variability below $J=0.1$. This limit is therefore used for analyses of BEST II data sets, for which it typically excludes 50\%--75\% of all stars. 
	\item The AoV algorithm is applied with $N=7$ harmonics to the selected subset ($J\geq 0.1$) in order to obtain power spectra.
	\item The improved selection method is applied to rank all investigated stars. Following the results from Section~\ref{sec:limitimprove4}, a master power spectrum is calculated, the number of empty boxes is counted for each test period, and each individual power spectrum is divided by the master power spectrum (Equation~(\ref{eq:psscale})). For the ranking, the frequency $\omega\subscript{max}^{(2)}$ is determined from the subset of non-systematic frequencies in the divided spectrum $\delta\Theta(\omega)$ by following Equations~(\ref{eq:n1})--(\ref{eq:n12}) and (\ref{eq:wmax2}) with the parameters $n_1=0$ and $n_2=10\%$. The corresponding maximum $\delta\Theta(\omega\subscript{max}^{(2)})$ (Equation~(\ref{eq:q12})) serves as the quantity $q$ for priorization (see example in Figure~\ref{fig:specexamples}). However, in order to improve the final period $\omega_f$, $\omega\subscript{max}^{(2)}$ is recalculated without exclusion of systematic frequencies from the master power spectrum, i.e., by setting $n_1\rightarrow\infty$, $n_2=10\%$ and applying Equations~(\ref{eq:n1})--(\ref{eq:n12}) and (\ref{eq:wmax2}) again.
	\item All light curves are folded with their respective final periods $\omega_f$ and analyzed visually in descending order of $q=\delta\Theta(\omega\subscript{max}^{(2)})$.
\end{enumerate}

\section{RESULTS}\label{sec:results}
In addition to the 350 variables already published in Paper~I, 279 stars in LRa02 were identified with clear periodic variability (114 in LRa02a and 165 in LRa02b). Furthermore, we identified 52 suspected periodic variable stars (17 in LRa02a and 35 in LRa02b). For the latter, the quality of the light curves is not sufficient to fully exclude systematic errors as sources of variability, the folded light curves are partly incomplete or the point-spread functions of two neighboring stars overlap each other such that the signal cannot be well separated. The number of detections in this work is compared with Paper~I in Table~\ref{tab:oldnew}.

\begin{table}[htpc]
\begin{center}\footnotesize
\caption{Variable Star Detections in BEST II Data Set LRa02 -- Summarized Counts for Paper~I \citep{LRa02} and This Work. \label{tab:oldnew}}
\begin{tabular}{lccc}
\tableline\tableline                       & LRa02a  & LRa02b  & Total \\
\tableline
Paper~I                & 173 (4) & 177 (1) & 350 (5) \\
This paper             & 114 (2) & 165 (5) & 279 (7) \\
This paper (suspected) &  17 (0) &  35 (0) &  52 (0) \\
\tableline
Total                  & 304 (6) & 377 (6) & 681 (12) \\
\tableline
\end{tabular}
\end{center}
\footnotesize\textbf{Note.} The number of previously known variables in the field confirmed by BEST II is given in brackets (included in first number).
\end{table}

\subsection{Application of the New Method to LRa02}\label{sec:results:newmethod}
A large number of 189 new variable stars was already identified in the BEST~II data set LRa02 by the first step of the reanalysis, the manual screening (see Section \ref{sec:reanalysis}). After the search procedure was tested and optimized using the results from Paper~I and this additional sample of detections, the most successful sorting method (see description in Section \ref{sec:limitimprove}) was finally applied to search the data set LRa02 once more.

\begin{figure*}[htpc]\centering
  \includegraphics[width=\linewidth]{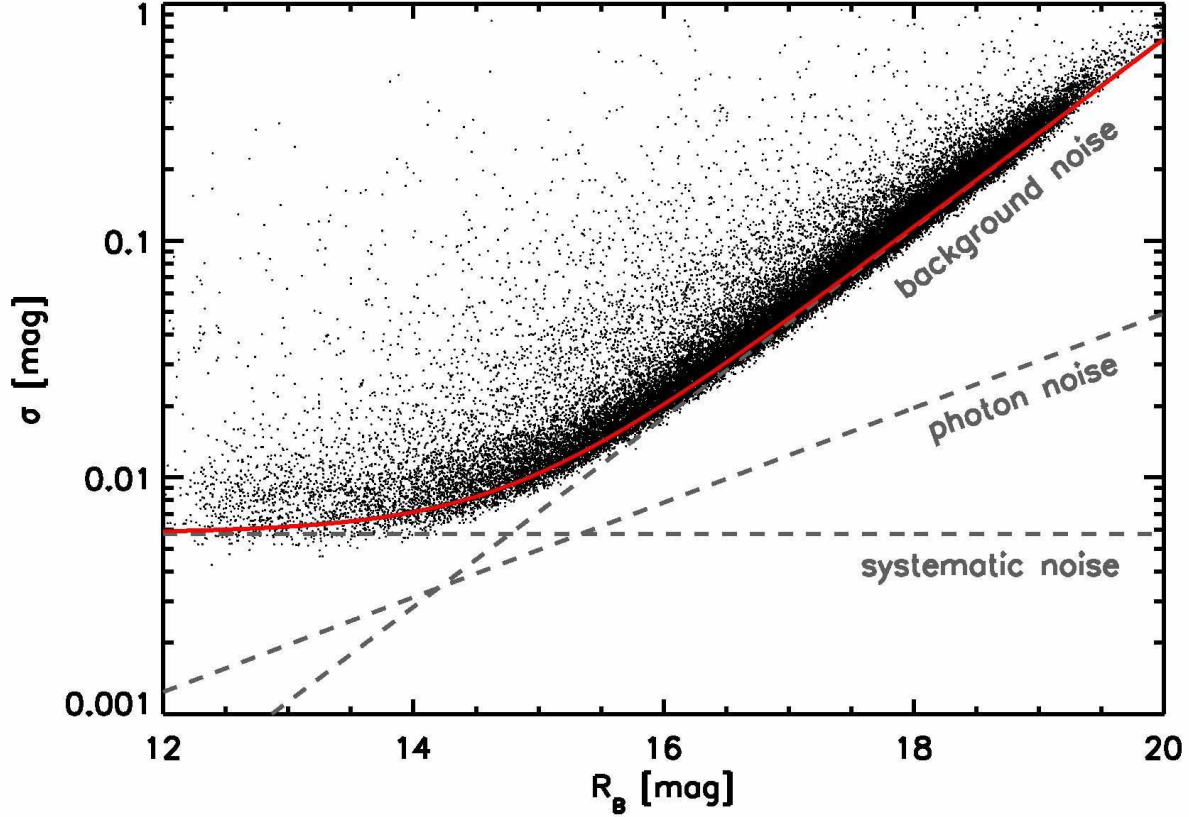}
\figcaption{\footnotesize Photometric quality of the BEST~II data set LRa02a (in line with LRa02b). The standard deviation $\sigma$ of each light curve is plotted vs. the BEST~II instrumental magnitude $R_B$ for each star. The red fit line indicates the lower noise limit $\sigma_B\superscript{min}(R_B)$ for this data set.\label{fig:rmsplot}}
\end{figure*}

In addition to the 350 variable stars published in Paper~I and the 189 found by the manual reanalysis, this improved search yielded another 135 previously unknown variable stars. Reasons why these went undetected by previous searches are as follows.
\begin{itemize}
	\item Systematic trends or aliases were found instead of the real periodicity.
	\item The AoV algorithm was run with $N=2$ harmonics for Paper~I and the manual reanalysis, but with $N=7$ for the latest search. This leads to a higher sensitivity for non-sinusoidal variations, which is particularly important for the detection of Algol type eclipsing binaries (at least 27 additional EA variables with long periods ($P>2$~days) can be attributed to the increase of~$N$, some of them being very eccentric).
	\item A total of 21 variable stars on the edge of the BEST II FOV with few datapoints were excluded by earlier reductions of the data set.
\end{itemize}
Due to an improved data quality (Figure~\ref{fig:rmsplot}) and increased sensitivity for non-sinusoidal events, we could also refine the periods for 17 of the variable stars published in Paper~I (see Table~4).

\begin{figure*}[t]
\includegraphics[width=0.32\textwidth]{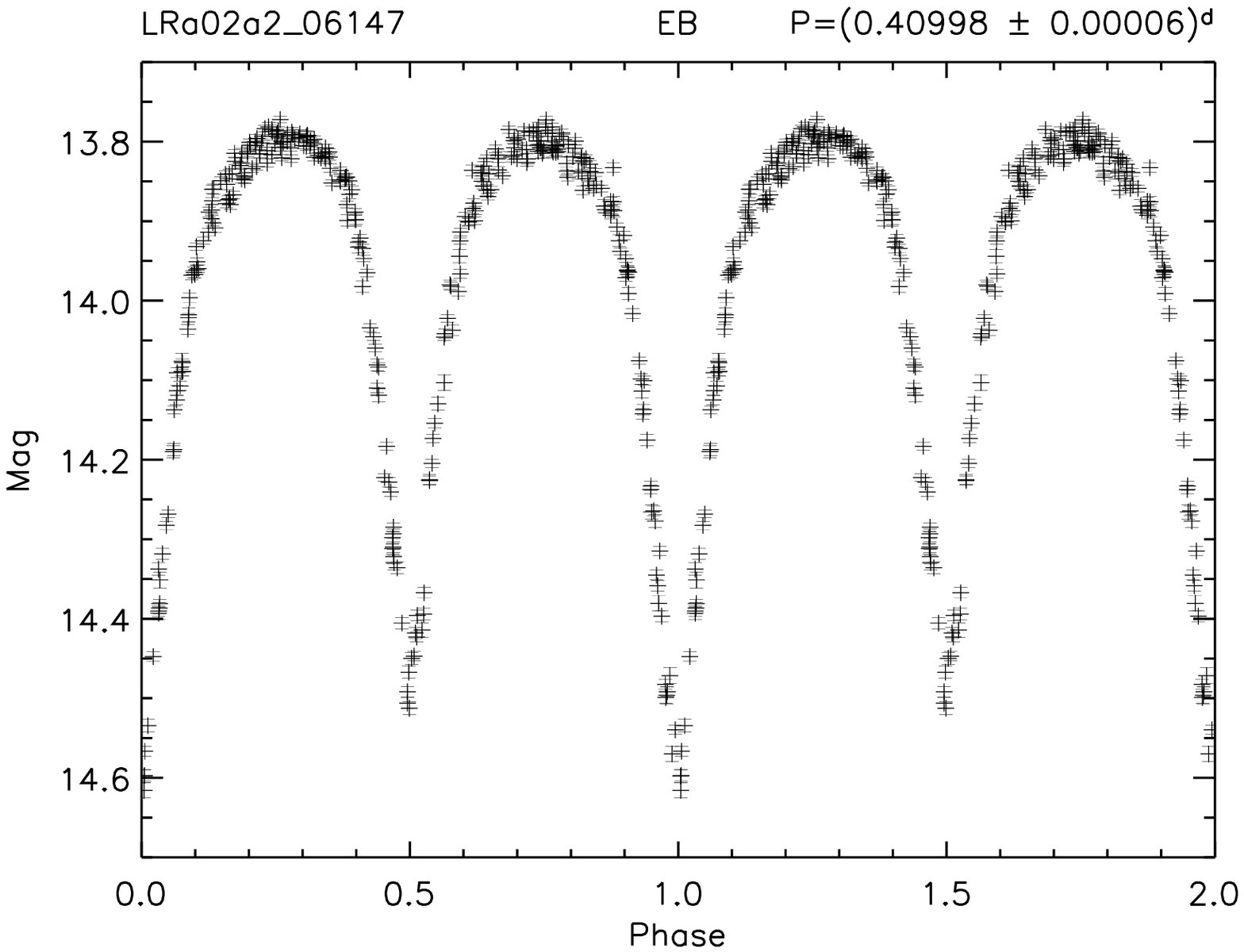}
\includegraphics[width=0.32\textwidth]{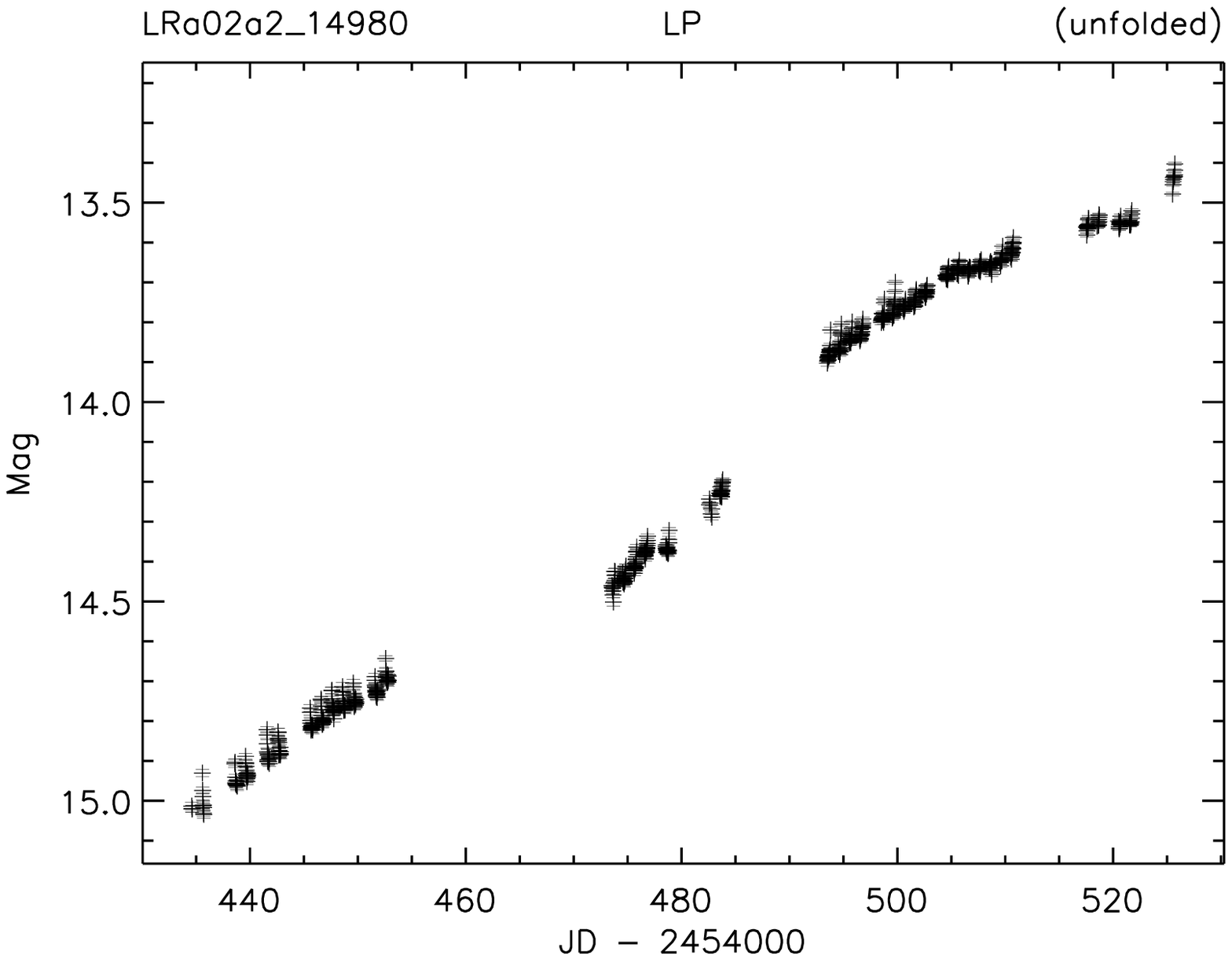}
\includegraphics[width=0.32\textwidth]{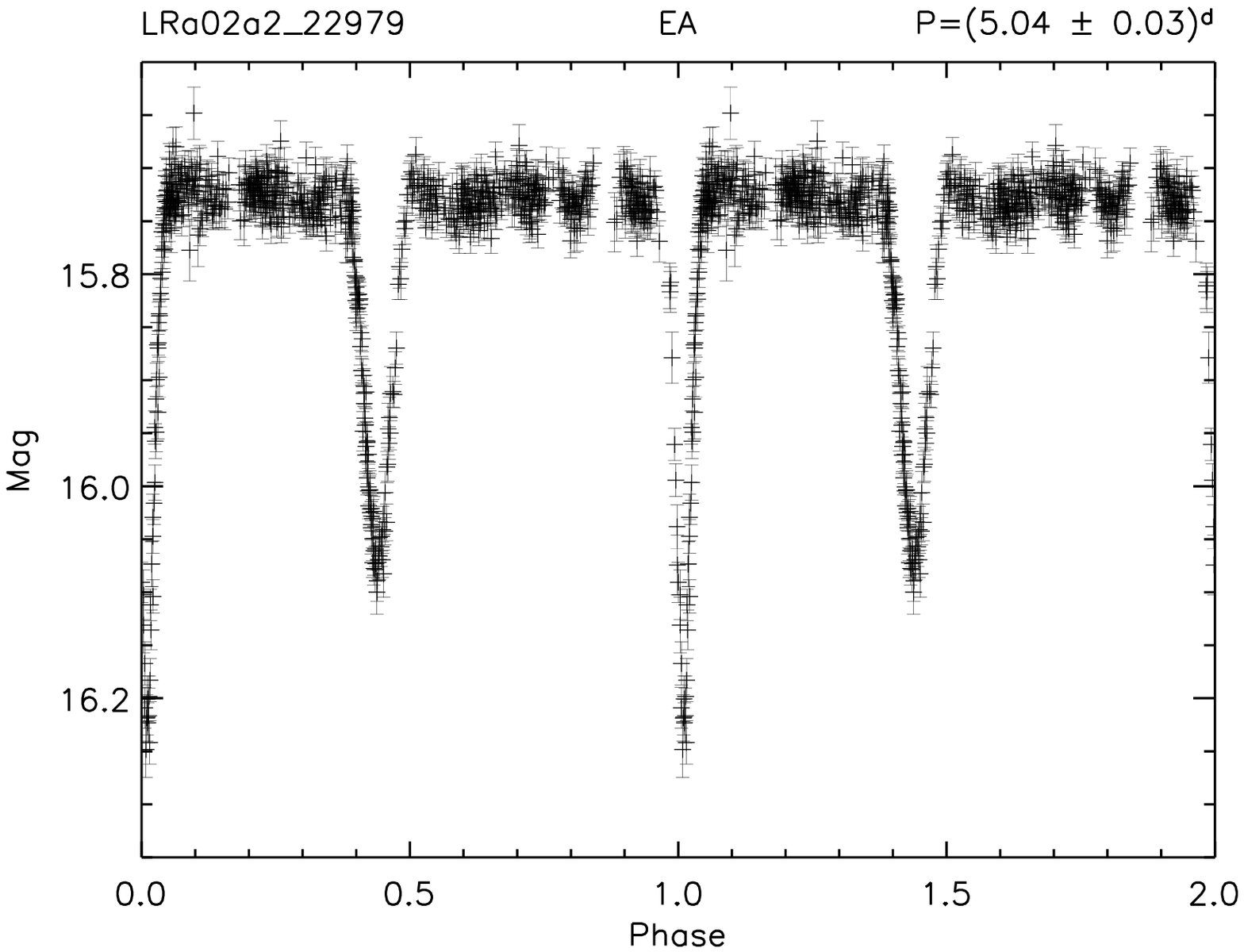}
\includegraphics[width=0.32\textwidth]{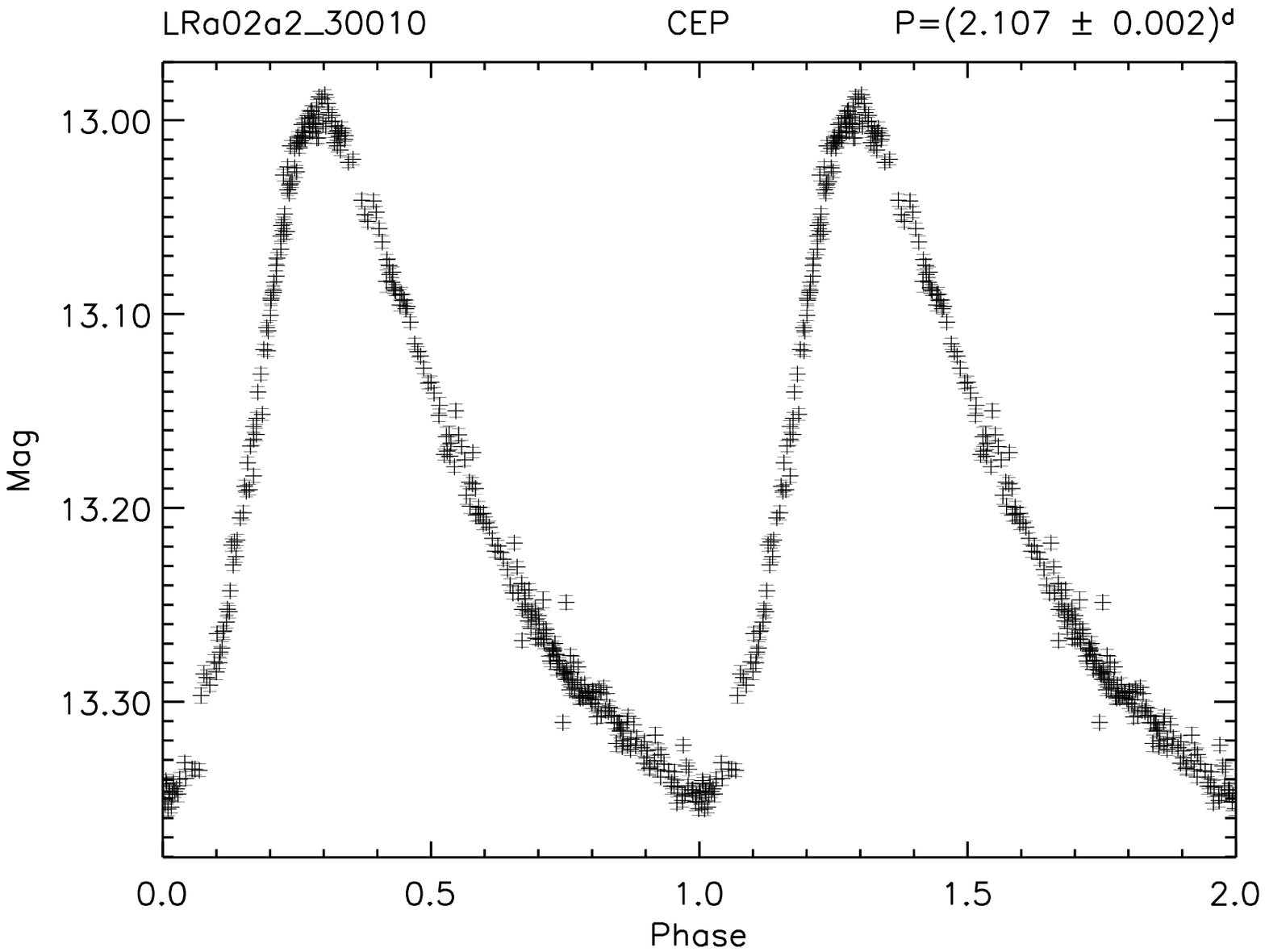}
\includegraphics[width=0.32\textwidth]{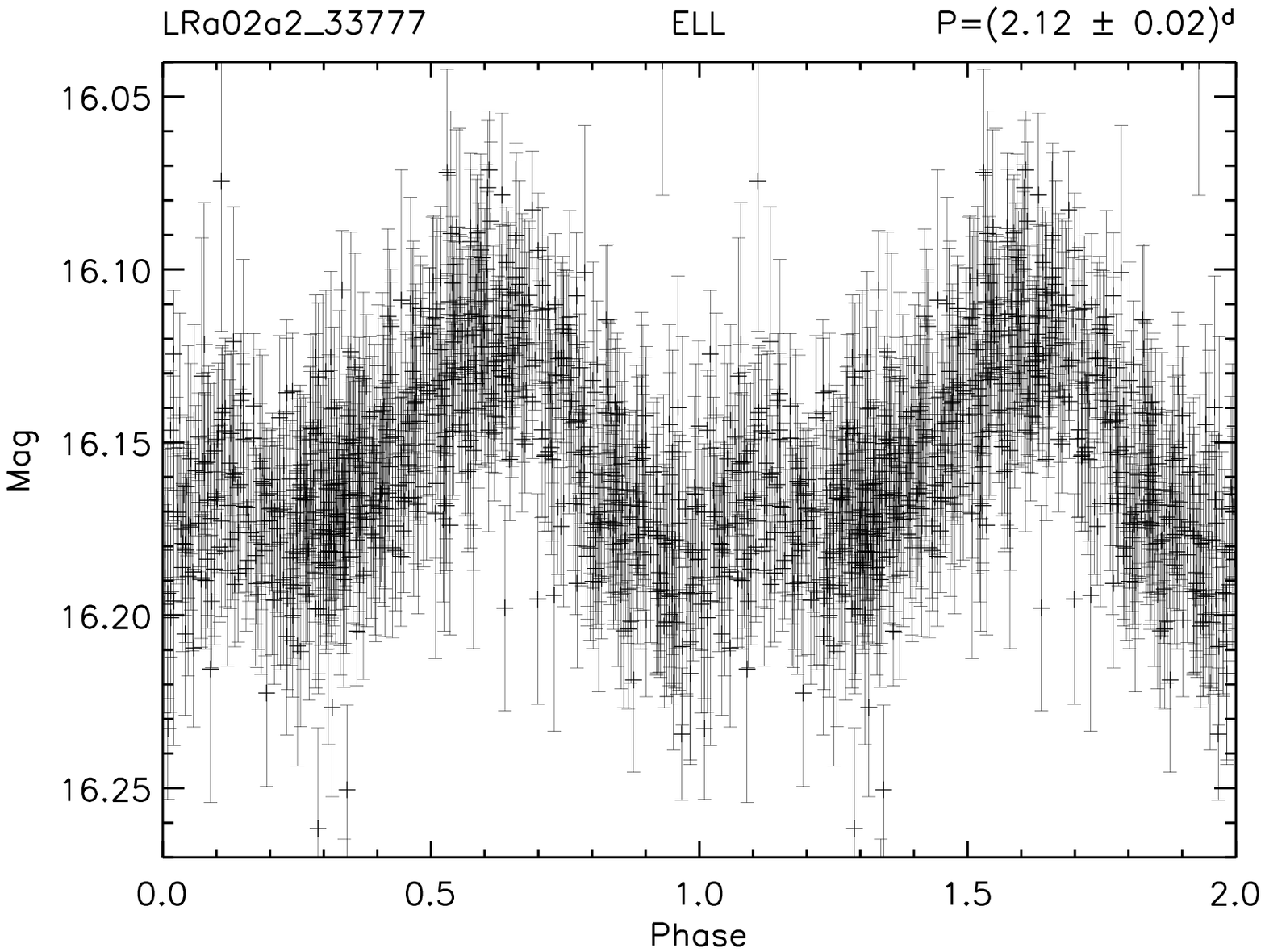}
\includegraphics[width=0.32\textwidth]{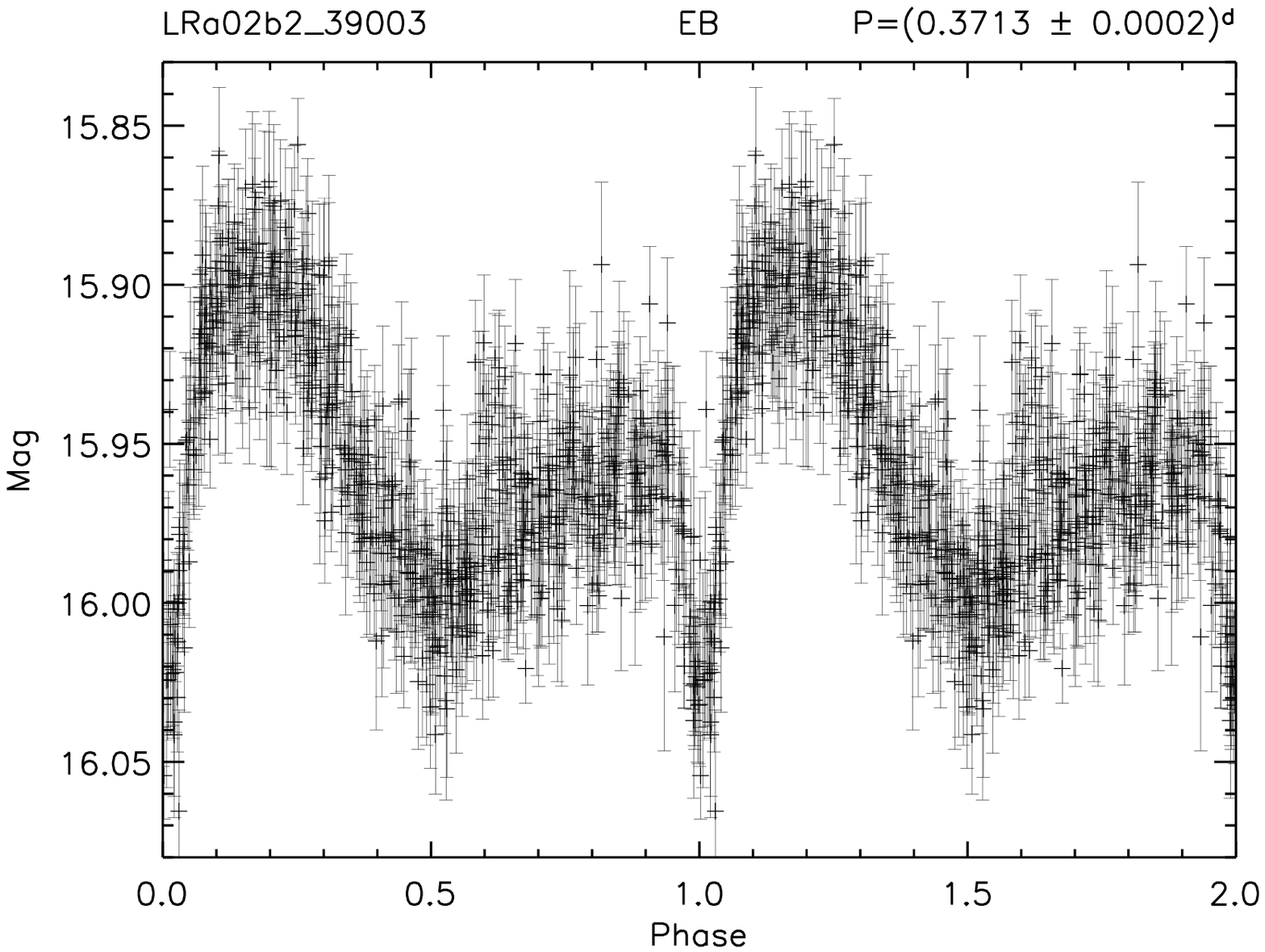}
\figcaption{\footnotesize Phase-folded light curves of variable stars detected in field LRa02 after reanalysis of the data set.
\newline (The complete figure set (331 images) is available in the electronic edition of the {\it Astronomical Journal})
\label{fig:lcs:new}}
\end{figure*}

\begin{figure*}[htpc]
\includegraphics[width=0.32\textwidth]{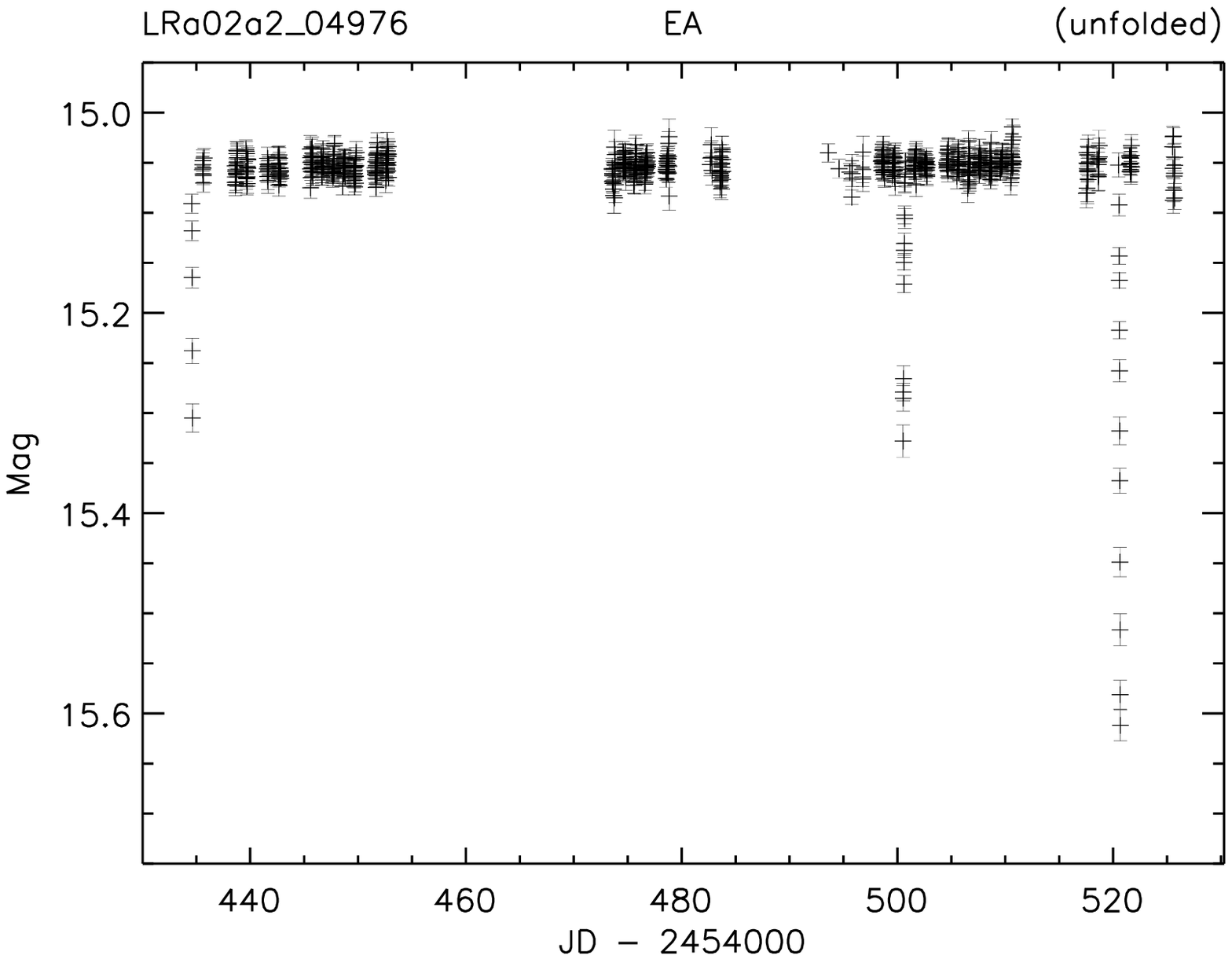}
\includegraphics[width=0.32\textwidth]{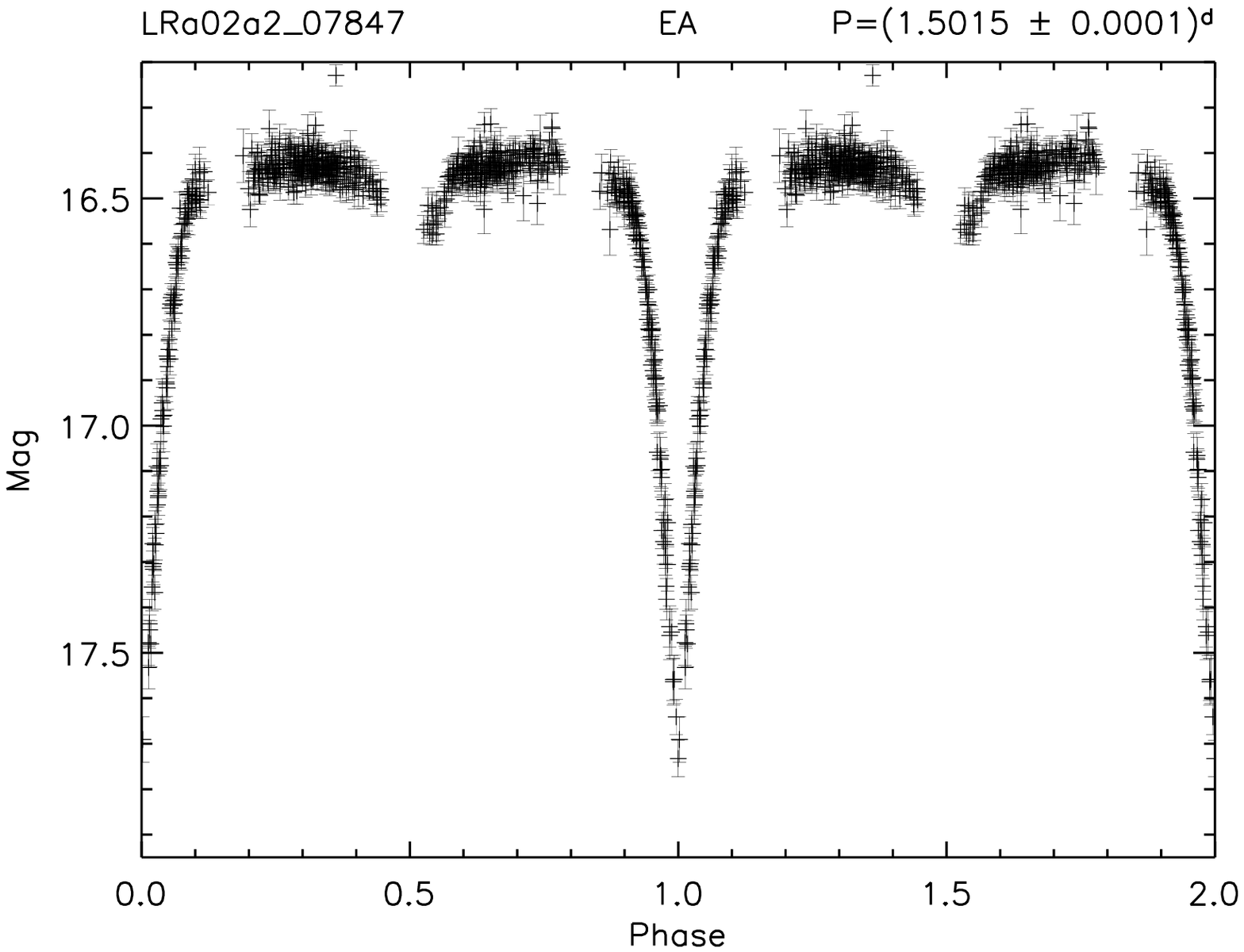}
\includegraphics[width=0.32\textwidth]{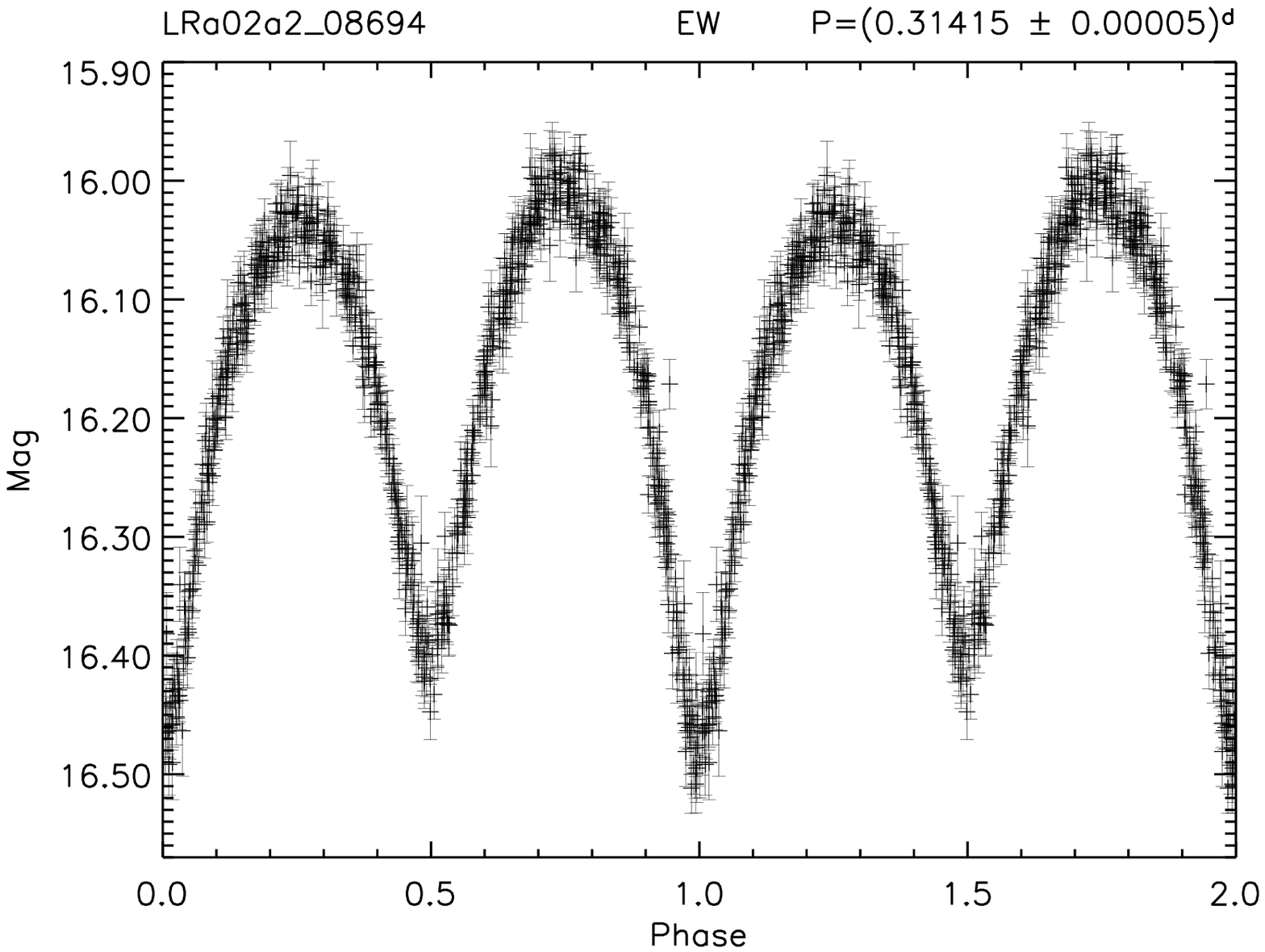}
\includegraphics[width=0.32\textwidth]{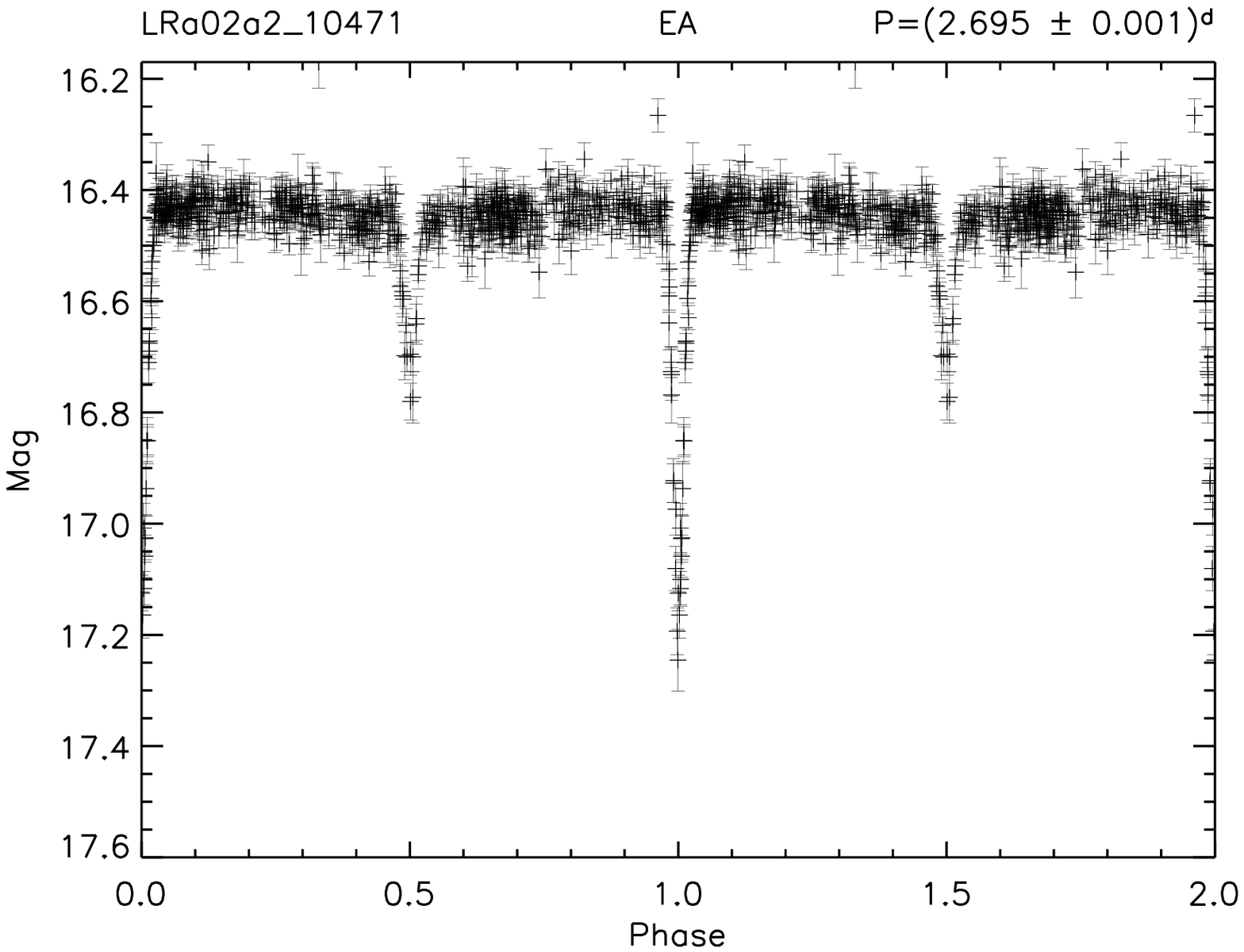}
\includegraphics[width=0.32\textwidth]{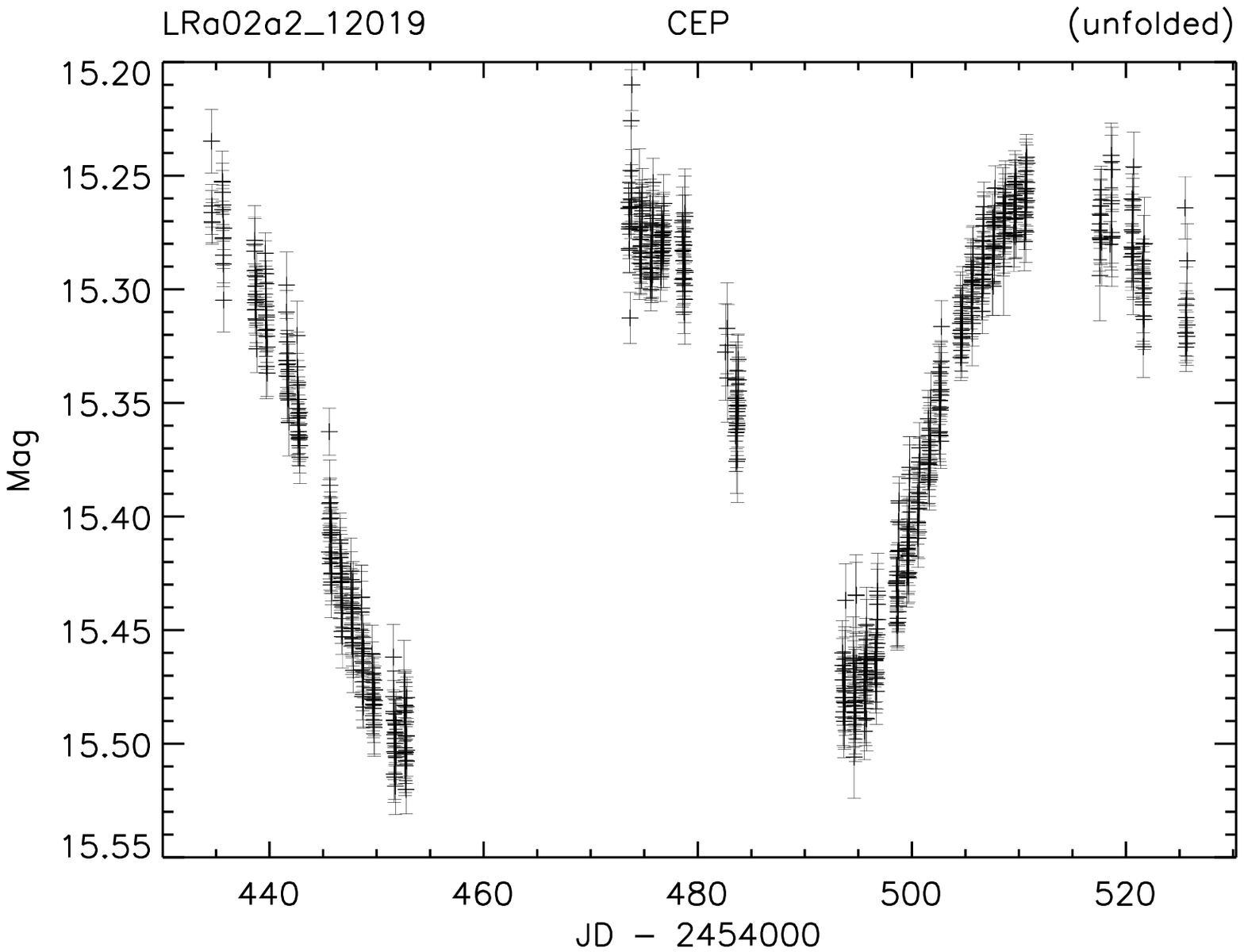}
\includegraphics[width=0.32\textwidth]{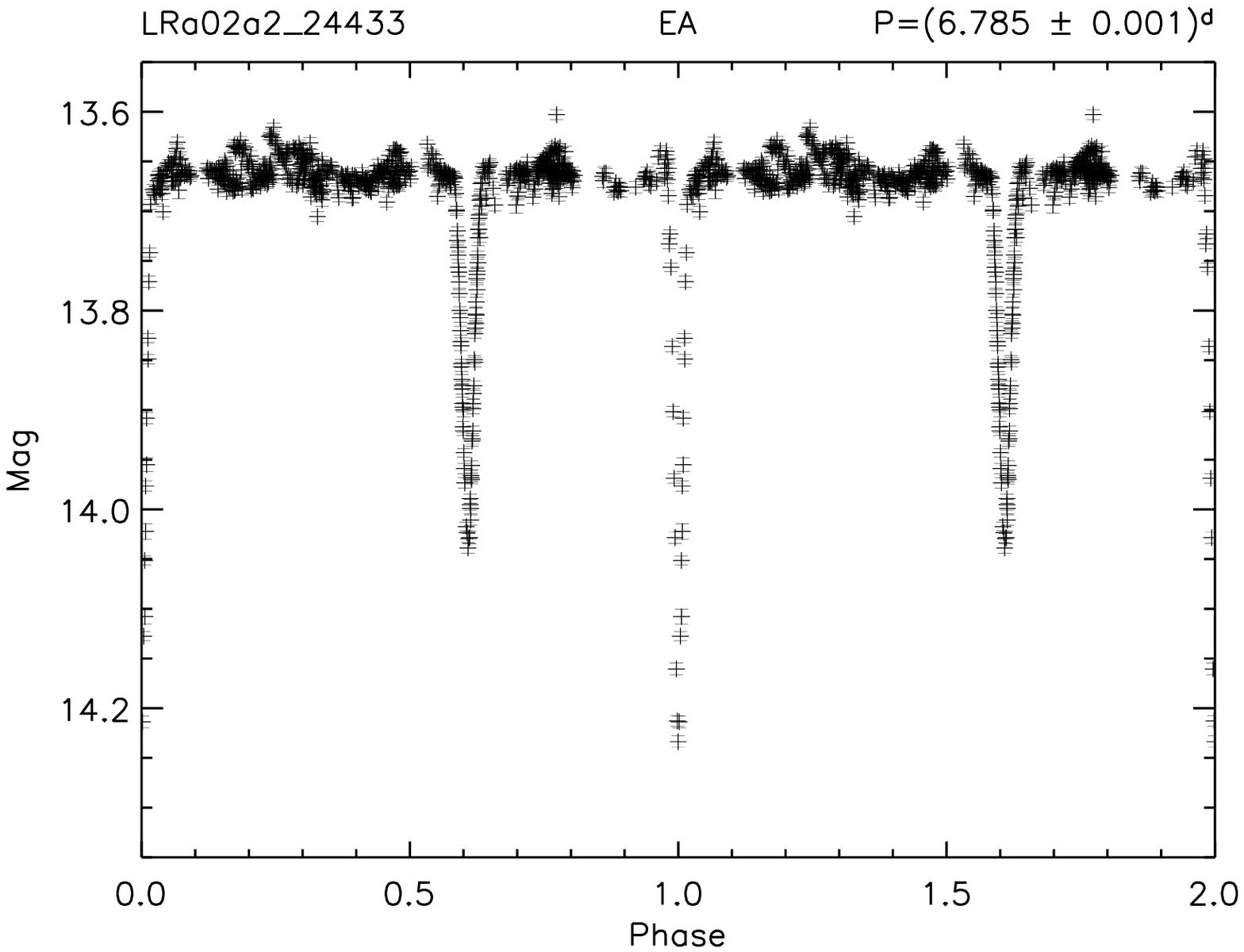}
\includegraphics[width=0.32\textwidth]{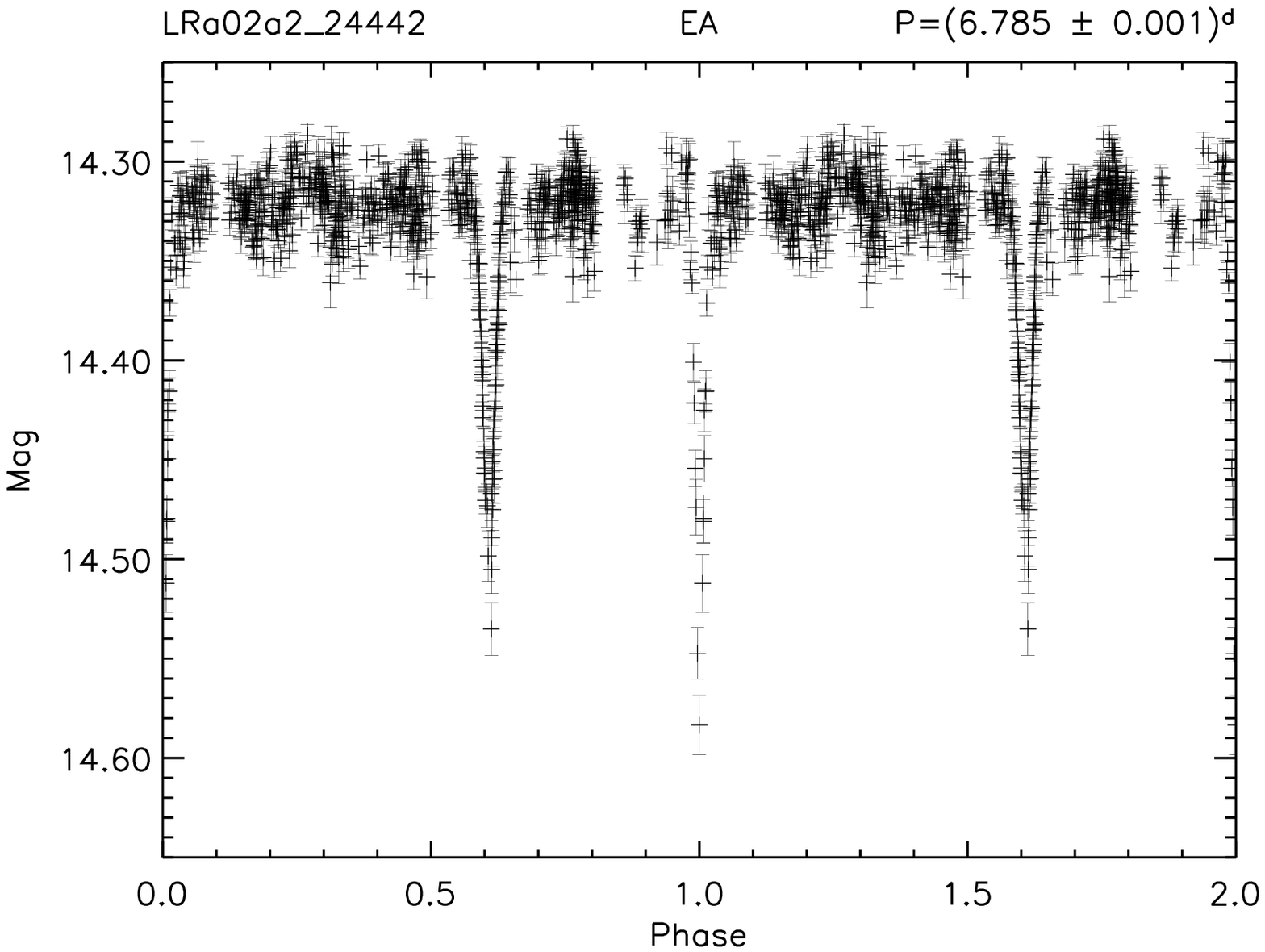}
\includegraphics[width=0.32\textwidth]{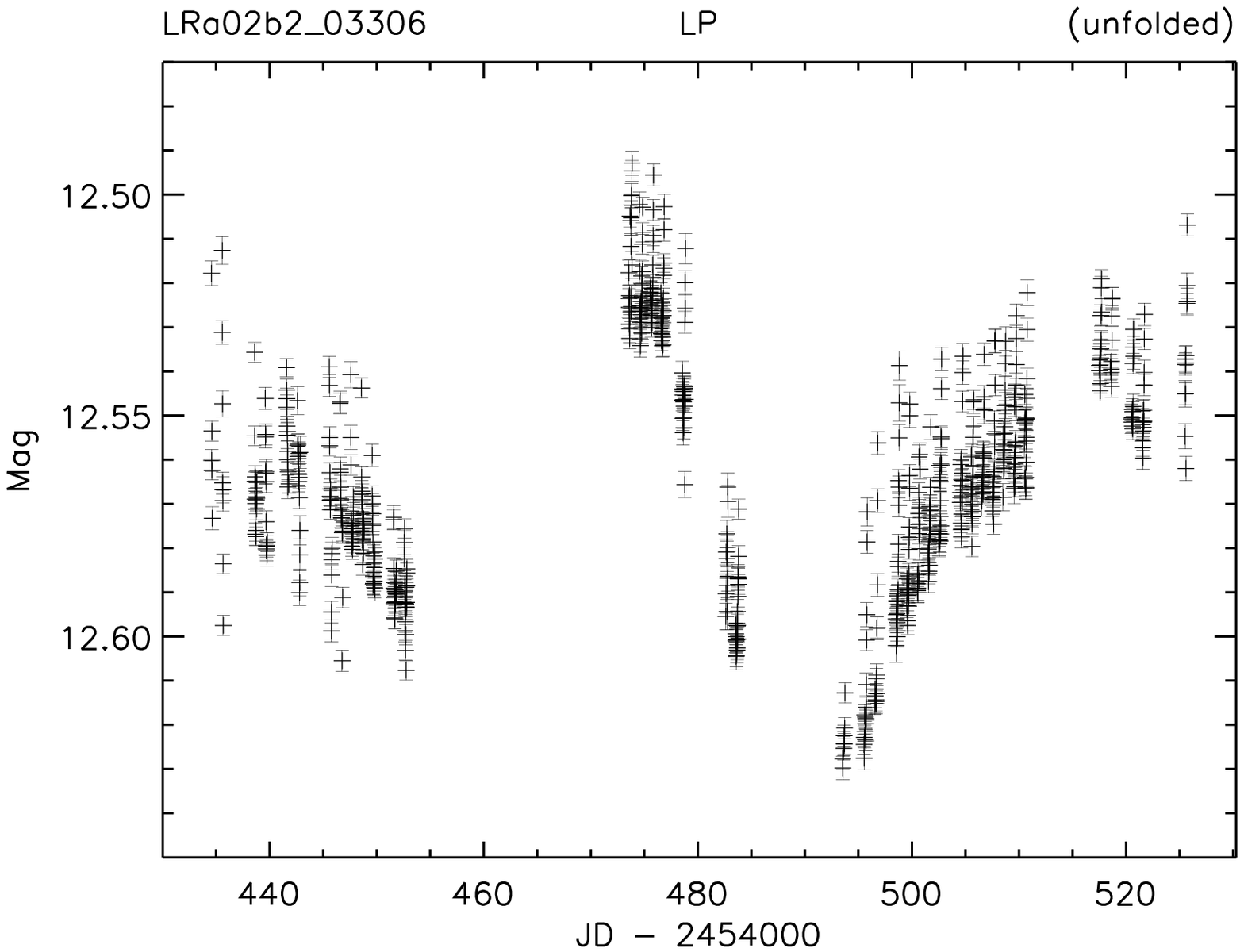}
\includegraphics[width=0.32\textwidth]{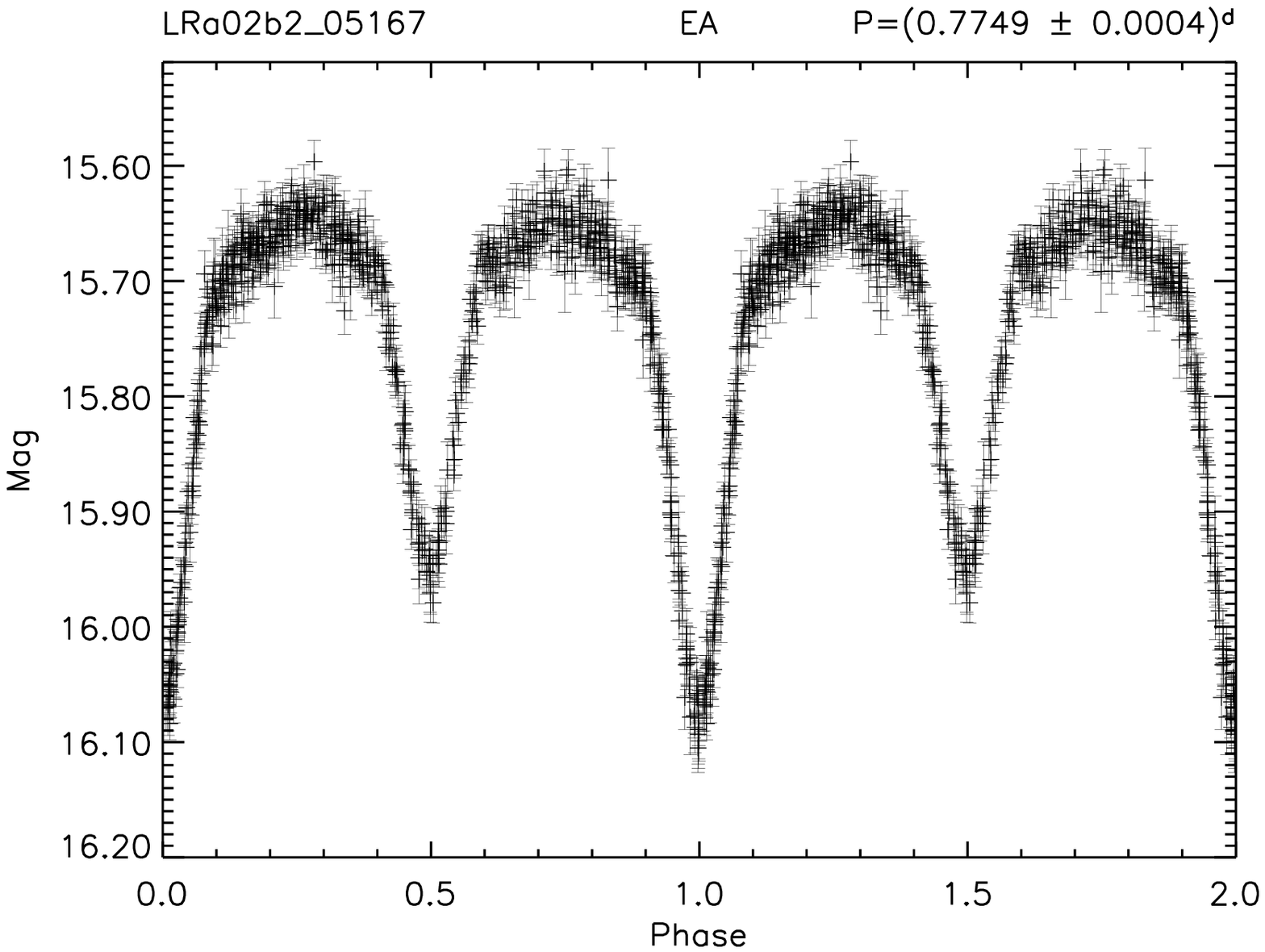}
\includegraphics[width=0.32\textwidth]{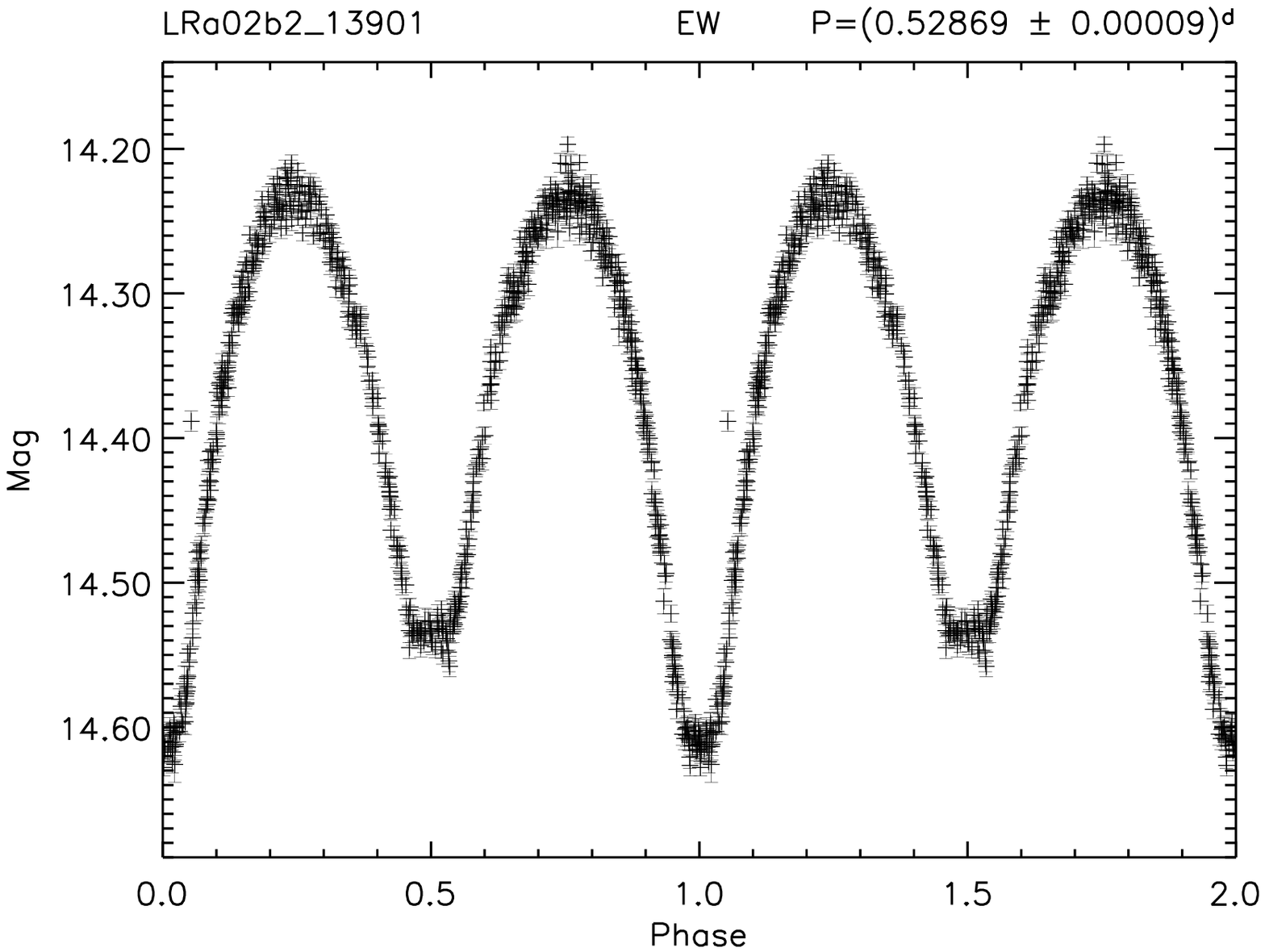}
\includegraphics[width=0.32\textwidth]{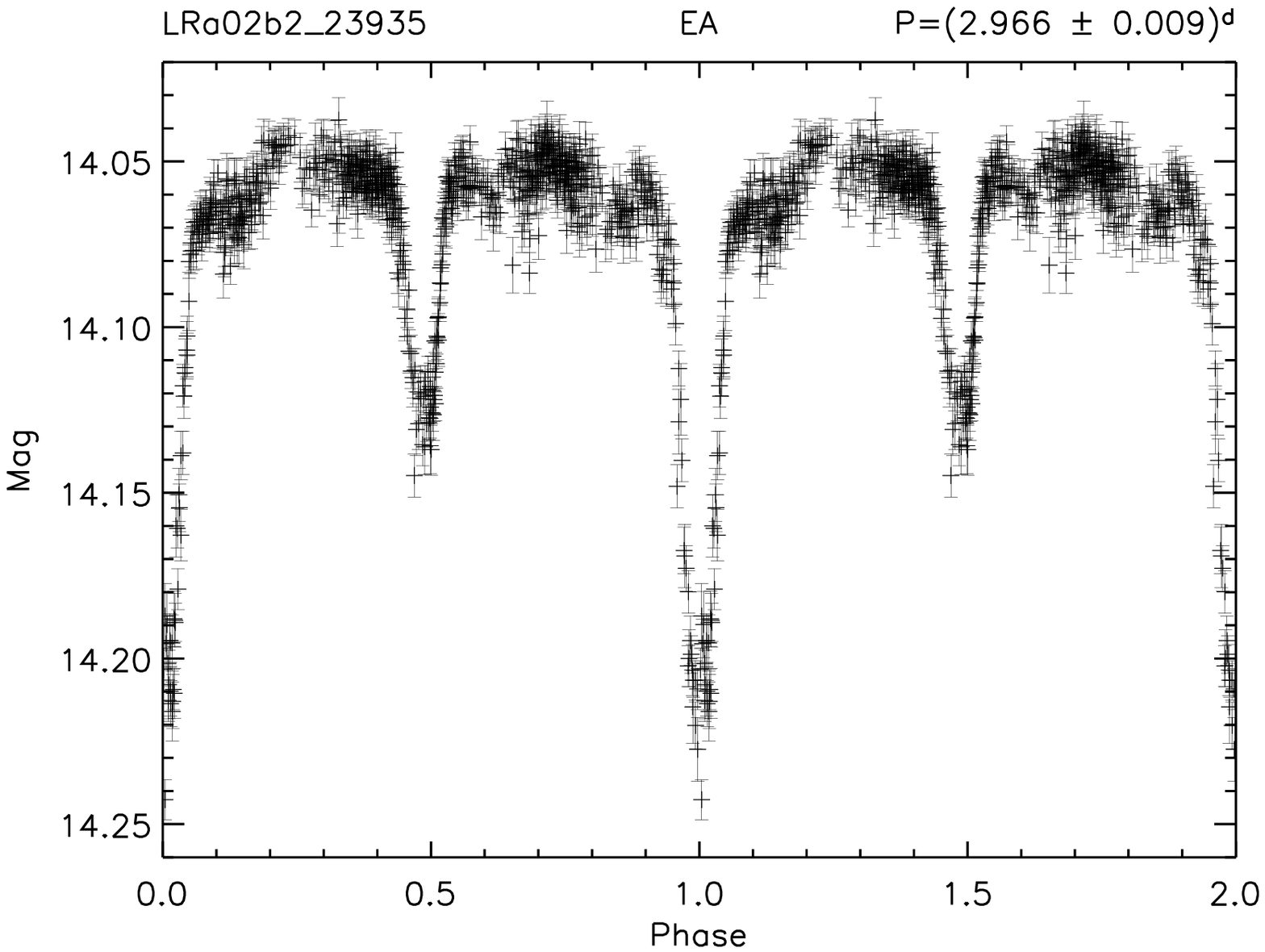}
\includegraphics[width=0.32\textwidth]{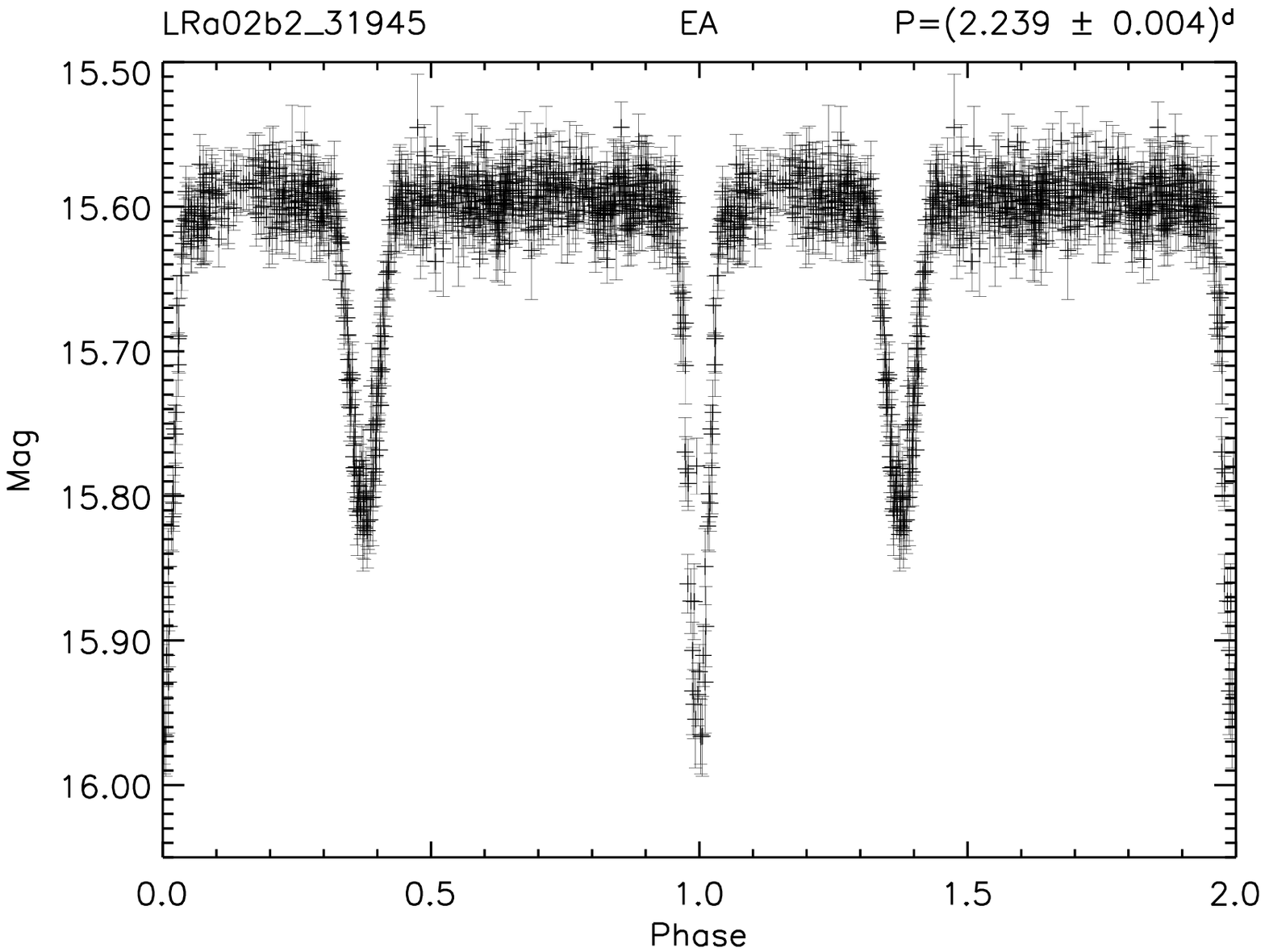}
\includegraphics[width=0.32\textwidth]{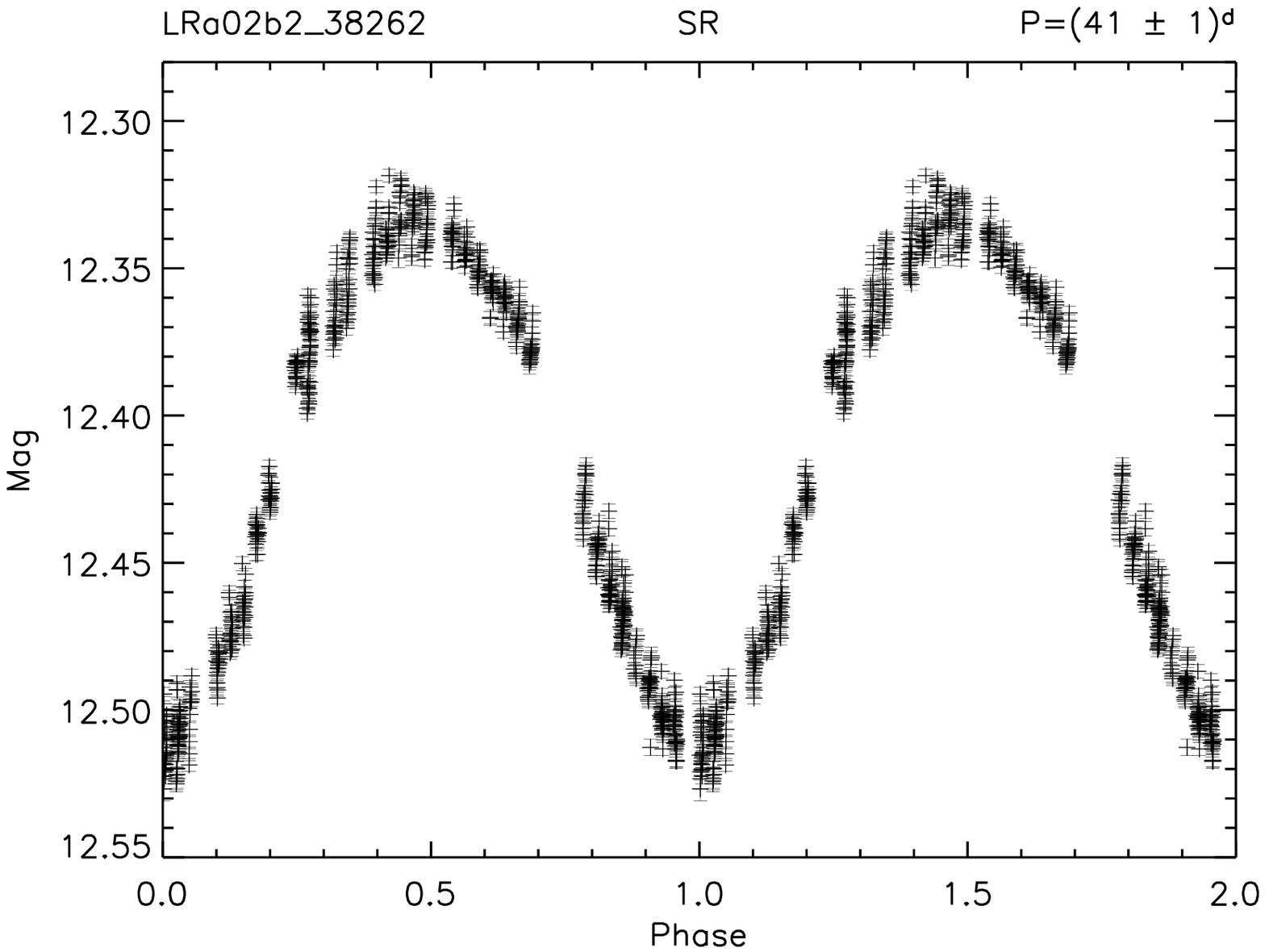}
\includegraphics[width=0.32\textwidth]{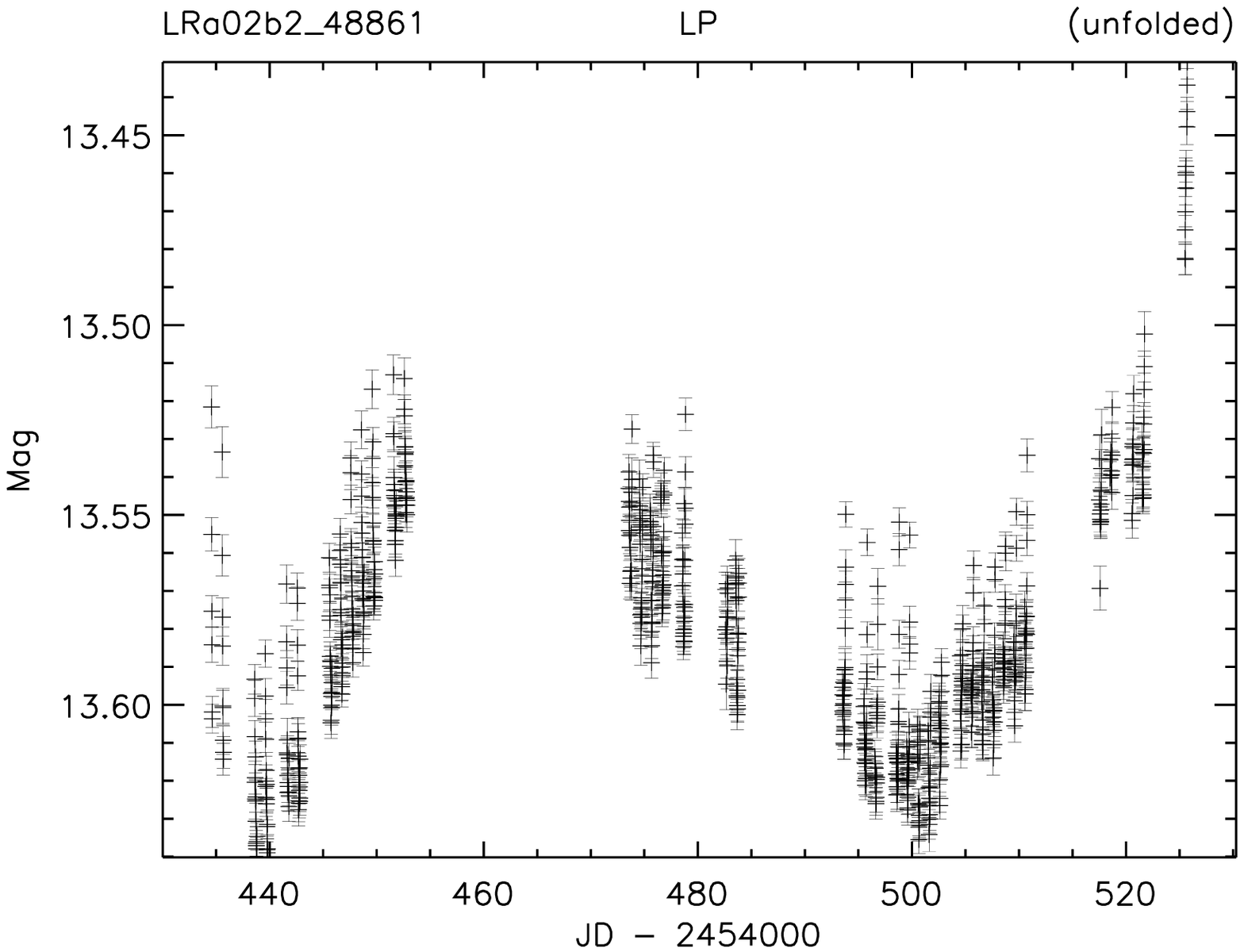}
\includegraphics[width=0.32\textwidth]{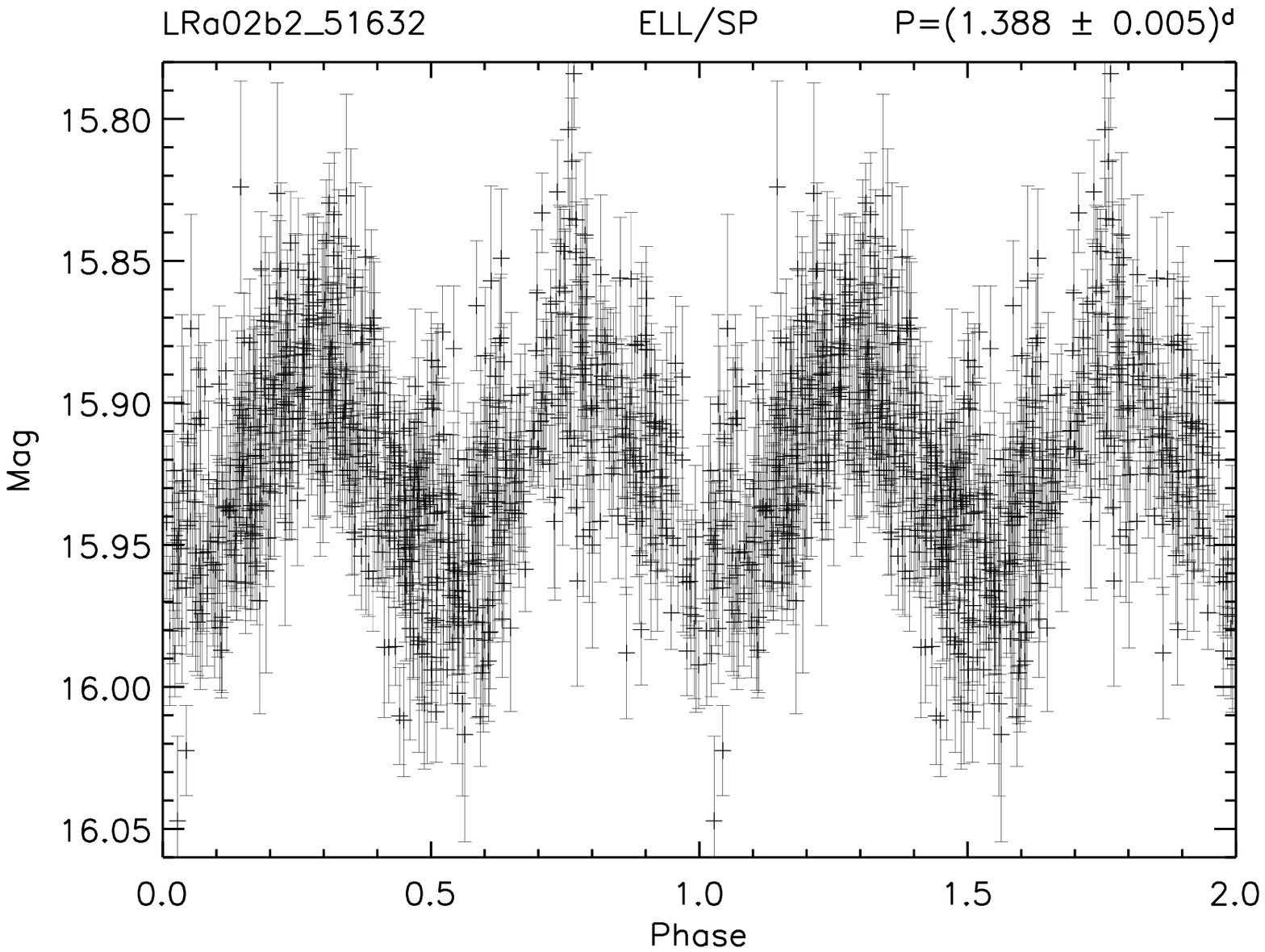}
\includegraphics[width=0.32\textwidth]{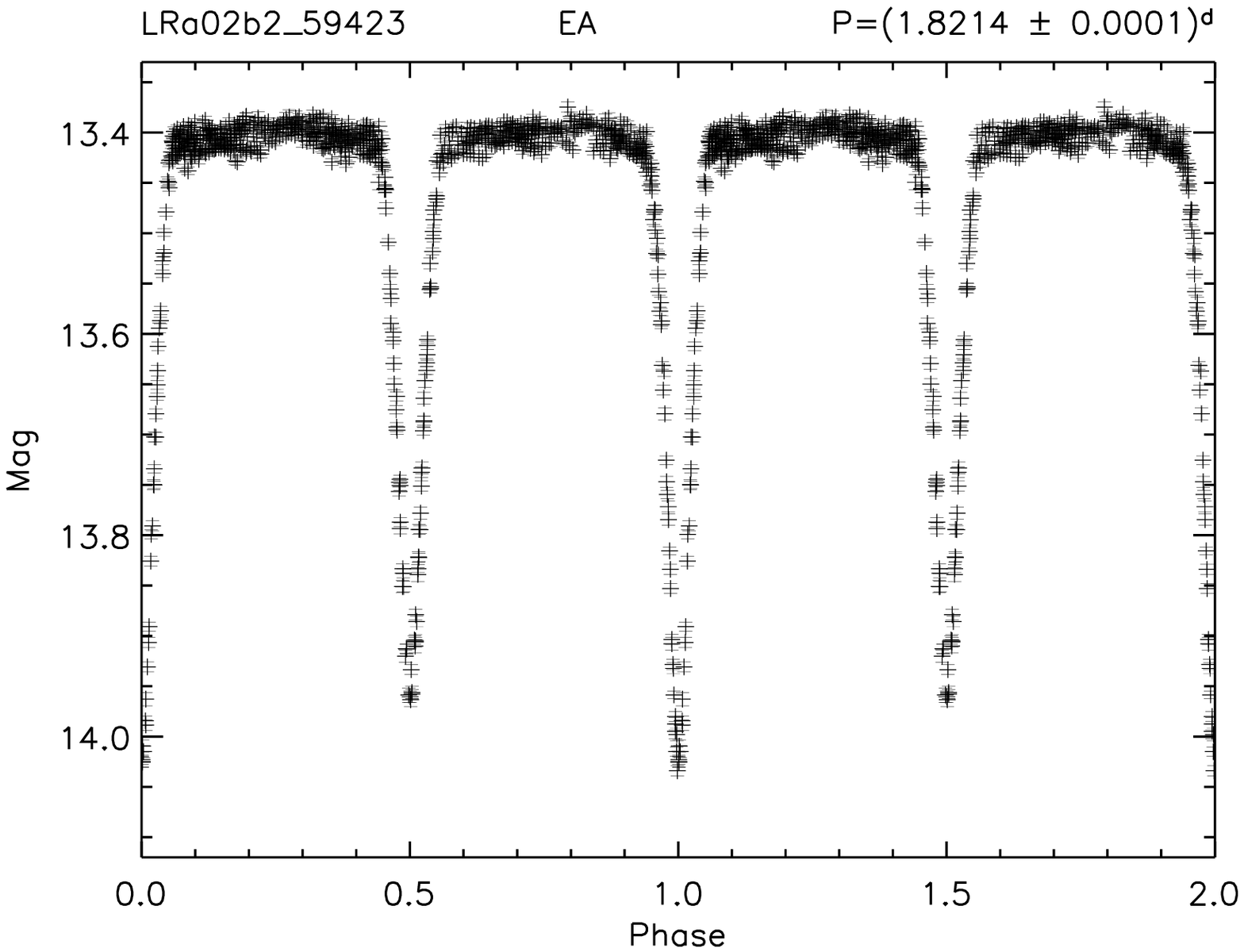}
\includegraphics[width=0.32\textwidth]{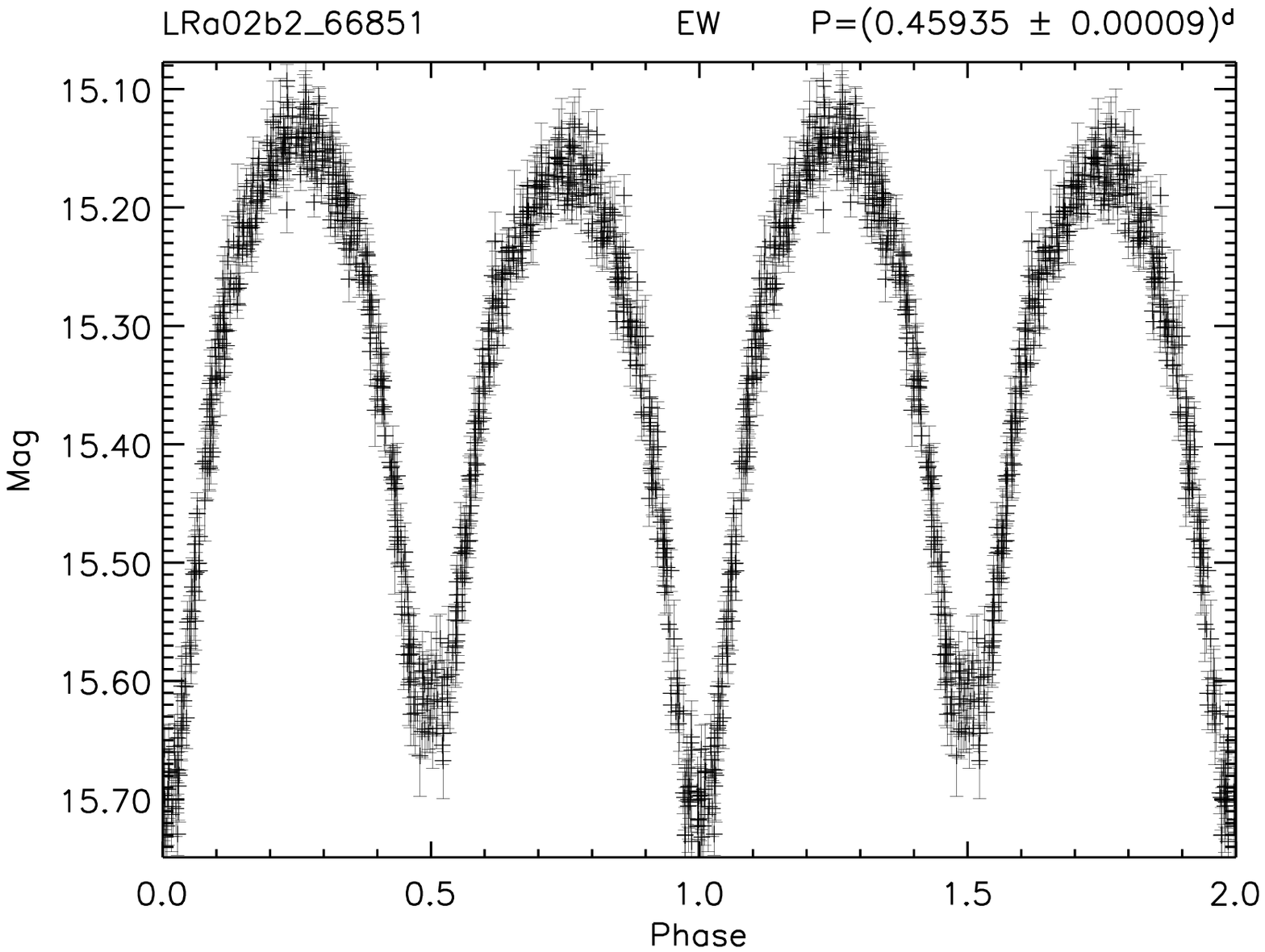}
\figcaption{\footnotesize Phase-folded light curves of known variable stars with revised parameters. 
\label{fig:lcs:revised}}
\end{figure*}

\subsection{Classification of Variability}
The newly discovered periodic variable stars are classified by the shape, amplitude, and period of their folded light curves according to the scheme used by the GCVS. We identified pulsating stars of the Delta Scuti type~(DSCT), RR Lyrae type~(RR), and Cepheid variables~(CEP). Eclipsing binary star systems were classified as Algol type~(EA), Beta Lyrae type~(EB), or W Ursae Majoris type~(EW). Stars having sinusoidal-like light curves and showing eclipses are marked as rotating ellipsoidal variables~(ELL), whereas some light curves exhibit characteristic features of spotted stars~(SP). In case the type of variability could not be determined from the light curve alone, we classified a star as VAR, meaning that further observations are needed to constrain the proper physical origin of stellar variability. Stars that are variable on time scales longer than the observational baseline are classified as long periodic~(LP). This class can also include non-periodic variables.

\subsection{Catalog of New Variables in LRa02}\label{sec:results:cat}
The newly identified periodic variable stars of the BEST II data set LRa02 are listed in Table~\ref{tab:vars1}. A set of stars from Paper~I with revised ephemerides is presented in Table~\ref{tab:vars2}. 
Due to the reanalysis, the internal numbering is not consistent with Paper~I, which is why IDs were given different prefixes (i.e., \textit{LRa02a2} and \textit{LRa02b2}, respectively). Corresponding IDs of the nearest Two Micron All Sky Survey (2MASS) object are given if the catalog coordinates do not differ from our astrometry by more than 2 arcsec. In some cases, two neighboring stars could be spatially resolved, but their photometric apertures overlap so that the actual source of variation is ambiguous. Such objects have been marked with a ``c'' (crowded), whereas suspected variables are indicated with an ``s'' flag. All instrumental magnitudes given are obtained without any filter so that their absolute value should only be considered as a rough estimation.\footnote{The zero point was shifted to minimize the residuals with the $R$ magnitude of USNO-A2, yielding an rms deviation with the catalog of about 0.5 mag for all matched stars.} Amplitudes and ephemerides ($T_0$ in rHJD=HJD-2{,}454{,}400) are the results of the AoV algorithm. However, after visual inspection, several periods have been refined manually (usually to rational multiples of the initial result).

\begin{sidewaystable*}[htpc]\scriptsize\centering
\caption{Catalog of Variable Stars Detected in CoRoT Field LRa02 after Reanalysis of the Data Set, Sorted by Internal BEST~II Identifiers. \label{tab:vars1}}
\begin{tabular}{lccccccccccl}
\hline\hline
BEST~ID & {Flag} & {2MASS~ID} & {$\alpha(J2000.0)$} & {$\delta(J2000.0)$} & 
{$R_B$ (mag)} &{T$_0$ [rHJD]} & {$P$ (d)} & {$A$ (mag)} & {$J$~Index} & {Type} & {Other Names}
\\\hline
\multicolumn{12}{c}{LRa02a}\\\hline
LRa02a2\_00759  &    & 06473611-0352264  &$06^h47^m36.1^s$  &$-03^\circ52'26\arcseccomma 6$  &18.11  &  34.638  &        $0.828 \pm 0.002$  &       $0.4 \pm 0.2$  & 0.306 &EA  &\\
LRa02a2\_01976  &    & 06483833-0312255  &$06^h48^m38.3^s$  &$-03^\circ12'25\arcseccomma 6$  &16.34  &  35.163  &      $0.8348 \pm 0.0006$  &     $0.30 \pm 0.05$  &  1.37 &EA/SP  &\\
LRa02a2\_03383  &    & 06481122-0346301  &$06^h48^m11.2^s$  &$-03^\circ46'30\arcseccomma 4$  &17.13  &  35.097  &        $1.294 \pm 0.005$  &     $0.12 \pm 0.07$  & 0.323 &CEP  &\\
LRa02a2\_04701  &    & 06472903-0432171  &$06^h47^m29.0^s$  &$-04^\circ32'17\arcseccomma 2$  &15.25  &  38.772  &            $8.6 \pm 0.2$  &     $0.04 \pm 0.02$  & 0.445 &ELL/SP  &\\
LRa02a2\_05225  &    & 06483988-0339178  &$06^h48^m39.9^s$  &$-03^\circ39'17\arcseccomma 9$  &16.51  &  34.937  &      $0.4350 \pm 0.0004$  &     $0.12 \pm 0.07$  & 0.812 &ELL/SP  &\\
LRa02a2\_05867  &    & 06475987-0416581  &$06^h47^m59.9^s$  &$-04^\circ16'58\arcseccomma 3$  &17.72  &  34.841  &      $0.5255 \pm 0.0005$  &       $0.4 \pm 0.2$  & 0.457 &EB  &\\
LRa02a2\_06108  &    & 06475271-0424518  &$06^h47^m52.7^s$  &$-04^\circ24'51\arcseccomma 9$  &17.00  &  34.601  &        $2.747 \pm 0.008$  &     $0.22 \pm 0.07$  & 0.469 &EA  &\\
LRa02a2\_06147  &    & 06492768-0308050  &$06^h49^m27.7^s$  &$-03^\circ08'05\arcseccomma 2$  &13.87  &  34.711  &    $0.40998 \pm 0.00006$  &     $0.73 \pm 0.03$  &  23.6 &EB  &\\
LRa02a2\_06168  &    & 06483769-0348472  &$06^h48^m37.7^s$  &$-03^\circ48'47\arcseccomma 3$  &16.15  &  35.256  &      $0.8230 \pm 0.0006$  &     $0.16 \pm 0.03$  & 0.807 &EA  &\\
LRa02a2\_06363  &$s$ & 06482994-0356402  &$06^h48^m29.9^s$  &$-03^\circ56'40\arcseccomma 3$  &15.80  &  35.944  &        $1.646 \pm 0.007$  &     $0.04 \pm 0.03$  & 0.548 &ELL  &\\
LRa02a2\_06993  &    & 06482849-0403248  &$06^h48^m28.5^s$  &$-04^\circ03'24\arcseccomma 8$  &13.82  &  34.705  &      $0.4248 \pm 0.0002$  &   $0.106 \pm 0.008$  &  5.75 &EW/ELL  &CoRoT 300002950\\
LRa02a2\_07010  &    & 06481669-0413092  &$06^h48^m16.7^s$  &$-04^\circ13'09\arcseccomma 3$  &14.56  &  35.923  &        $2.309 \pm 0.006$  &     $0.04 \pm 0.02$  & 0.658 &EA  &CoRoT 300002493\\
LRa02a2\_07056  &    & 06484660-0349153  &$06^h48^m46.6^s$  &$-03^\circ49'15\arcseccomma 5$  &17.20  &  34.934  &      $0.5425 \pm 0.0005$  &     $0.22 \pm 0.06$  & 0.838 &EW  &\\
LRa02a2\_07090  &    & 06493379-0311214  &$06^h49^m33.8^s$  &$-03^\circ11'21\arcseccomma 3$  &17.24  &  34.925  &      $0.5427 \pm 0.0007$  &     $0.15 \pm 0.07$  & 0.374 &EW  &\\
LRa02a2\_07148  &    & 06480066-0427255  &$06^h48^m00.7^s$  &$-04^\circ27'25\arcseccomma 5$  &16.74  &  34.676  &    $0.35786 \pm 0.00008$  &     $0.60 \pm 0.05$  &  1.62 &EW  &\\
\hline
\end{tabular}
\begin{flushleft}\textbf{Notes.} Stars have been matched with the closest 2MASS object within a radius of 2 arcsec. Coordinates are given for epoch J2000.0 and were derived by an astrometric match of CCD to USNO-A2 coordinates. Given magnitudes are instrumental and reflect the CCD sensitivity as observations have been obtained without filter. Overlapping apertures of neighboring stars can lead to contaminated light curves -- such cases are marked with a ``$c$'' flag. Suspected variables are marked by ``$s$''. Amplitudes $A$ and ephemerides (variability period $P$ and times of minimum brightness $T_0$, given in rHJD=HJD-2454400) are the results of the AoV algorithm.
\newline (This Table is available in its entirety in machine-readable and Virtual Observatory (VO) forms in the online journal. A portion is shown here for guidance regarding its form and content.)
\end{flushleft}
 
\vspace{2\baselineskip}
\caption{Catalog of Periodic Variable Stars with Revised Parameters Compared to Paper~I \citep{LRa02}. \label{tab:vars2}}
\begin{tabular}{lccccccccccl}
\hline\hline
BEST~ID & {Flag} & {2MASS~ID} & {$\alpha(J2000.0)$} & {$\delta(J2000.0)$} & 
{$R_B$ (mag)} &{T$_0$ [rHJD]} & {$P$ (d)} & {$A$ (mag)} & {$J$~Index} & {Type} & {Other Names}
\\\hline
LRa02a2\_04976  &$k$   & 06473121-0432567  &$06^h47^m31.2^s$  &$-04^\circ32'56\arcseccomma 9$  &15.05  & \nodata  &       \nodata             &         \nodata      & 0.964 &EA  &lra2a\_00269\\
LRa02a2\_07847  &$k$   & 06483188-0407577  &$06^h48^m31.9^s$  &$-04^\circ07'57\arcseccomma 9$  &16.45  &  39.649  &      $1.5015 \pm 0.0001$  &       $1.3 \pm 0.1$  &  4.04 &EA  &lra2a\_00416\\
LRa02a2\_08694  &$k$   & 06491193-0342336  &$06^h49^m11.9^s$  &$-03^\circ42'33\arcseccomma 8$  &16.14  &  34.694  &    $0.31415 \pm 0.00005$  &     $0.46 \pm 0.03$  &  4.50 &EW  &lra2a\_00450\\
LRa02a2\_10471  &$k$   & 06500331-0315474  &$06^h50^m03.3^s$  &$-03^\circ15'47\arcseccomma 4$  &16.44  &  35.159  &        $2.695 \pm 0.001$  &       $0.8 \pm 0.1$  & 0.943 &EA  &lra2a\_00531\\
LRa02a2\_12019  &$k$   & 06485506-0423379  &$06^h48^m55.1^s$  &$-04^\circ23'38\arcseccomma 1$  &15.33  & \nodata  &       \nodata             &         \nodata      &  6.12 &CEP  &lra2a\_00601, CoRoT 110826631\\
LRa02a2\_24433  &$ck$  & 06520050-0326327  &$06^h52^m00.5^s$  &$-03^\circ26'32\arcseccomma 8$  &13.66  &  98.695  &        $6.785 \pm 0.001$  &       $0.6 \pm 0.1$  &  6.82 &EA  &lra2a\_01126, CoRoT 110677259\\
LRa02a2\_24442  &$ck$  & 06520077-0326255  &$06^h52^m00.8^s$  &$-03^\circ26'25\arcseccomma 7$  &14.32  &  98.695  &        $6.785 \pm 0.001$  &     $0.26 \pm 0.01$  &  1.69 &EA  &lra2a\_01127\\
LRa02b2\_03306  &$k$   & 06474385-0541125  &$06^h47^m43.8^s$  &$-05^\circ41'12\arcseccomma 7$  &12.57  & \nodata  &       \nodata             &         \nodata      &  7.77 &LP  &lra2b\_01648\\
LRa02b2\_05167  &$k$   & 06484535-0458261  &$06^h48^m45.4^s$  &$-04^\circ58'26\arcseccomma 2$  &15.69  &  34.740  &      $0.7749 \pm 0.0004$  &     $0.43 \pm 0.04$  &  3.92 &EA  &lra2b\_01600, CoRoT 300003628\\
LRa02b2\_13901  &$k$   & 06484069-0537534  &$06^h48^m40.7^s$  &$-05^\circ37'53\arcseccomma 6$  &14.33  &  35.085  &    $0.52869 \pm 0.00009$  &     $0.39 \pm 0.02$  &  14.9 &EW  &lra2b\_01437, CoRoT 300003441\\
LRa02b2\_23935  &$k$   & 06491567-0550387  &$06^h49^m15.7^s$  &$-05^\circ50'38\arcseccomma 9$  &14.06  &  35.241  &        $2.966 \pm 0.009$  &     $0.16 \pm 0.02$  &  2.53 &EA  &lra2b\_01257, CoRoT 110657689\\
LRa02b2\_31945  &$k$   &                   &$06^h50^m23.9^s$  &$-05^\circ26'29\arcseccomma 5$  &15.60  &  34.697  &        $2.239 \pm 0.004$  &     $0.24 \pm 0.04$  &  1.42 &EA  &lra2b\_01080\\
LRa02b2\_38262  &$k$   & 06505364-0527405  &$06^h50^m53.6^s$  &$-05^\circ27'40\arcseccomma 8$  &12.43  &  41.500  &               $41 \pm 1$  &     $0.18 \pm 0.01$  &  28.1 &SR  &lra2b\_00968, CoRoT 110839568\\
LRa02b2\_48861  &$k$   & 06512387-0549233  &$06^h51^m23.9^s$  &$-05^\circ49'23\arcseccomma 6$  &13.58  & \nodata  &       \nodata             &         \nodata      &  5.93 &LP  &lra2b\_00738\\
LRa02b2\_51632  &$k$   & 06511964-0605018  &$06^h51^m19.7^s$  &$-06^\circ05'02\arcseccomma 2$  &15.92  &  35.282  &        $1.388 \pm 0.005$  &     $0.08 \pm 0.05$  & 0.879 &ELL/SP  &lra2b\_00687\\
LRa02b2\_59423  &$k$   & 06522734-0545516  &$06^h52^m27.3^s$  &$-05^\circ45'51\arcseccomma 8$  &13.41  &  34.730  &      $1.8214 \pm 0.0001$  &     $0.62 \pm 0.01$  &  14.4 &EA              &lra2b\_00469\\
LRa02b2\_66851  &$k$   & 06531961-0540496  &$06^h53^m19.6^s$  &$-05^\circ40'49\arcseccomma 6$  &15.29  &  34.760  &    $0.45935 \pm 0.00009$  &     $0.57 \pm 0.03$  &  6.44 &EW  &lra2b\_00323\\
\end{tabular}
\end{sidewaystable*}

The corresponding phase-folded light curves can be found in Figures~\ref{fig:lcs:new} and \ref{fig:lcs:revised}. For stars classified as LP, light curves are shown unfolded and no ephemerides are given in Tables~\ref{tab:vars1} and~4. Light curves in a machine-readable format as well as finding charts are available upon request.

\subsection{Comparison with Known Variables}
We searched the VSX and the GCVS for previously known variable stars within a radius of 10 arcsec around each of our new detections. A total number of seven variables was found in these catalogs. For comparison, the new BEST~II results for these are included in Table~\ref{tab:vars1} and marked with ``k'' as known variables.

The three stars CoRoT 110742676, NSVS 12579155, and NSVS\,12585233 have periods longer than 50 days. BEST~II confirms their long-time periodicity, but the phase coverage of their cycles is insufficient to confirm the periods quantitatively. For the four eclipsing binaries ASAS J064835-0534.3, DY\,Mon, [KEE2007]\,1318, and [KEE2007] 1334, both the classifications and periods are completely confirmed. 

The latter two have first been detected by \citet{IR01} in the CoRoT IR01 field with BEST, for which the chosen FOV shows a small overlap with the BEST~II field LRa02a. However, the precision in the periods of these two binaries could be significantly improved because the LRa02 data set covers a much larger time period (41 compared to 12 nights) and the photometric quality of BEST~II is better.

\subsection{Comparison with CoRoT}
The LRa02 data sets of BEST~II and CoRoT are, except for the fact that they point to the same field, completely independent of each other and thus provide a great opportunity to compare the scientific results of the two surveys. 

A step in the scientific analysis of CoRoT data consists of an automatic stellar variability classification \citep{debosscher2007,debosscher2009}. This method was also applied to the CoRoT observations of field LRa02, and the results are meanwhile -- together with the full light curves -- publicly available through the CoRoT archive.\footnote{CoRoT data are available to the community from the CoRoT archive: http://idoc-corot.ias.u-psud.fr/.}

The CoRoT LRa02 data set contains 11{,}448 targets, from which 10{,}392 (91\%) are matching a BEST~II target within a maximum angular distance of $1''$. Because LRa02 was not covered completely by BEST~II (Figure~1 in Paper~I), 454 CoRoT targets are located outside of the BEST~II FOV. Furthermore, the magnitude ranges do not overlap completely, so that 425 bright CoRoT stars are saturated on the BEST~II CCD. The remaining 177 CoRoT targets are within the FOV and right magnitude range, but have no BEST~II counterpart due to technical issues such as blooming. In the same way, BEST~II observed a total of 93{,}943 stars in both pointings that have not been given a CoRoT mask or are located outside the CoRoT FOV. 

\begin{figure*}[htpc]
\includegraphics[width=0.5\textwidth]{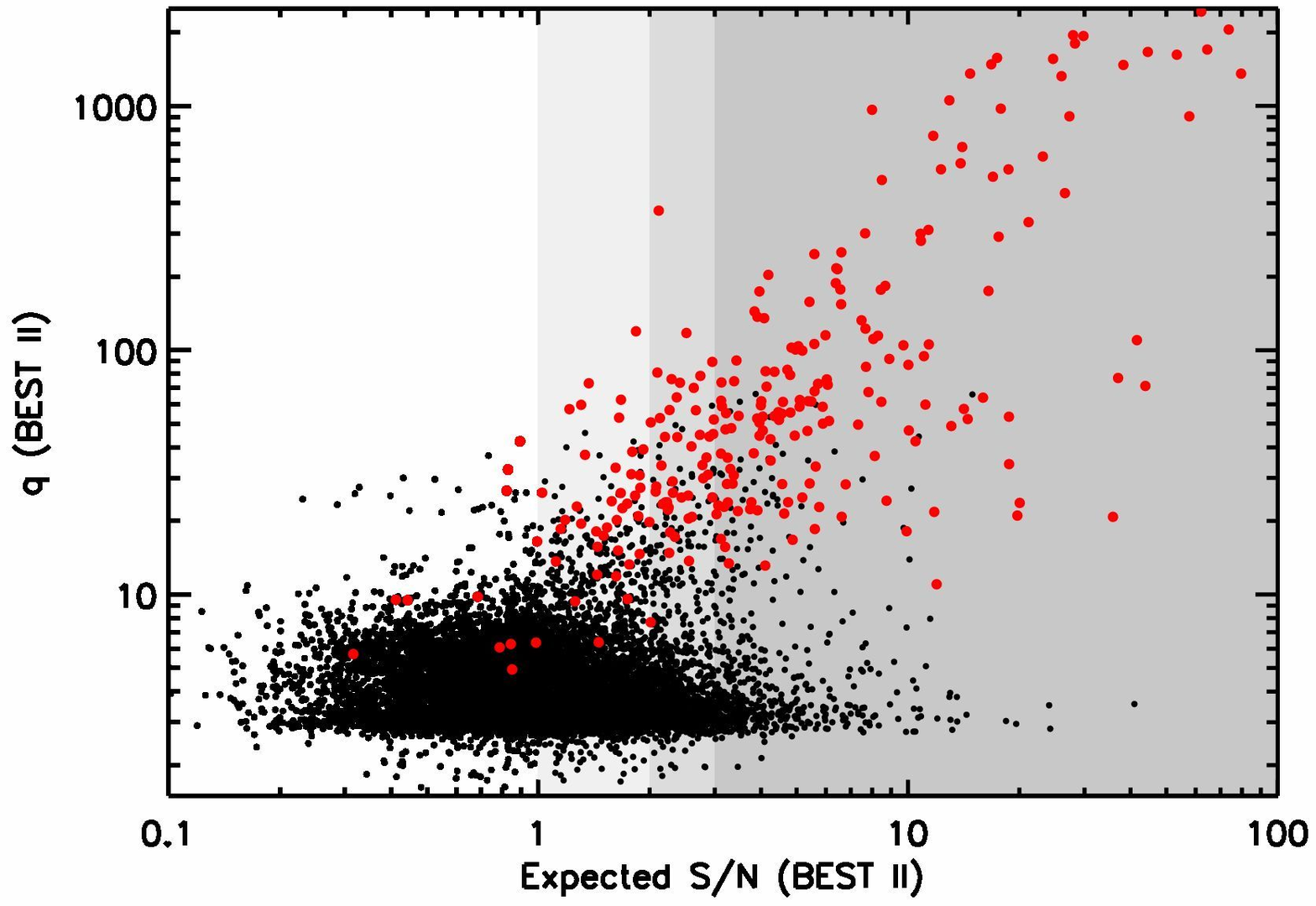}
\includegraphics[width=0.5\textwidth]{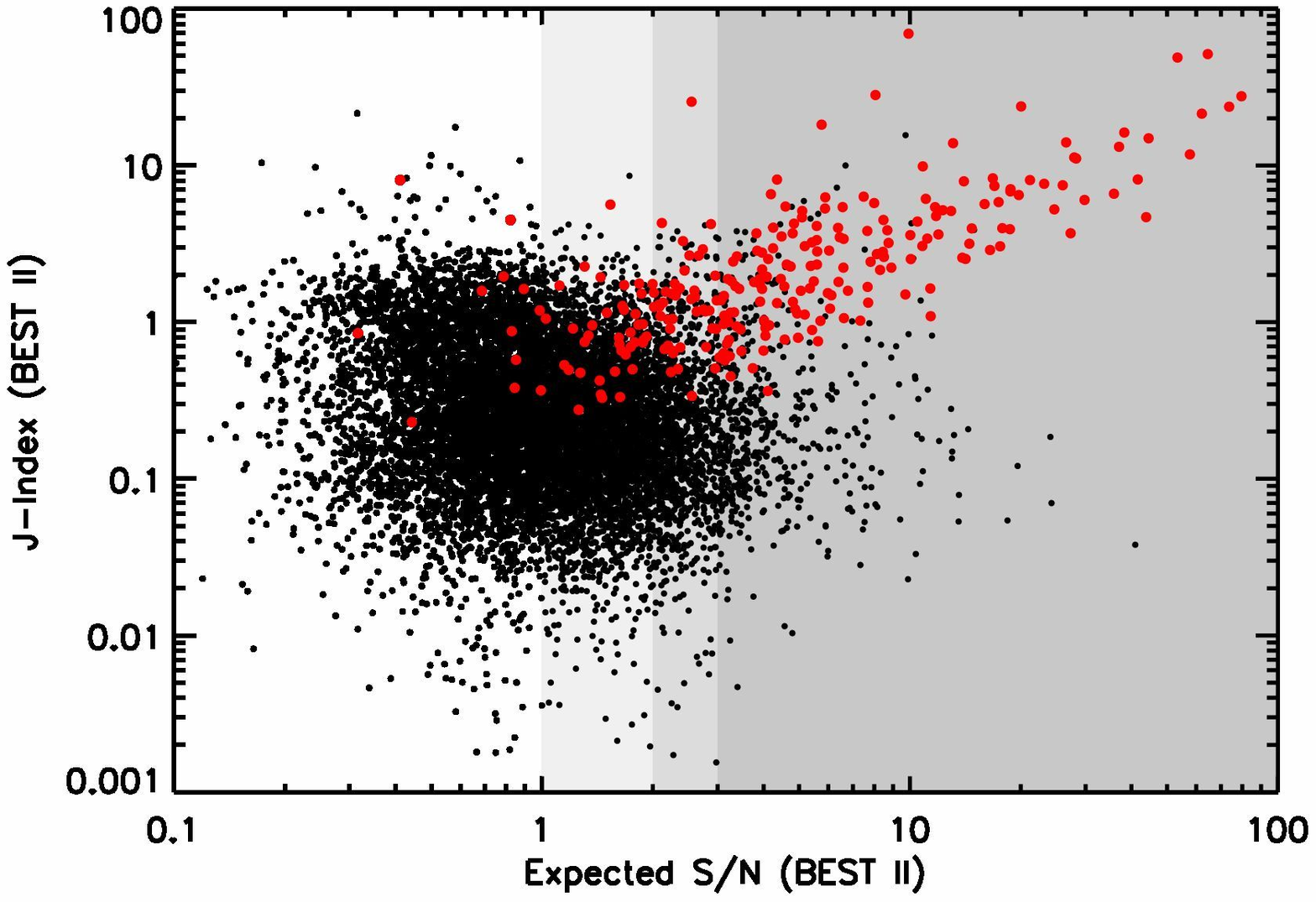}
\figcaption{\footnotesize Detection efficiency of BEST~II for the field LRa02. Shown are 10{,}392 stars that are measured by both BEST~II and CoRoT. The expected BEST~II S/N of CoRoT amplitudes is plotted on the $x$-axis (Equation~(\ref{eq:snrcorotbest})). On the $y$-axis, the left plot shows the suggested new quantity for variability ranking $q=\delta\Theta(\omega\subscript{max}^{(2)})$ (Equation~(\ref{eq:q12})) and the right plot the Stetson $J$~index for comparison. Variable stars detected in Paper~I or this work are marked red. The area of possible BEST~II detections is indicated in gray.
\label{fig:detefficiency}}
\end{figure*}

From the 681 variable stars presented together in Paper~I and this work, 262 variables (190 from Paper~I, 72 from this work) match a CoRoT target. The corresponding CoRoT IDs are shown for the new detections in Tables~\ref{tab:vars1} and~4. 

CoRoT observed the LRa02 field about 1 year after BEST~II, and the observing times do not overlap. Because this work focuses on the improvement of variable star detection for ground-based telescopes, CoRoT and BEST~II light curves have not been combined. However, we note that such a combination might yield improved ephemerides. BEST~II light curves of interesting objects will be provided on request.

\subsubsection{Detection Efficiency of BEST~II and the New Search Algorithm}
For stars both observed with CoRoT and BEST~II, it is possible to investigate the performance of BEST~II more in detail. The automatic classification \citep{debosscher2007} provides information about the amplitude of variation for CoRoT targets. Because the satellite has a much higher photometric precision, most of these amplitudes are well below the detection limit of BEST~II. However, the knowledge of stellar variability with higher precision can be used to evaluate the detection efficiency of BEST~II. Of particular interest are two questions: first, how many stars with sufficiently high variation in the CoRoT data set have been detected as variables from BEST~II data. Second, if these are clearly distinguished by the new detection algorithm from the rest of stars having variabilities below the threshold of BEST~II.

The full amplitude $A_C$ of variation was derived from the fit coefficients given in the CoRoT classification. As this contains each a low- and high-frequency entry for every CoRoT target, the amplitude was calculated as the maximum of both. The value $A_C$ itself contains no information on whether the signal can be detected by BEST~II or not, which strongly depends on the magnitude of a given star. Therefore, the quantity
\begin{equation}\label{eq:snrcorotbest}
  \mbox{S/N}=A_C/\sigma_B\superscript{min}(R_B)
\end{equation}
is used to estimate the variability signal to noise (S/N). The noise $\sigma_B\superscript{min}(R_B)$ gives the photometric precision achievable with BEST~II in the given data set for a star of magnitude $R_B$. It was determined by a fit to the $\sigma$-magnitude plot of the field (Figure~\ref{fig:rmsplot}).

Figure~\ref{fig:detefficiency} shows the expected S/N for all stars that are contained in both data sets. From 680 stars with CoRoT amplitudes that should be visible in the BEST~II data ($\mbox{S/N}>3$), 162 were detected as variable stars in Paper~I and this work. 448 stars are expected to have a large $\mbox{S/N}>3$, but show no significant variability in the BEST~II data set ($q<9$). 70 stars with both expected and measured variability ($\mbox{S/N}>3$, $q>9$) were not detected by BEST~II.

We have checked CoRoT light curves with significant variability in the CoRoT classification, but that are not variable in the BEST~II data. Many of them show strong instrumental effects \citep[hot pixels;][]{Auvergne2009} which obviously mislead the automatic classification algorithm. However, from the 448 stars in this region, only 88 show a probability larger than 95\% to belong to any class so that most can be considered false alarms.

Stars that are expected to be variable ($\mbox{S/N}>3$) and show variation in the BEST~II light curve ($q>9$), but were not detected as variable stars after visual inspection have been re-inspected. Such targets have been missed due to the smaller phase coverage of BEST~II, or because the real noise of individual light curves is underestimated using $\sigma_B\superscript{min}$ (e.g., due to higher order extinction effects for very red stars).

The distinction between variable and non-variable stars works very well using the new ranking quantity $q=\delta\Theta(\omega\subscript{max}^{(2)})$ (see Section~\ref{sec:limitimprove}). If the limit $q=9$ is chosen to separate variables from the bulk of non-variable stars, 92\% of all matched stars are found below the limit. Only seven variables have $q<9$, but their light curves and low S/N indicate rather false detections than too low variability values. From all 780 stars with $q>9$, one third belongs to the set of variable star detections. For comparison, Figure~\ref{fig:detefficiency} also shows the $J$ index vs.~the expected S/N. The plot shows clearly that the separation between real and artifical variability is much weaker. The strength of the new ranking is particularly clear in the regime of $1<\mbox{S/N}<3$, i.e., close to the detection limit of BEST~II.

\subsubsection{Comparison of Classifications}
\begin{figure*}[htpc]
\includegraphics[width=0.3\textwidth]{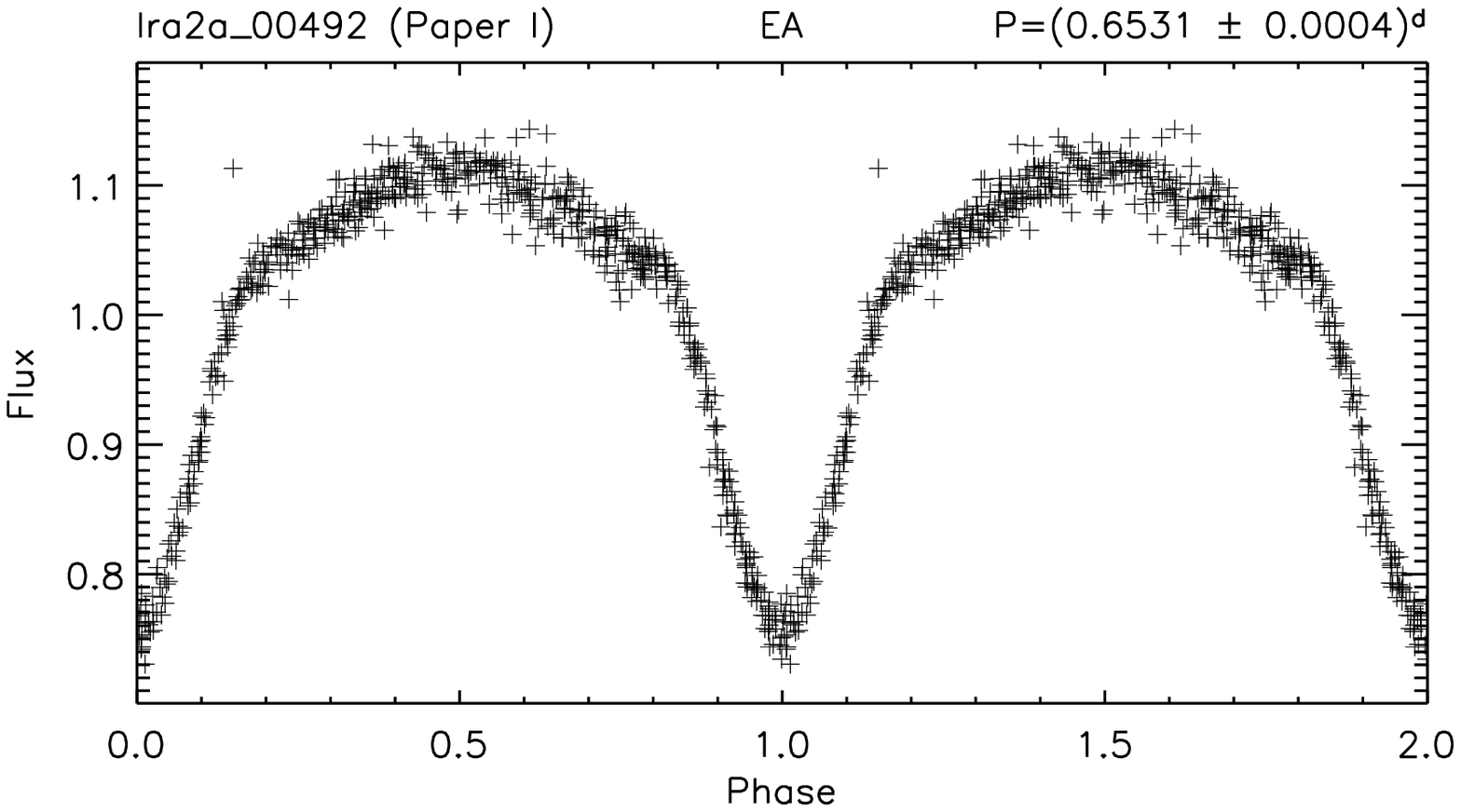}
\includegraphics[width=0.3\textwidth]{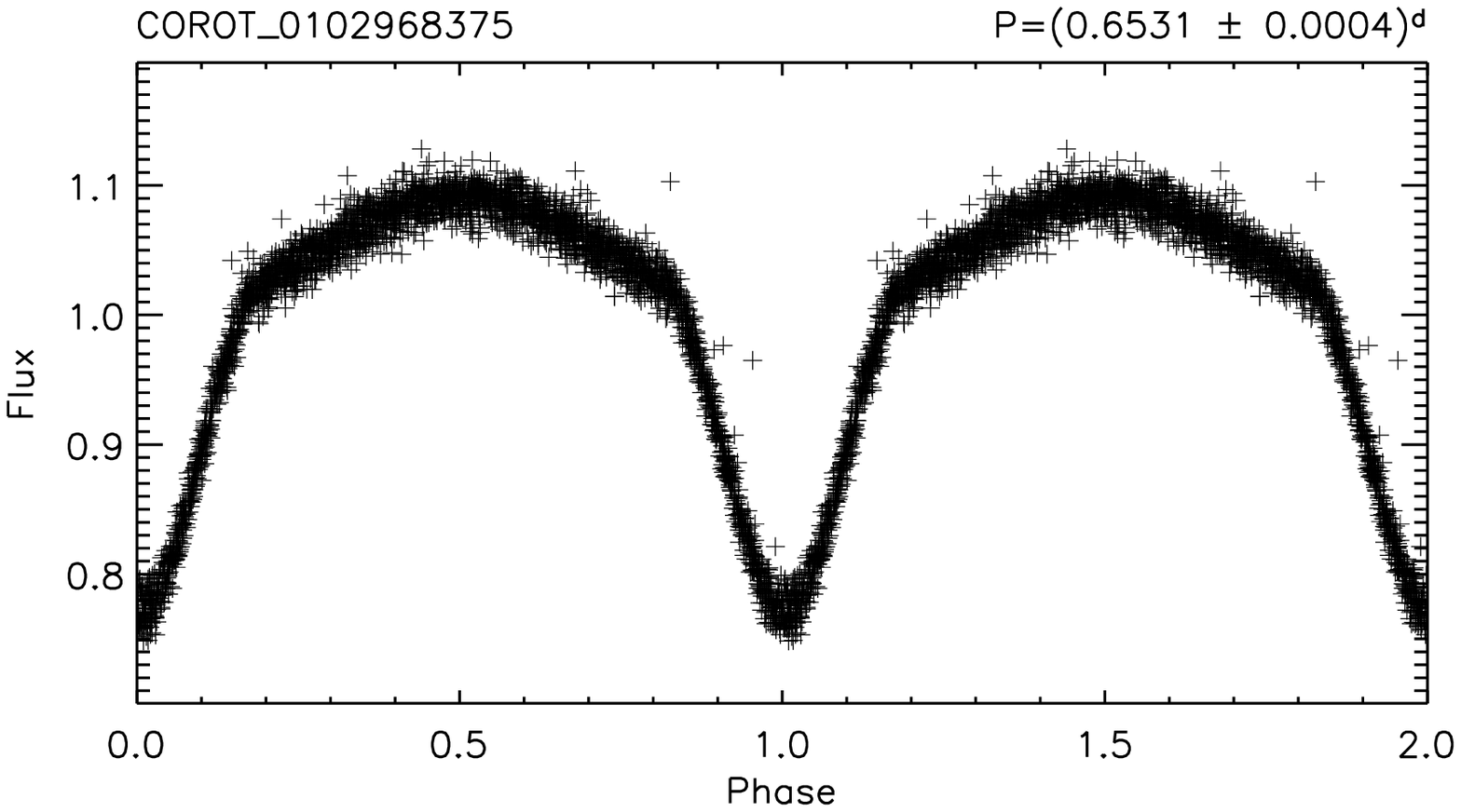}
\includegraphics[width=0.3\textwidth]{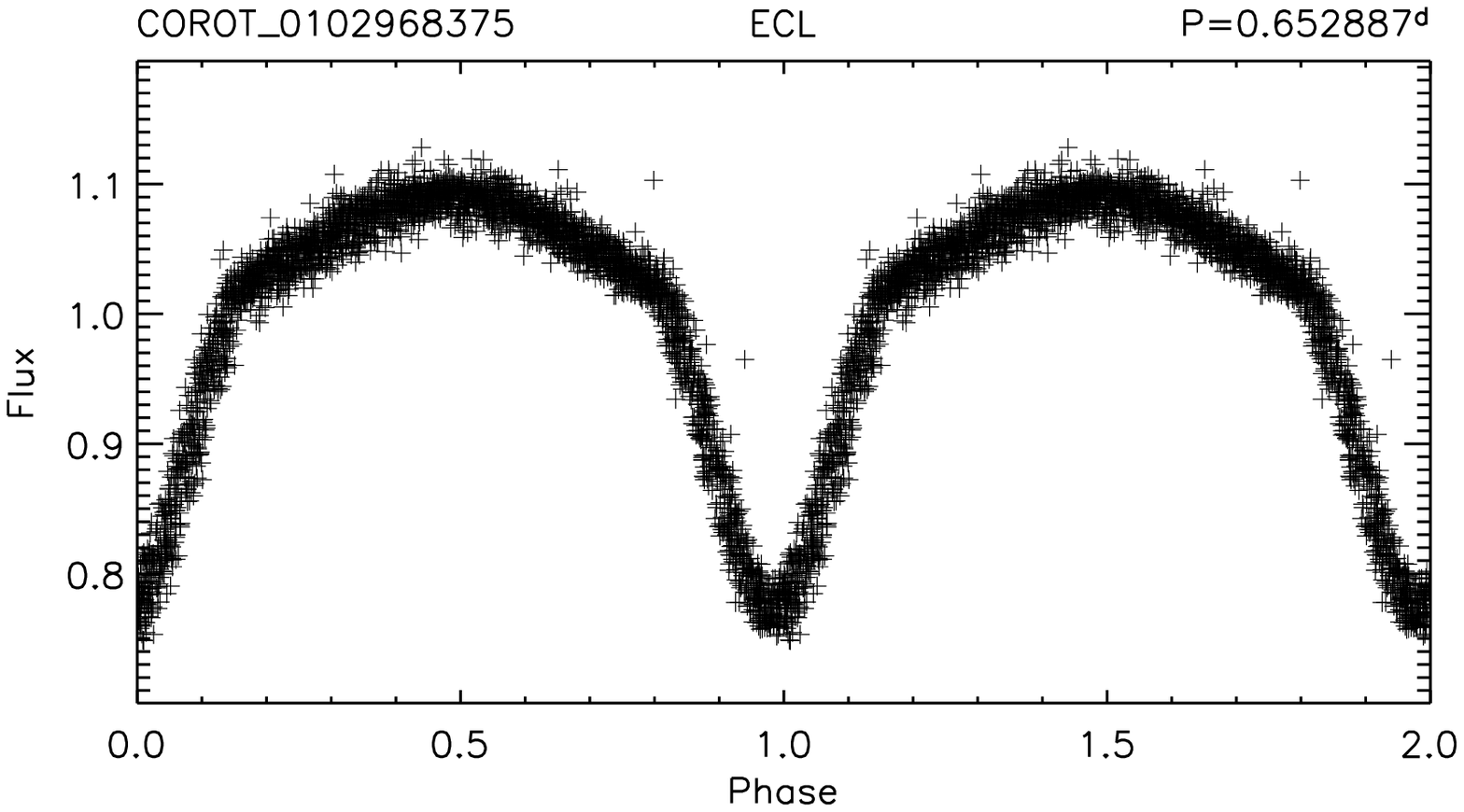}
(a)
\includegraphics[width=0.3\textwidth]{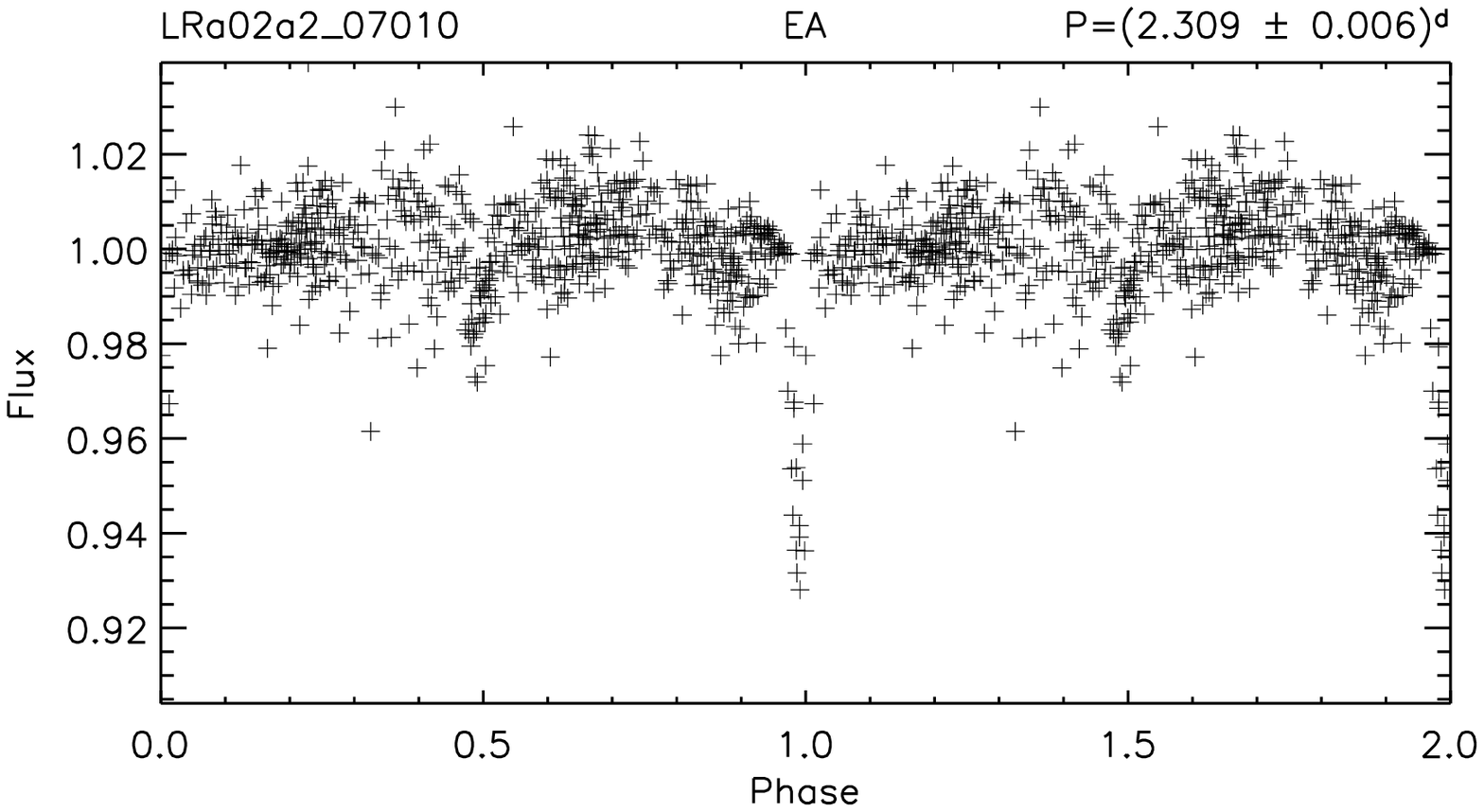}
\includegraphics[width=0.3\textwidth]{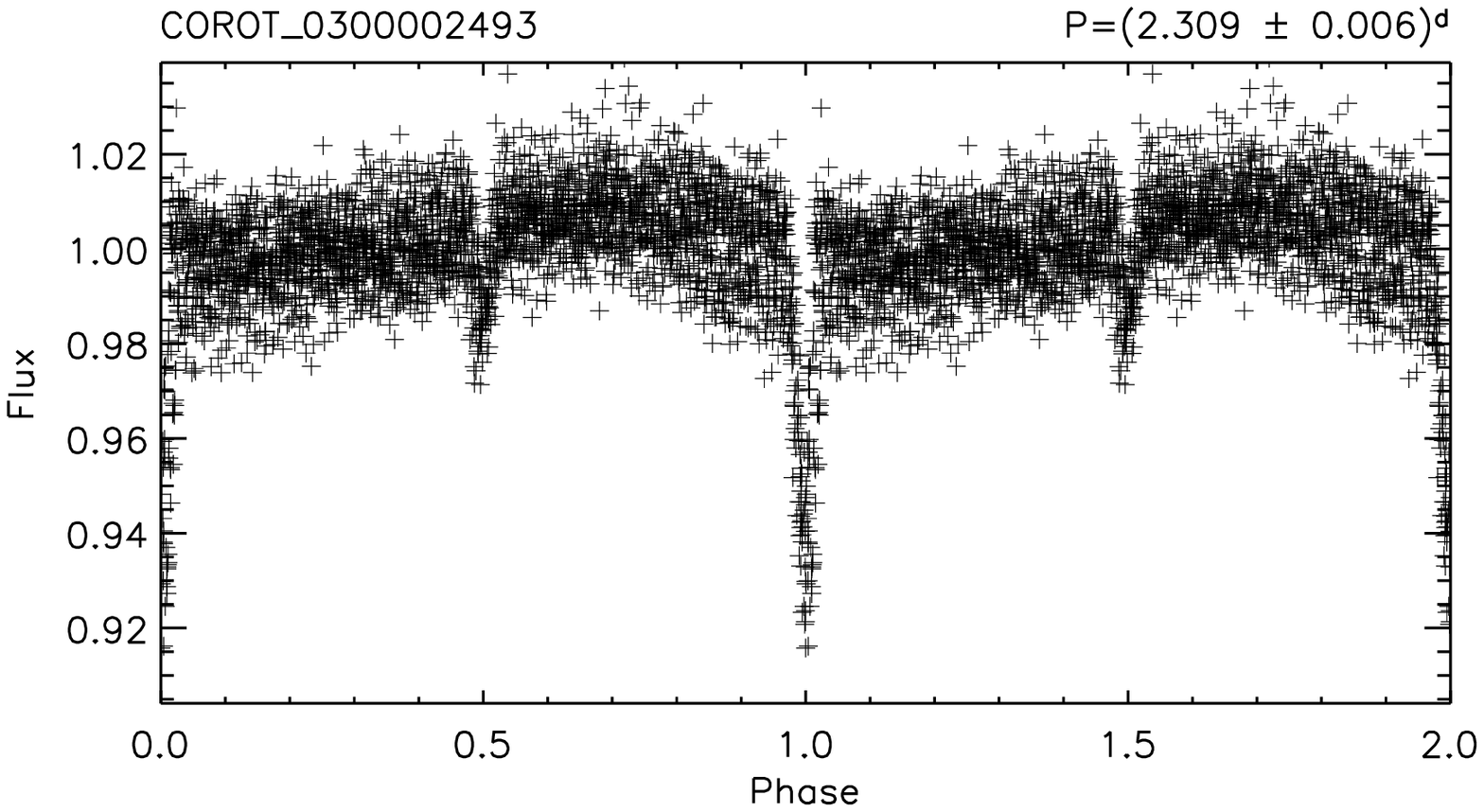}
\includegraphics[width=0.3\textwidth]{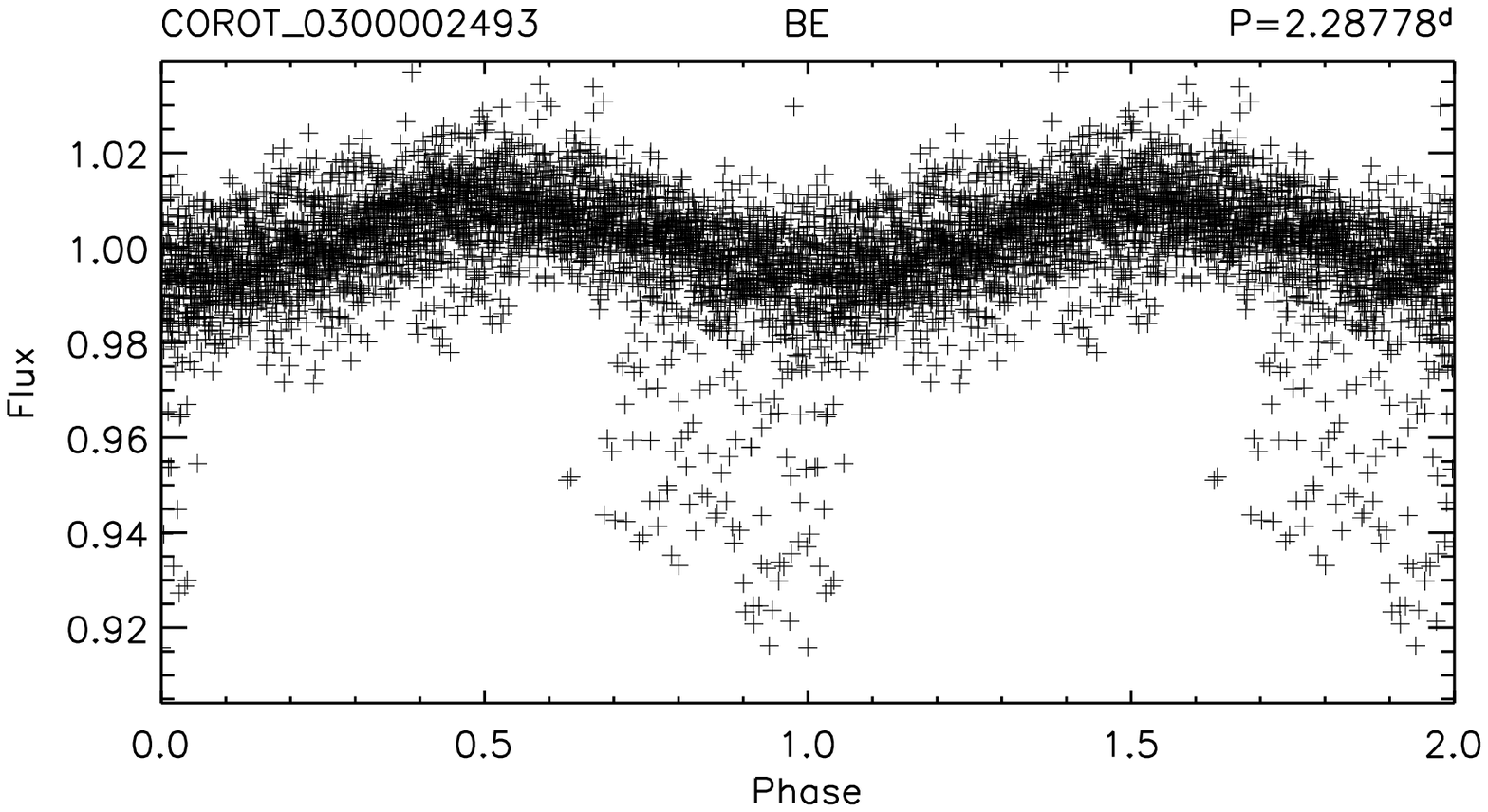}
(b)
\includegraphics[width=0.3\textwidth]{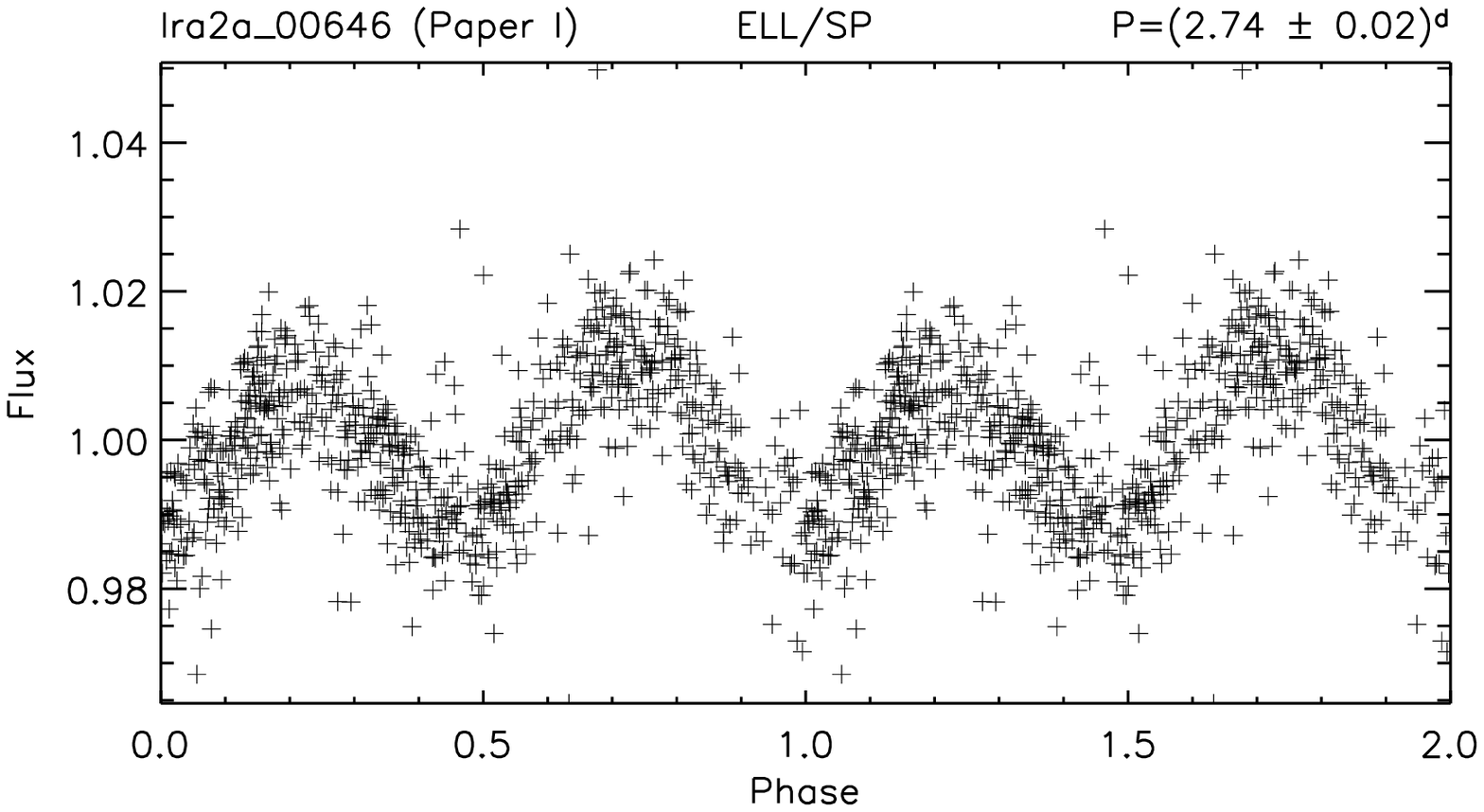}
\includegraphics[width=0.3\textwidth]{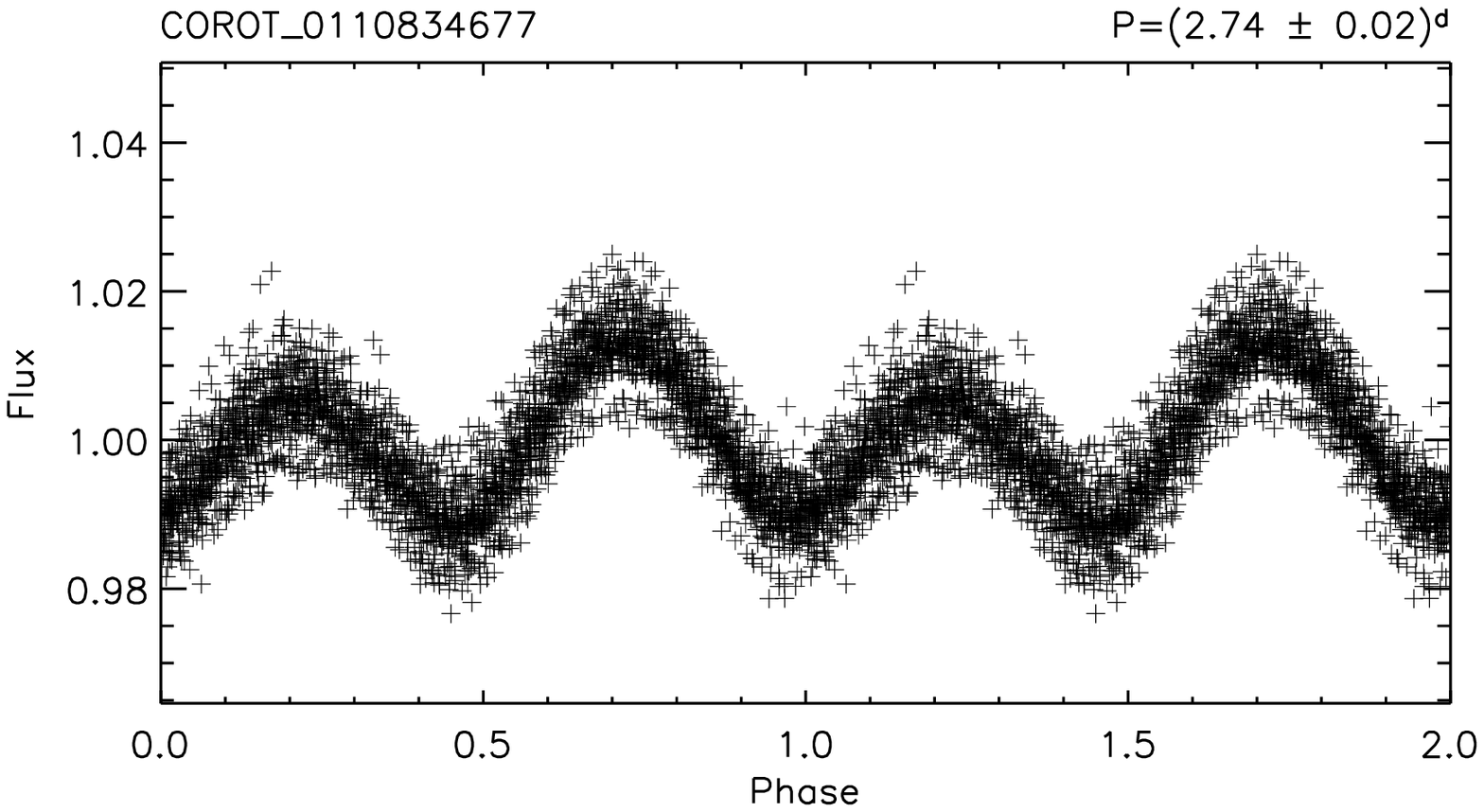}
\includegraphics[width=0.3\textwidth]{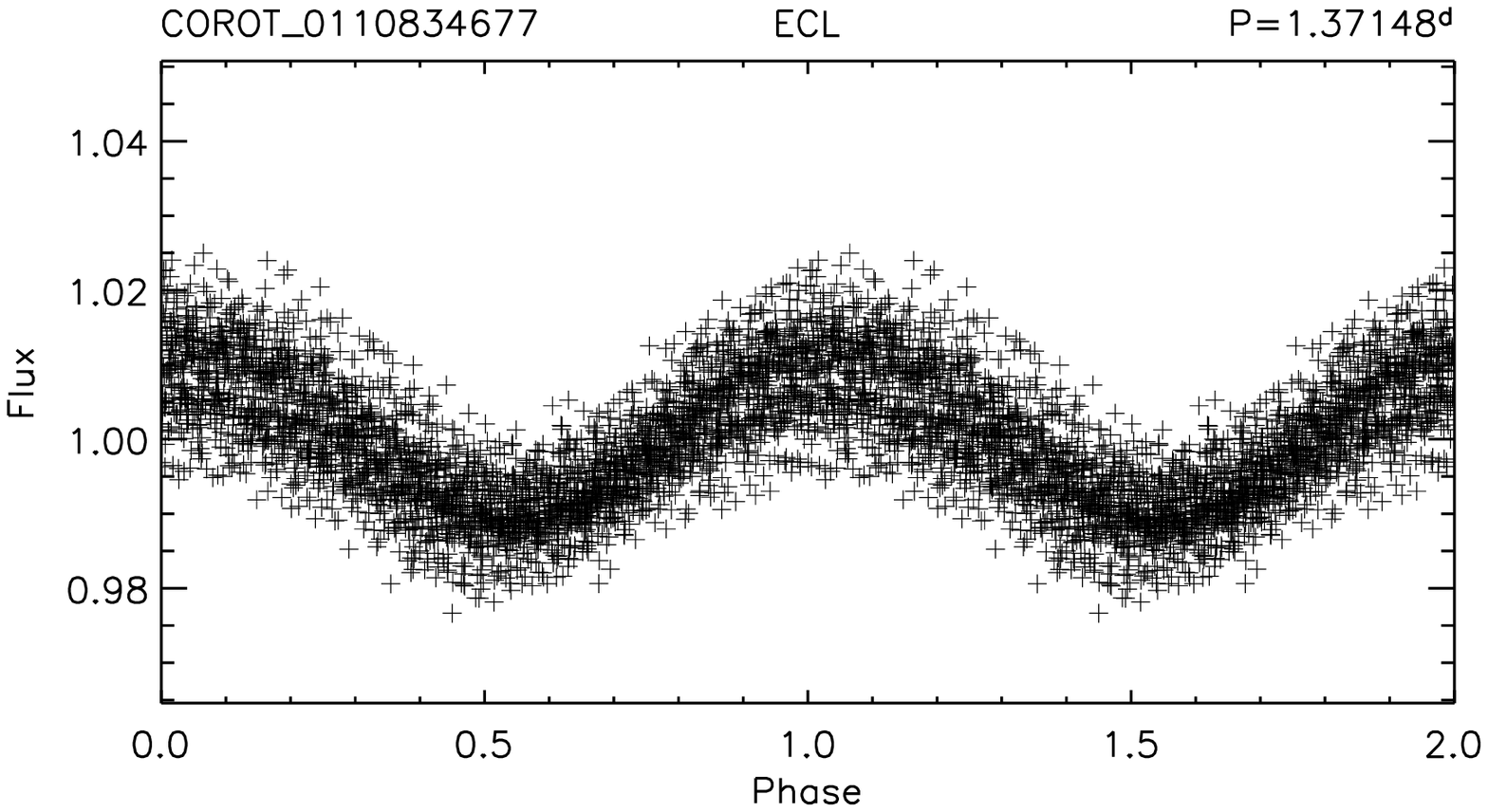}
(c)
\includegraphics[width=0.3\textwidth]{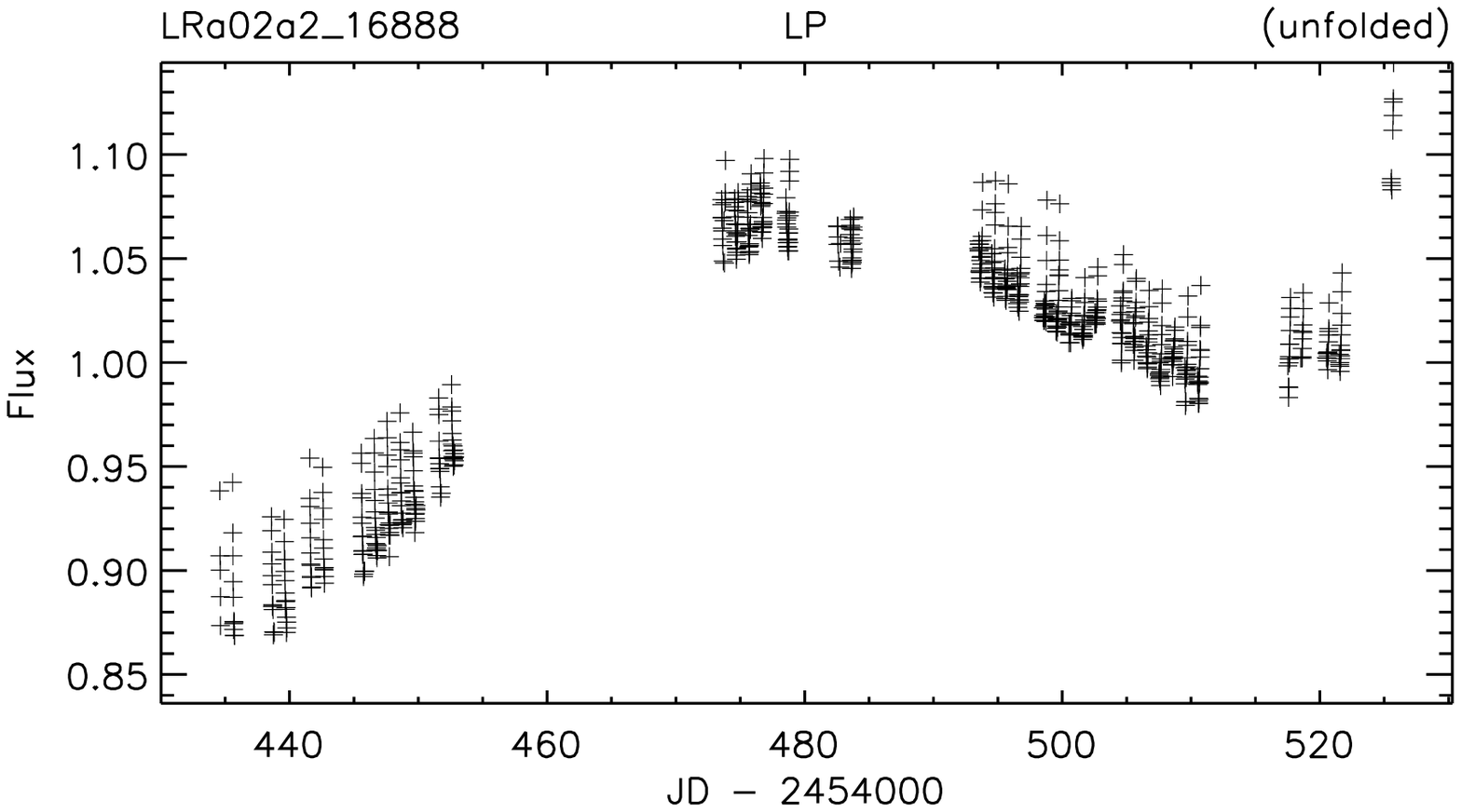}
\includegraphics[width=0.3\textwidth]{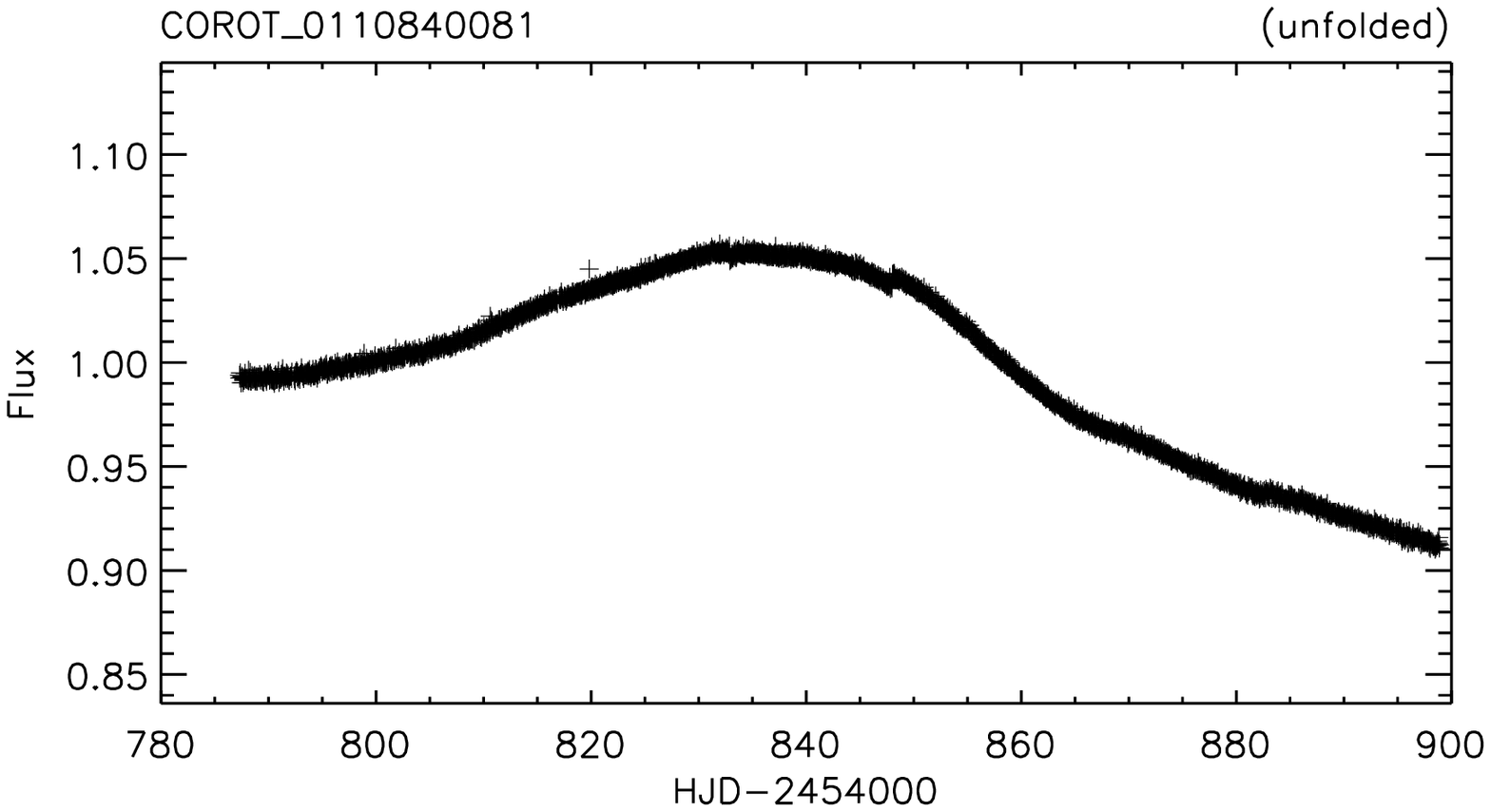}
\includegraphics[width=0.3\textwidth]{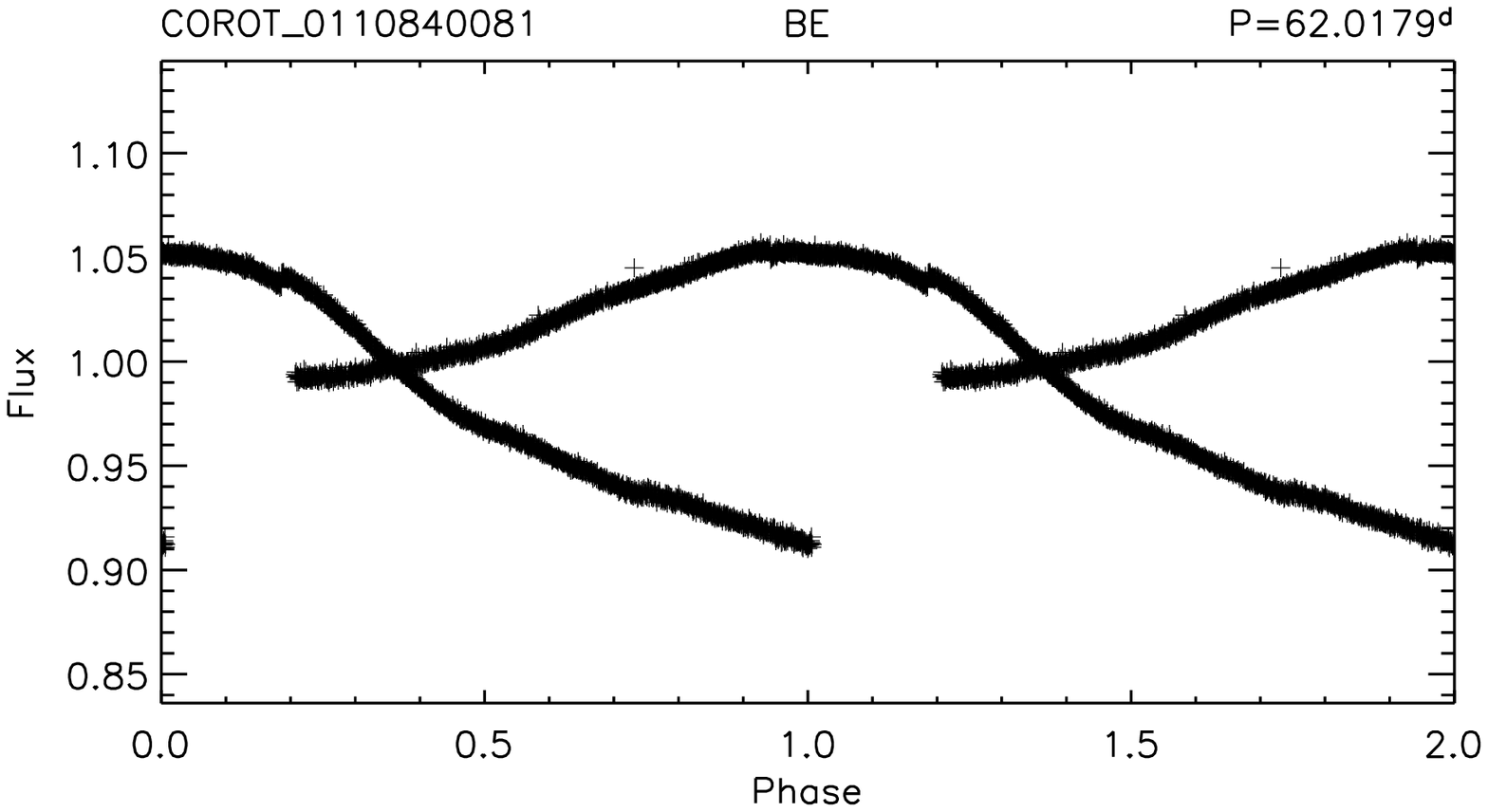}
(d)
\includegraphics[width=0.3\textwidth]{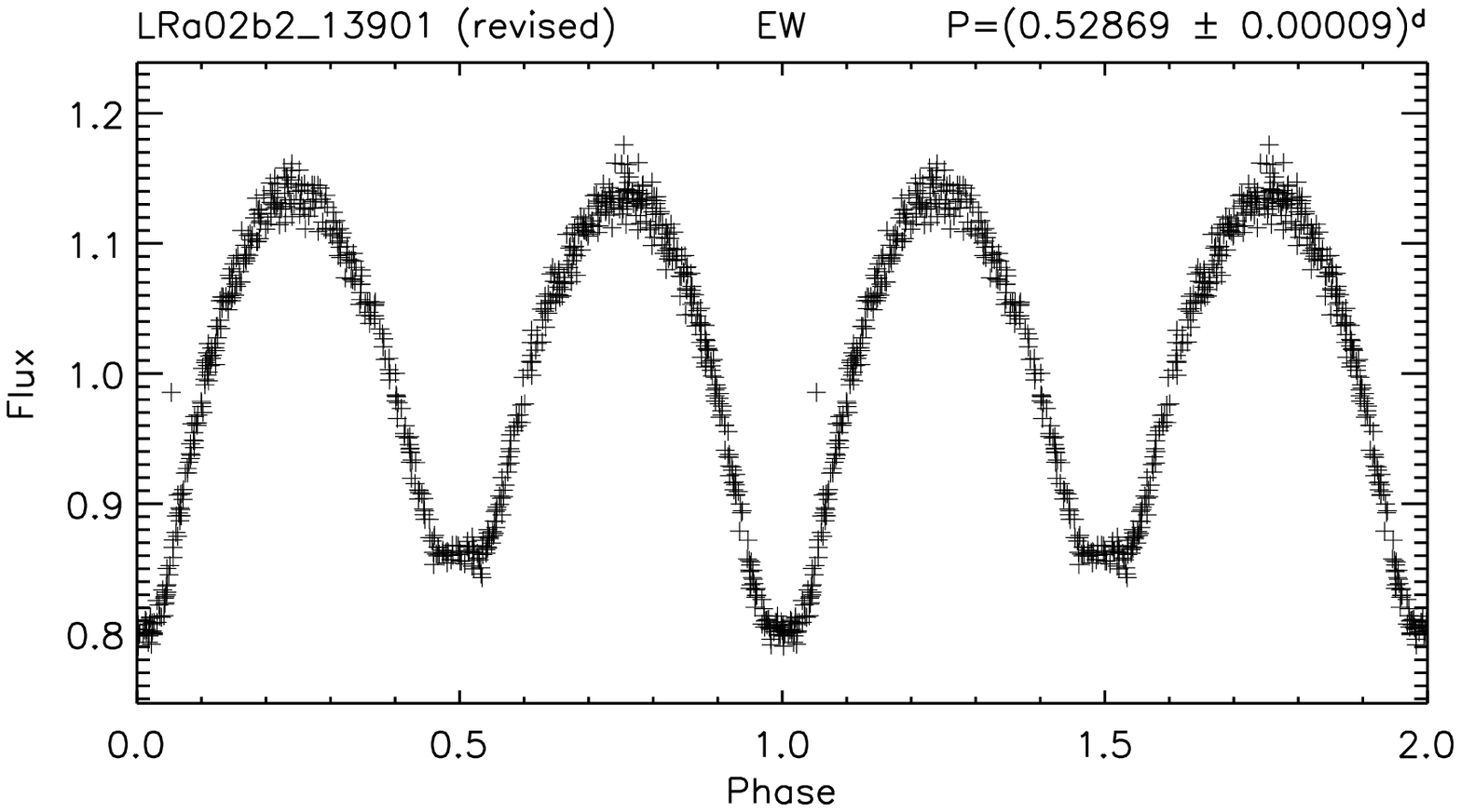}
\includegraphics[width=0.3\textwidth]{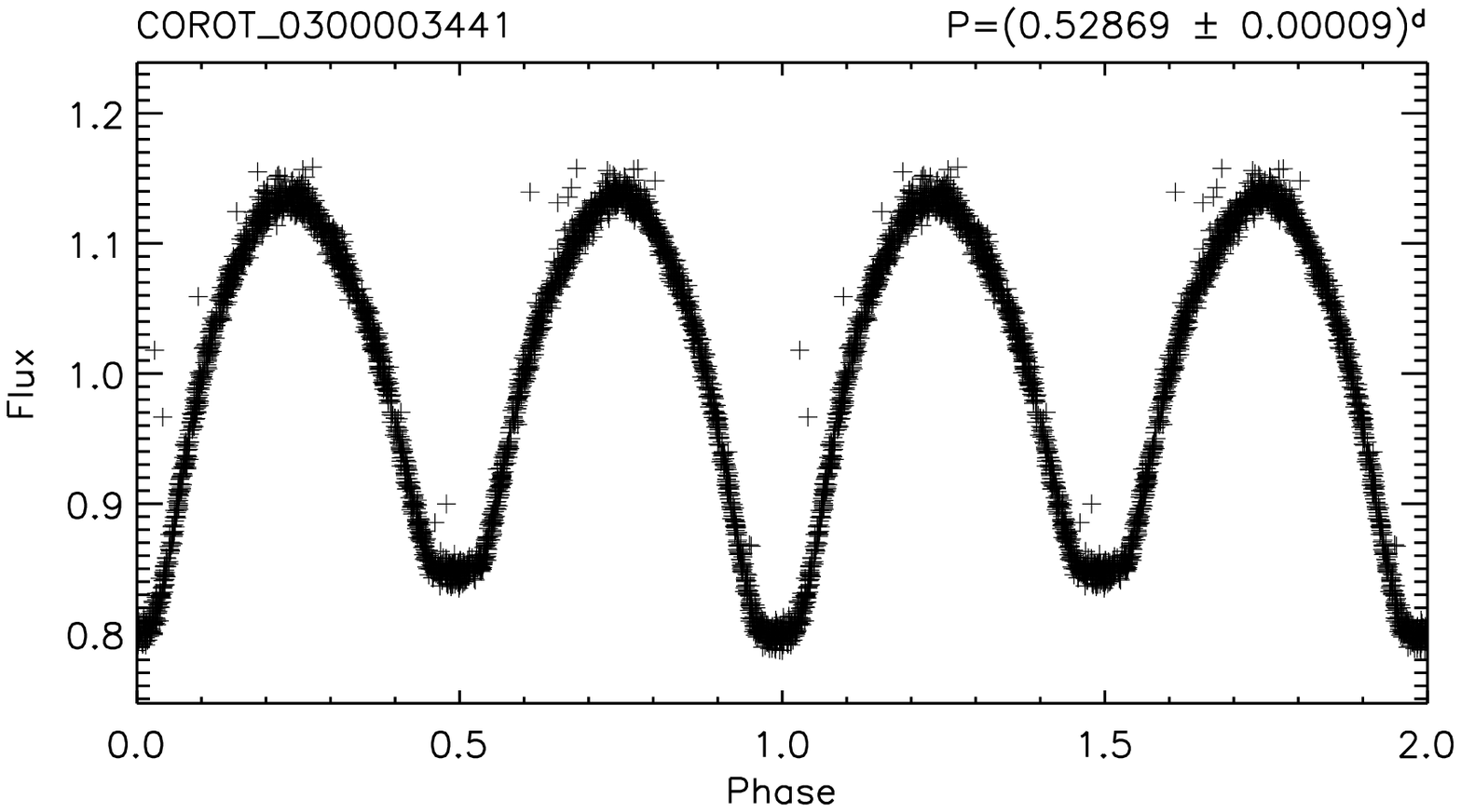}
\includegraphics[width=0.3\textwidth]{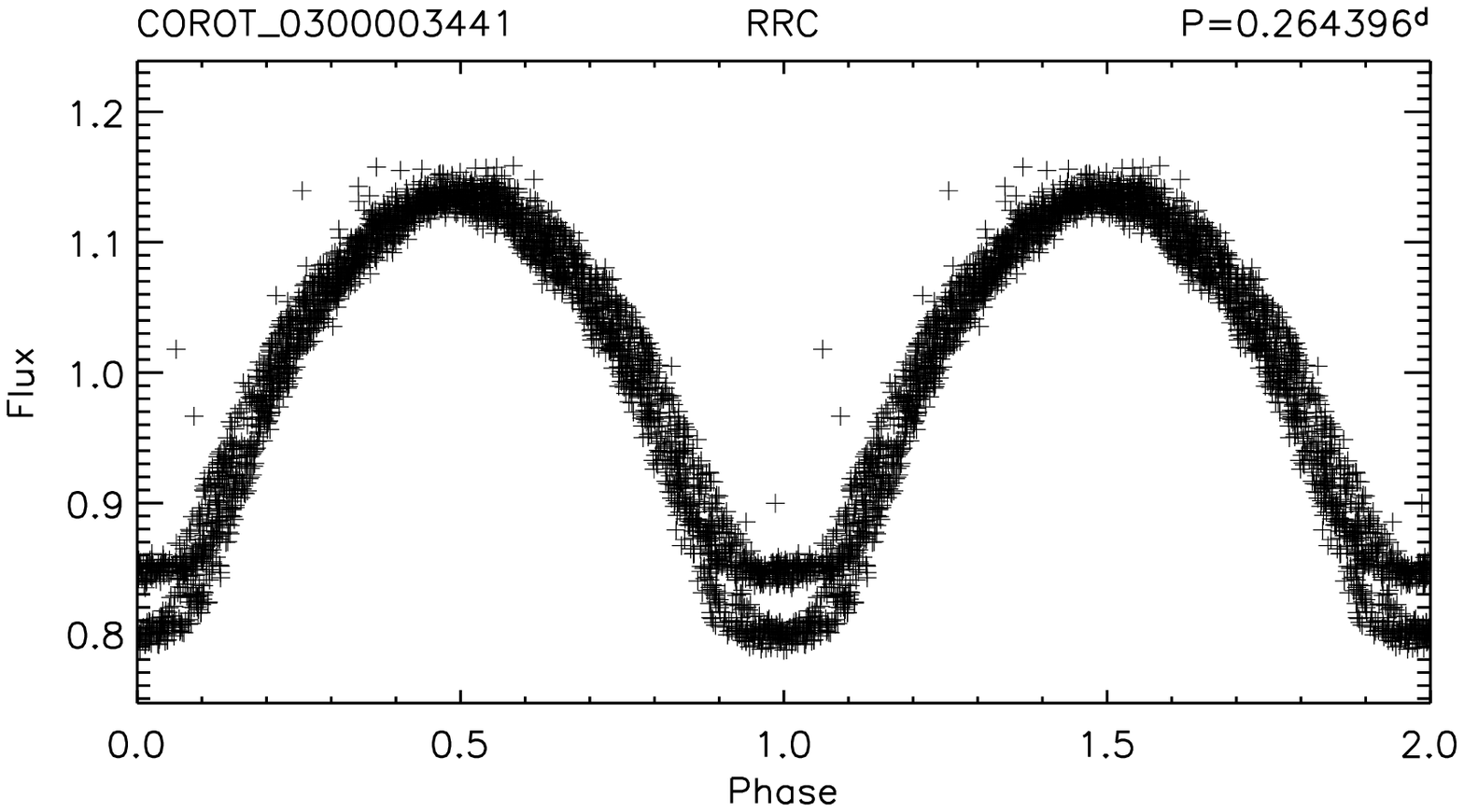}
(e)
\includegraphics[width=0.3\textwidth]{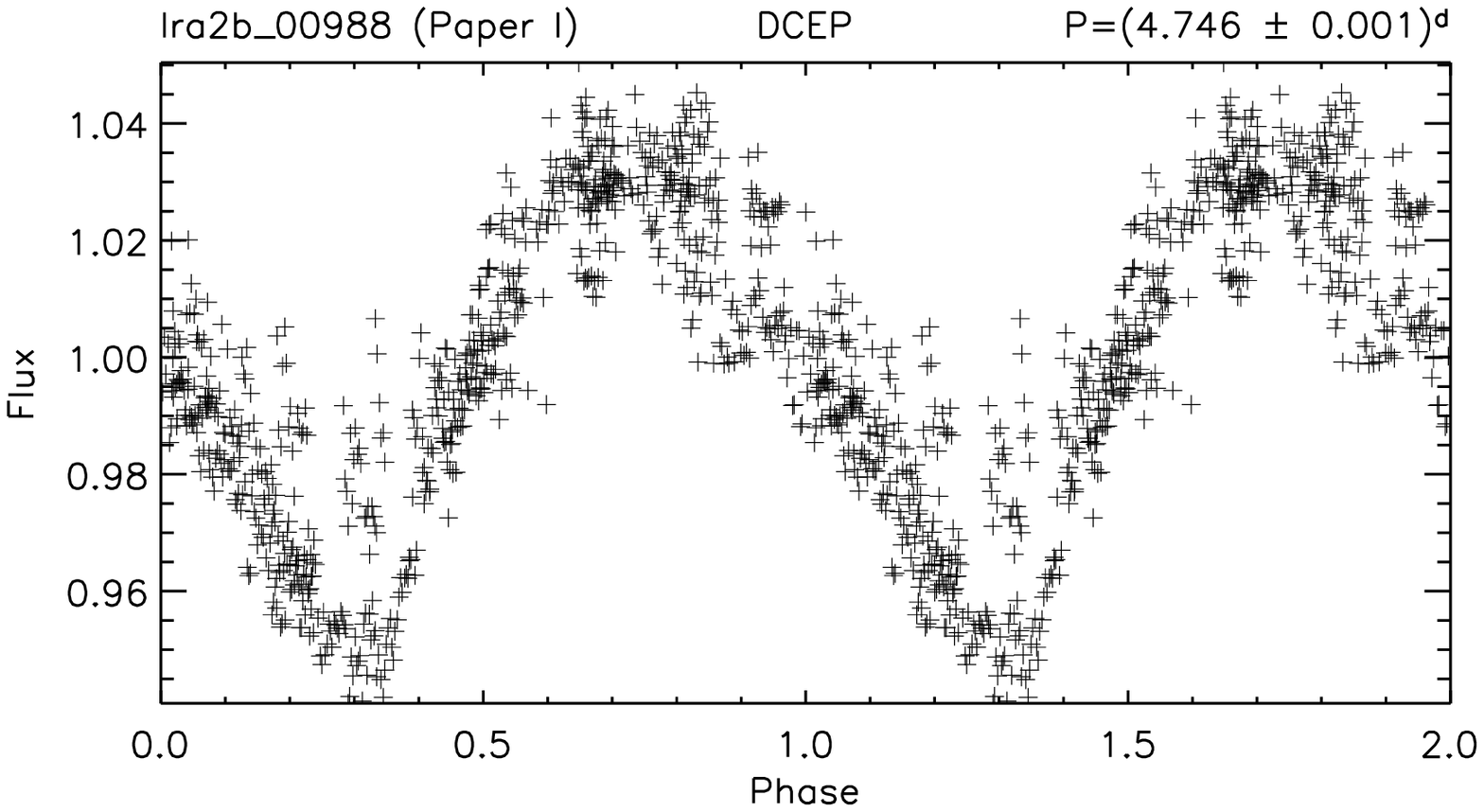}
\includegraphics[width=0.3\textwidth]{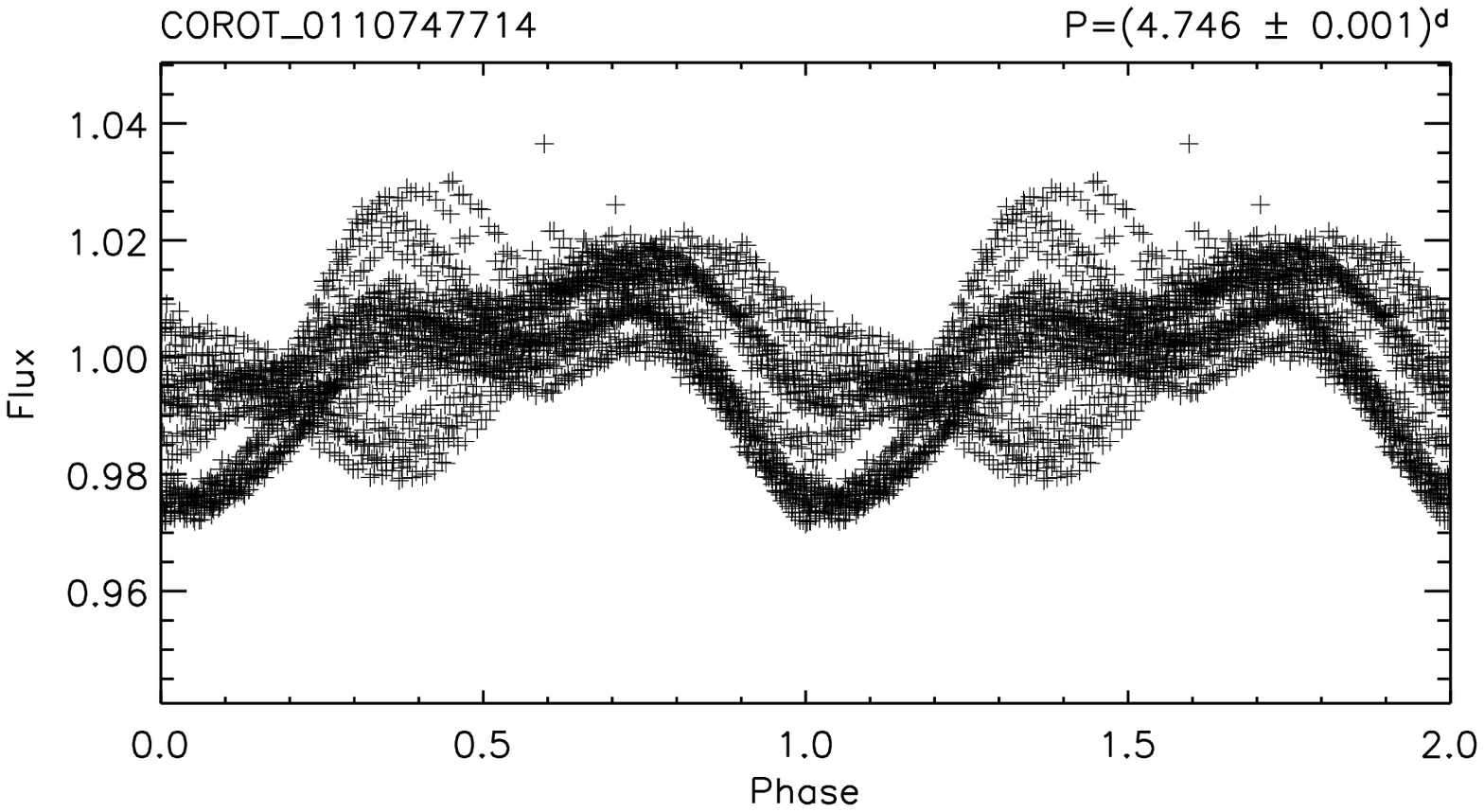}
\includegraphics[width=0.3\textwidth]{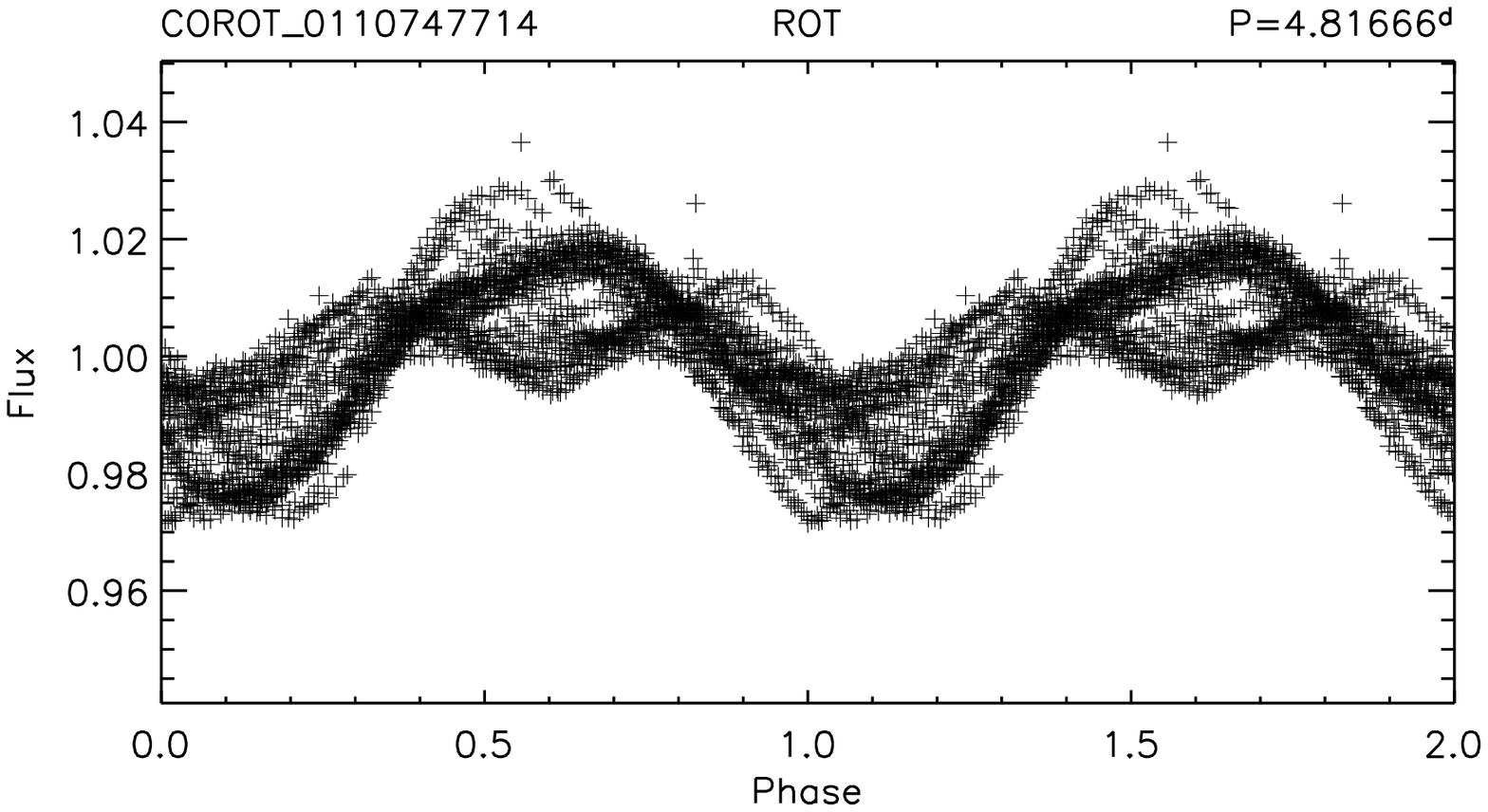}
(f)
\figcaption{\footnotesize Examples for variable stars identified both by BEST~II and CoRoT. Each row shows light curves of the same star: in the first column, the BEST~II light curve is folded with the period from BEST~II; in the second column, the CoRoT light curve folded with the BEST~II period; and in the third column, the CoRoT light curve folded with the CoRoT period. Unfolded light curves are shown for long periodic variables.\label{fig:lcs:corot}}
\end{figure*}

The variable star detections of BEST~II were compared in detail with the automatic CoRoT classification for the 262 matched variables. The overall agreement between both methods is very good; details regarding the period and magnitude determination as well as the classifications obtained by BEST~II and CoRoT are given in the remainder of this section. Figure~\ref{fig:lcs:corot} shows some instructive examples of variable stars in agreement (a) and with differences in the determined periods and/or classifications ((b)--(f)).

\begin{figure}[t]\centering
  \includegraphics[width=\linewidth]{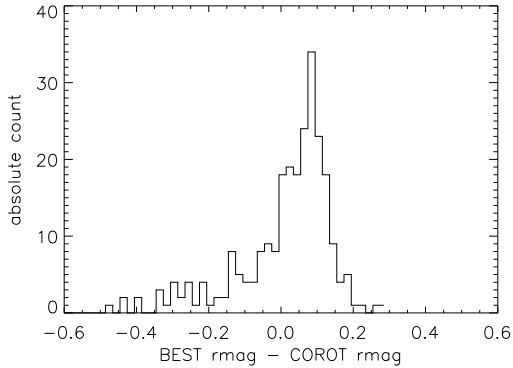}
  \caption{\footnotesize Histogram of differences between CoRoT and BEST~II magnitudes (for matched detections, without very long periodic variables).}
  \label{fig:bestcorot:mag}
\end{figure}

The mean magnitudes of matched stars are in reasonably good agreement (Figure~\ref{fig:bestcorot:mag}). Only a few very long-term variables -- like the example in Figure~\ref{fig:lcs:corot}(d) -- show differences in the order of 1~mag and above because BEST~II and CoRoT observed during different phases; the remaining majority differs by only $(0.059\pm0.158)$ mag.

\begin{figure}[t]\centering 
  \includegraphics[width=\linewidth]{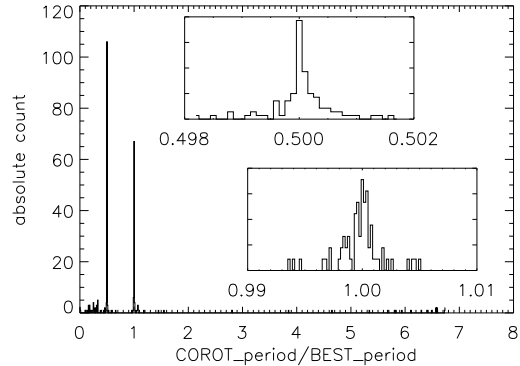}
  \caption{\footnotesize Histogram of CoRoT and BEST~II period ratio for matched variable star detections. The insets show the two main peaks enlarged (normalized).}
  \label{fig:bestcorot:per}
\end{figure}

Periods determined by BEST~II have been compared with the main frequency from the automatic CoRoT characterization. For 72.5\% of the matched stars, the periods are equal or integral ($n=1,\ldots, 5$) multiples of each other to a precision of at least 1\%. The histogram of period ratios in Figure~\ref{fig:bestcorot:per} shows that most detections have been identified in the CoRoT data with half the period compared to BEST~II. This is because many periods are doubled during the visual inspection of BEST~II light curves in order to show full cycles (e.g., Figures~\ref{fig:lcs:corot}(c) and~\ref{fig:lcs:corot}(e)), in particular for W~Ursae Majoris eclipsing binaries (EW). 

The classes from visual inspection of BEST~II light curves match the automatic classification of the CoRoT data set well. From the 262 variables present in both data sets, 196 stars have BEST~II variability classes that are consistent with either the short- or long-periodic classification in the CoRoT data set. Note that the variability classes used by BEST~II and CoRoT are slightly different: for example, BEST~II distinguishes within the CoRoT class ECL between the eclipsing binary types EA, EB, and EW, while the CoRoT scheme includes, e.g., slowly pulsating B-stars, which are simply identified as VAR within the BEST~II study. All such consistent refinements are considered a classification match. 


The 64 stars with a clear disagreement in the variability classifications were checked carefully by reviewing the light curves from both BEST~II and CoRoT. For 26 stars, the variability classes obtained by BEST~II appear more realistic. Most of such cases are LP variable stars (e.g., Figure~\ref{fig:lcs:corot}(d)) that have been identified as such in the BEST~II data set by visual inspection. For these cases, even the longer CoRoT baseline does not cover a full cycle, so that the period and classification obtained from the CoRoT pipeline are not conclusive. Furthermore, some eclipsing binaries are clearly misclassified by the automatic CoRoT procedure. Figure~\ref{fig:lcs:corot}(e) shows an example of a W~Ursae Majoris type eclipsing binary that was identified as an RR Lyrae pulsator -- most likely because it was detected with half of its physical period from the CoRoT data set. For three stars similar to the example in Figure~\ref{fig:lcs:corot}(b), narrow eclipse events were not detected by the CoRoT analysis. Most of the stars with implausible variability types were classified as BE by the automatic classification, which was described by \citet{debosscher2009} as a ``trash'' class regarding its wide parameter spread. For 16 stars like the example in Figure~\ref{fig:lcs:corot}(f), the CoRoT classifications are in better agreement with the measurements. For most of these cases, this clearly results from the better photometric quality of the satellite data. For 22 cases, the photometric data itself are insufficient to choose between the CoRoT and BEST~II classifications (e.g., EW/ELL). No automatic classification data were available for the two CoRoT targets 110833621 and 300001413.

\section{SUMMARY AND DISCUSSION}
The CoRoT target field LRa02 was observed with BEST~II during 41 nights from 2007 November to 2008 February (see Paper~I). We reanalyzed the data set in order to improve the detection method and to maximize the number of detections. In addition to the 350 periodic variable stars already published earlier (Paper~I), we present a catalog of 272 new variables and 52 stars with suspected variability. For seven known variables both the classifications and periods are confirmed, and the periods could significantly be improved for two of them. Revised ephemerides are presented for 17 variable stars from Paper~I.

From a manual inspection of all light curves without any preselection criterion, it turned out that most of the new periodic variable stars went undetected in Paper~I because the applied variability criterion using the Stetson $J$~index was too restrictive. But because the $J$~index is heavily biased by systematic effects, a smaller cutoff limit leads to very high false alarm rates ($\approx 98\%$) and is therefore not a practical alternative. However, although the $J$~index is not capable to distinguish between systematic and stellar variability, it can still be used to exclude non-variable stars from the analysis: no variable star is falsely rejected if light curves with low variability indices $J<0.1$ are sorted out in both data sets of this study. This separation can be particularly useful if the full AoV process is too time consuming for a whole data set.

The deep characterization of the data set enabled us to compare and train different automatic methods for an improved variability ranking. In particular, a master power spectrum was calculated as the mean of all individual AoV spectra. This method proved a valuable tool for exclusion of systematic frequencies and hence the ranking of real variability. The best algorithm found separates variable stars very effectively from the non-variable background population and in parallel recovers their frequencies well. The new ranking method is particularly superior to the $J$ index in regimes where the amplitude of variation becomes comparable to the noise level, i.e., close to the detection limit of the photometric system. It shows an almost equal performance for both independent subsets LRa02a and LRa02b, so that it should be easily applicable to other data sets.

Finally, the results from both Paper~I and this work have been compared with the publicly available CoRoT data set. BEST~II obtained light curves for 91\% of all CoRoT targets in LRa02. From all 681 variable stars observed with BEST~II in the field, 262 are matched with CoRoT targets. Stellar amplitudes measured by CoRoT show that BEST~II detects variability efficiently in its parameter range. The results of period determination and variability classification are found to be in very good overall agreement, confirming the validity of the measurements, data reduction, and scientific analysis of both projects. 196 stars show consistent classifications. CoRoT yields a better classification in case of small variability for 16 targets, e.g., Be stars. On the other hand, the final visual screening of folded BEST~II light curves proves to be an important step in the analysis. For 26 stars it was found to yield a more realistic classification, in particular for eclipsing binaries and stars with variability on time scales comparable to the observational baseline.

\acknowledgments
\textbf{Acknowledgments.} This work was funded by Deutsches Zentrum f\"ur Luft- und Raumfahrt and partly by the Nordrhein-Westf\"alische Akademie der Wissenschaften. We kindly thank Julian Petrasch for his great commitment of scanning all folded light curves. Furthermore, we appreciate valuable comments on the initial manuscript by the anonymous referee. Our research made use of the 2MASS, USNO-A2 and GCVS catalogs, the AAVSO variable star search index and the SIMBAD database, operated at CDS, Strasbourg, France.

\appendix
\section{Calculation of sorting parameter}\label{sec:xi}
Let $X_*=\{x_i\}$ with $i\in \{1,\ldots, N_*\}$ be the group of all stars in a data set. The selection of variable stars -- e.g., by sorting all stars by a ranking quantity $q$ -- can then be considered a permutation $\widetilde{p}_v: X_* \rightarrow X_*$ that splits $X_*$ into a part $X_v \subset X_*$ containing all $N_{v}$ variable stars and another part with the rest:
\begin{eqnarray*}
	&\widetilde{p}_v(x_i) \in X_v \ \ \ \mbox{for}\ \ \ i\leq N_v \\
  &\widetilde{p}_v(x_i) \notin X_v \ \ \ \mbox{for}\ \ \ i> N_v \ .
\end{eqnarray*}
A check of the first $N_{v}$ within $\{\widetilde{p}_v(x_i)\}$ would thus reveal all variable stars in the data set.

Unfortunately, such an \textit{optimal} sorting $\widetilde{p}_v$ is usually unknown. In practice, a given permutation $p_v$ aims at a similar
splitting of variable and non-variable stars, but contaminates both groups with false positives. The number of identified variable stars $N_v'\leq N_v$ thus depends on the number of stars $N_{c*}\leq N_*$ that are actually checked:

\begin{eqnarray*}
N_v'\left(p_v,\,N_{c*}\right) &=& \sum_{i=1}^{N_{c*}}{\delta_v (p_v(x_i))} \\
\mbox{with}\ \ \ \ \delta_v (x_i) &:=& 
  \begin{cases} 
    1 & x_i \in X_v \\ 
    0 & \mbox{otherwise} 
  \end{cases} \ .
\end{eqnarray*}

A given variable star selection $p_v$ can be compared directly with the optimal procedure $\widetilde{p}_v$. The number of missed variable stars is
$$\overline{N}_v'\left(p_v,\,N_{c*}\right) =  N_v'\left(\widetilde{p}_v,\,N_{c*}\right) - N_v'\left(p_v,\,N_{c*}\right),$$
whereby
$$N_v'\left(\widetilde{p}_v,\,N_{c*}\right) = \min(N_{c*},\ N_v).$$ 

For a comparison between different approaches, it is useful to evaluate the performance of $p_v$ as a whole. For that, 
one can define the quality parameter
$$\xi=\frac{  \sum_{N_{c*}=1}^{N_*}{\overline{N}_v'\left(p_v,\,N_{c*}\right)}}   { N_v(N_*-N_v)  } \ .$$
The parameter $\xi$ sums missed variable stars for all values of $N_{c*}$. The denominator accounts for normalization such that $\xi=0$ for $\widetilde{p}_v$ (best selection) and $\xi=1$
for its counterpart (worst selection).

\end{document}